\documentclass{aa}  
\usepackage{graphicx}
\usepackage{txfonts}
\usepackage[colorlinks=true,citecolor=blue]{hyperref}
\usepackage{newtxtext,newtxmath}
\usepackage[T1]{fontenc}
\usepackage{amsmath}
\usepackage{amssymb}
\usepackage{color}
\usepackage{hyperref}
\usepackage{pdflscape}
\usepackage{supertabular}
\usepackage[toc,page]{appendix}
\usepackage{multicol}

\usepackage{subfig}
\usepackage{soul}
\usepackage[dvipsnames]{xcolor}
\usepackage{placeins}

\newcommand{\orcid}[1]{\href{https://orcid.org/#1}{\includegraphics[width=10pt]{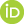}}}

\begin{document} 
\title{The ANDICAM-SOFI Near-infrared and Optical type Ia Supernova (ASNOS) sample: Description and data release}
\titlerunning{ASNOS I}
\authorrunning{Phan et al.}
\author{
Kim Phan\inst{1,2}\thanks{E-mail: phan@ice.csic.es}\orcid{0000-0002-1296-6887},
Llu\'is Galbany\inst{2,1}\orcid{0000-0002-1296-6887},
Tom\'as E. M\"uller-Bravo\inst{3,4}\orcid{0000-0003-3939-7167},
Subhash Bose\inst{5}\orcid{0000-0002-5571-1833},
Christopher R. Burns$^{6}$\orcid{0000-0002-5221-7557},
Maximilian D. Stritzinger\inst{5}\orcid{0000-0002-5571-1833},
Camilla T. G. Sørensen\inst{5}\orcid{0009-0002-4012-5329},
Chris Ashall$^{7}$\orcid{0000-0002-5221-7557},
Francisco J. Castander\inst{2,1}\orcid{0000-0001-7316-4573},\\
Cristina Jiménez Palau\inst{2,1}\orcid{0000-0002-4374-0661},
Joel Johansson$^{8}$\orcid{0000-0001-5975-290X},
Joseph P. Anderson$^{9}$\orcid{0000-0003-0227-3451},
Ken. C. Chambers$^{10}$\orcid{0000-0001-6965-7789},\\
Mariusz Gromadzki$^{11}$\orcid{0000-0002-1650-1518}, 
Priscila J. Pessi$^{8}$\orcid{0000-0002-8041-8559}, and
Ting-Wan Chen$^{12}$\orcid{0000-0002-1066-6098}
}
\institute{
Institut d’Estudis Espacials de Catalunya (IEEC), E-08034 Barcelona, Spain. 
\and
Institute of Space Sciences (ICE, CSIC), Campus UAB, Carrer de Can Magrans, s/n, E-08193 Barcelona, Spain.
\and
School of Physics, Trinity College Dublin, The University of Dublin, Dublin 2, Ireland
\and 
Instituto de Ciencias Exactas y Naturales (ICEN), Universidad Arturo Prat, Chile
\and
Department of Physics and Astronomy, Aarhus University, Ny Munkegade 120, DK-8000 Aarhus C, Denmark
\and
Observatories of the Carnegie Institution for Science, 813 Santa Barbara St, Pasadena, CA, 91101, USA
\and
Institute for Astronomy, University of Hawai‘i at Manoa, 2680 Woodlawn Drive, Hawai‘i, HI 96822, USA
\and
Oskar Klein Centre, Department of Physics, Stockholm University, AlbaNova, SE-10691 Stockholm, Sweden
\and
European Southern Observatory, Alonso de Crdova 3107, Casilla 19 Santiago, Chile
\and
Institute for Astronomy, University of Hawaii, 2680 Woodlawn Dr., Honolulu, HI 96825
\and
Astronomical Observatory, University of Warsaw, Al. Ujazdowskie 4,
00-478 Warszawa, Poland
\and
Graduate Institute of Astronomy, National Central University, 300 Jhongda Road, 32001 Jhongli, Taiwan
}
\date{Received \today; accepted XXX}

\abstract
{
Type Ia supernovae (SNe Ia) provide the most robust means of measuring extragalactic distances.
While most of the effort has focused on increasing the number of SNe Ia observed in the optical, near-infrared (NIR) observations remain scarce despite their advantages, that is, reduced dust extinction and a more intrinsic standard candle behavior, requiring little to no empirical corrections. Here, we present ASNOS (ANDICAM-SOFI Near-infrared and Optical type Ia Supernova), a dataset with sample size of 1,482 epochs in the $BVRIYJH$ filters from the ANDICAM instrument on the 1.3-meter SMARTS telescope at Cerro Tololo Inter-American Observatory, along with 125 $JHK$ epochs from the SOFI instrument on the 3.58-meter New Technology Telescope on the La Silla Observatory. Additionally, we incorporate optical forced photometry from the Zwicky Transient Facility and the Asteroid Terrestrial-impact Last Alert System. The sample comprises 41 SNe Ia in total, including 29 normal events, eight 1991T-like objects, and four peculiar subtypes, all located at redshifts $z < 0.085$. This paper provides a detailed overview of the ASNOS sample selection, data reduction, SN photometry, host-galaxy spectral energy distribution construction, both global and local, and SN light-curve fitting using three methods: SALT3-NIR, SNooPy, and BayeSN. A companion paper will present the cosmological analysis.
}
\maketitle


\section{Introduction}

Type Ia supernovae (SNe Ia) are thermonuclear explosions that occur in stellar systems containing at least one white dwarf (WD). The WD is supported by electron degeneracy pressure, which counteracts gravitational collapse. Several progenitor scenarios have been proposed for SNe~Ia, the most common being the single-degenerate scenario \citep{1973ApJ...186.1007W}, involving a WD and a main-sequence companion; the double-degenerate scenario \citep{1984ApJ...277..355W}, consisting of two WDs; as well as configurations involving triple systems \citep{Thompson_2011}, where two WDs and a tertiary star can interact to induce a direct collision  between the WDs \citep{Kushnir2013}, and systems with WDs and post-AGB companions \citep{Livio2003,Kashi2011}. In each of these scenarios, the explosion may be triggered either by mass accretion from a companion or by stellar mergers. The explosion mechanisms include central or outer carbon ignition and surface helium ignition. These combination of progenitor scenarios and explosion mechanisms result in either near-Chandrasekhar-mass ($M_{Ch}$) or sub-$M_{Ch}$ progenitor WDs exploding as SNe~Ia \citep{2000ARA&A..38..191H}. SNe~Ia can reach a luminosity around $10^{43}$ erg~s$^{-1}$ at peak, and their light curves are highly homogenous, making them ideal for cosmological distance measurements. They have been used to demonstrate the accelerated expansion of the Universe \citep{Perlmutter_1999,Riess_1998}, as well as to measure the local expansion rate \citep[e.g.,][]{2022ApJ...934L...7R,2023A&A...679A..95G}.

In most cases, SNe~Ia are observed at optical wavelengths, with more than 6,000 publicly available observed objects \citep{Scolnic_2022,rigault2024ztfsniadr2,descollaboration2024darkenergysurveycosmology}. The peak $B$-band brightness of SNe~Ia correlates with the post peak decline rate 15 days after maximum $\Delta m_{15}$ \citep{1977SvA....21..675P,1993ApJ...413L.105P} and with the $B-V$ color at peak \citep{1996ApJ...473...88R,1998A&A...331..815T}, allowing them to be used as standardizable candles after applying this  correction. However, analyses of SNe~Ia in the near-infrared (NIR) have shown that they are nearly natural standard candles because the NIR light is less affected by dust and they exhibit less intrinsic  scatter in their peak luminosities \citep{Krisciunas_2004, 2012MNRAS.425.1007B, 2018A&A...615A..45S, 2019hst..prop15889J, 2022A&A...665A.123M}. Unfortunately, observing in the NIR is challenging due to a brighter sky background and the fact that host galaxies are often brighter than the SNe, making it challenging to accurately determine their  brightness. The largest publicly available samples of SNe~Ia with NIR light curves include those from the \textit{Carnegie Supernova Project} (CSP-I: \citealt{Contreras_2010,Stritzinger_2011,2017AJ....154..211K}; CSP-II: \citealt{2019PASP..131a4001P,2019PASP..131a4002H}), the Harvard Center for Astrophysics (CfA; \citealt{2015ApJS..220....9F}), the RATIR sample \citep{2021ApJ...923..237J}, SweetSpot \citep{2018AJ....155..201W}, RAISIN \citep{Jones_2022}, DEHVILS \citep{Peterson_2023}, and the Hawai‘i Supernova Flows \citep{Do_2024}. Ongoing projects such as the Supernovae in the InfRAred Avec Hubble (SIRAH; Jha et al. in prep.; \citealt{2020sea..confE..37G}) are expected to increase the sample size by a few dozen in the coming years. Currently, the total number of publicly available SNe~Ia with NIR $JH$-band observations is around 300, significantly fewer than the number publicly available at optical wavelengths.

In this paper, our aim is to expand the number of SNe observed in the NIR by presenting 41 SNe with optical (\textit{BVRI}) and NIR ($YJHK_s$) data obtained using the A Novel Dual Imaging CAMera (ANDICAM; \citealt{2003SPIE.4841..827D}) instrument mounted on the 1.3-meter SMARTS telescope at the Cerro Tololo Inter-American Observatory (CTIO). The main goal is to provide an expanded sample of SNe Ia with NIR observations in order to do precision cosmology. 

Additionally, we include $JHK_s$ images taken with the Son OF ISAAC(SOFI) instrument, mounted on the 3.58-meter New Technology Telescope (NTT) at La Silla Observatory. We describe the methods used to reduce the data and calculate the brightness of the SNe at various epochs to construct light curves.  Then the light-curve fitting tools SNooPy \citep{2014ApJ...789...32B}, BayeSN \citep{grayling2024scalablehierarchicalbayesninference}, and SNCosmo \citep{2022zndo....592747B} with the SALT3-NIR template \citep{Pierel_2022} are used to fit our data and determine light-curve parameters. Furthermore,  global and local host galaxy parameters are obtained by performing photometry with Hostphot \citep{2022JOSS....7.4508M} and conducting stellar population synthesis with Prospector \citep{2017ApJ...837..170L,Johnson_2021}. This paper is the first of two and serves as a data reduction description and release paper, while the second paper will focus on studying Hubble diagram residuals using a combination of our data and other datasets available in the literature. The methods applied here will serve as the basis for our Aarhus-Barcelona cosmic FLOWS project, where we currently have more than 700 observations of SNe~Ia  in NIR.


\section{Data sample}\label{sec:sample}

\subsection{Target selection and observing strategy}

Our observing campaign lasted three semesters (2018A-2019A), during which the objective was to observe approximately 12 SNe per semester. To select targets for observation, we monitored newly classified objects in the Transient Name Server (TNS\footnote{\href{https://www.wis-tns.org}{https://www.wis-tns.org}}) and ensured that each object would be observable from CTIO for about two months at magnitudes brighter than 18 mag. Although our initial goal was to select nearby objects in the Hubble flow with redshifts between $z =  0.01$ and 0.04 to accurately measure distances to their host galaxies, a few SNe at higher redshifts (but still within the Hubble flow), were included. The objective was to obtain 10 epochs in each band ($JH$) for each SN, covering a time span from $-$10 to +50 days past the optical peak, with observations approximately every six days. This cadence would allow us to determine the peak magnitude in $JH$ and capture the behavior around the secondary peak in the near-infrared. Unfortunately, the optical camera malfunctioned around April 2019, limiting our ability to obtain optical images. 
As a result, the last five SNe in the sample only contained $I$-band images.

\subsection{ANDICAM images}

The images were obtained using the SMARTS telescope at CTIO with the ANDICAM camera, which had been in operation for approximately 20 years, with its last observation taken in July 2019. We obtained simultaneous optical (BVRI) and NIR (YJH) images, with pixel scales of 0.371 and 0.271 arcsec/pixel and field-of-view (FoV) of 6$\times$6 and 2.4$\times$2.4 arcmin$^2$ for the optical and NIR channels, respectively. In total, 41 SNe were observed with ANDICAM, of which 29 were classified as normal SNe~Ia, 8 as 1991T-like SN, 2 as 1991bg-like SN, 1 as SNe~Ic-broad Line, and 1 as a 2002cx-like SN. A total of 33 out of 41 SNe had $Y$-band images, and the median number of epochs in $YJH$ was 12. The optical images were taken with an exposure time of 200 sec, while the NIR images were acquired in sets of five at different dither positions by 10 or 20 arcsec, each with an exposure time of 20 sec. In the case of the $Y$-band images, only one set of dithered images was obtained, whereas two sets were taken for $JH$. A few $K$-band images were also obtained, but these were too noisy to be useful and were therefore excluded from this study. Additionally, we obtained images of standard stars on the same nights as our SN observations.

Table~\ref{tab:sntable} presents the list of SNe observed with ANDICAM, along with their sub-types, right ascensions (RA), declinations (Dec), discovery groups, classification groups, all obtained from TNS, which reports new astronomical transients such as supernova candidates. To match host galaxy and redshift, a thorough search was made in the NASA/IPAC Extragalactic Database (NED\footnote{\href{https://ned.ipac.caltech.edu/}{https://ned.ipac.caltech.edu/}}), the SIMBAD astronomical database\footnote{\href{https://simbad.cds.unistra.fr/simbad/}{https://simbad.cds.unistra.fr/simbad/}}, and the second data release of the Zwicky Transient Facility (ZTF DR2; \citealt{2025A&A...694A...1R}). In the last column we also included the number of epochs in \textit{YJH}. 

\begin{figure*}[!t]
\centering
\includegraphics[width=\textwidth]{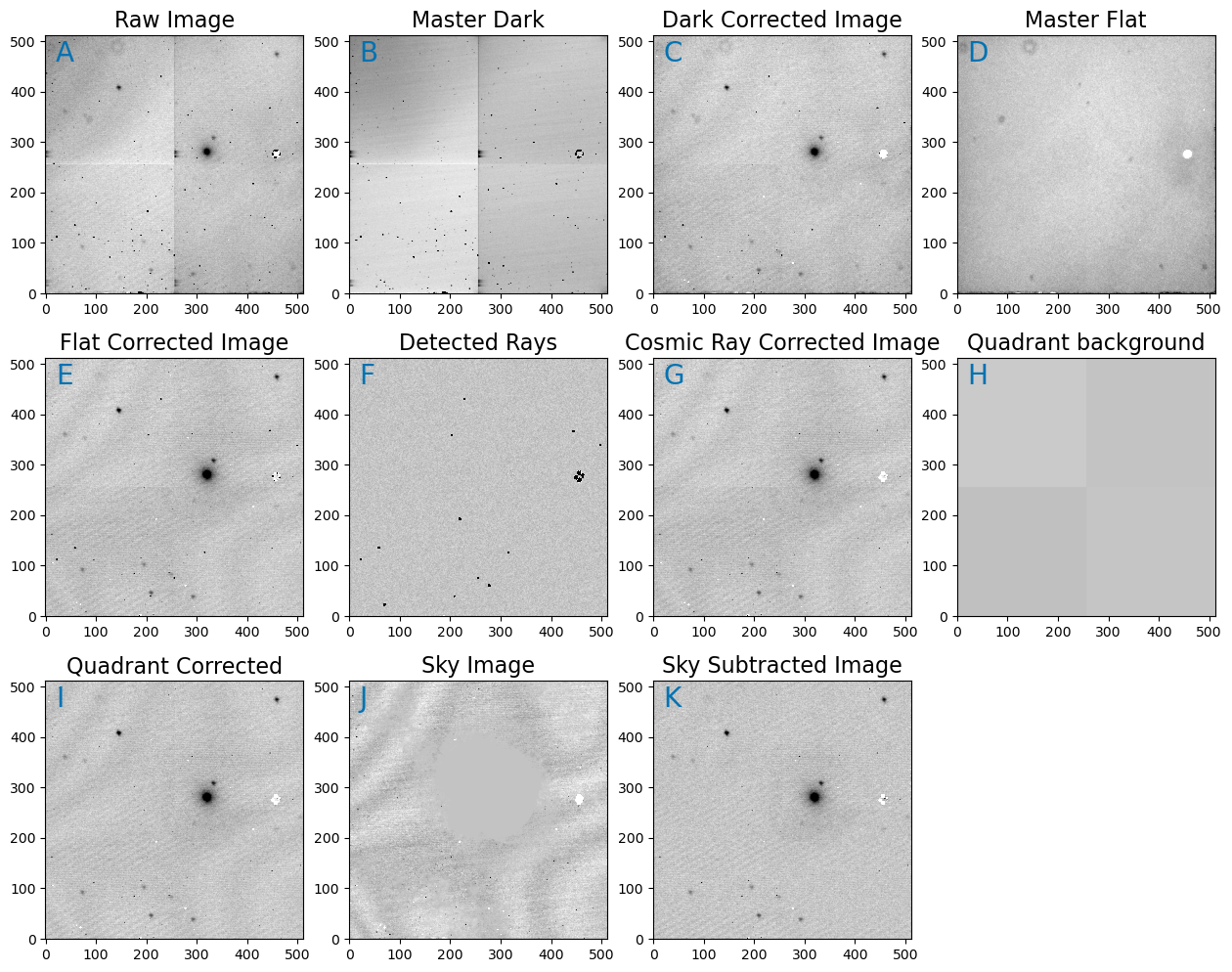}
  \caption{
  Example of raw image reduction for ANDICAM $J$-band image. Top row (left to right): Single raw image of SN~2019so, master dark, dark reduced raw image using the two previous images and master flat. Middle row: Flat and dark reduced image, detected cosmic rays, cosmic ray corrected image and quadrant background level. Bottom row: Image corrected for quadrant features, sky image, and sky subtracted image. The sky image was evaluated from the remaining set of dithered images taken on the same night. The host galaxy was masked out for each single image when constructing the sky image.
  }
  \label{fig:dark_and_flat_correction}
\end{figure*}

\subsection{Additional NIR SOFI images}

Images taken with the SOFI instrument on the NTT in La Silla were obtained through the Public European Southern Observatory Spectroscopic Survey of Transient Objects (PESSTO) collaboration \citep{2015A&A...579A..40S} and typically consisted of 1 to 5 epochs of $JHK_s$ images. The pixel scale is 0.25 arcsec/pixel, and the images cover a FoV of 5$\times$5 arcmin$^2$.
A summary of images collected with SOFI is also included in Table~\ref{tab:sntable}.

\subsection{Additional optical data}

In addition to our optical ANDICAM images, we also obtained optical data from the Asteroid Terrestrial-impact Last Alert System (ATLAS; \citealt{Tonry_2018}) and the Zwicky Transient Facility (ZTF; \citealt{Bellm_2018}). ATLAS is a highly efficient system for detecting potentially hazardous asteroids, as well as tracking and discovering transients, with a two-day cadence in the cyan band ($c$; 4200–6500 \AA) and orange band ($o$; 5600–8200 \AA). Currently, ATLAS observes the whole sky with a cadence of one day between declinations of $-50$ and +50 and two days in the polar regions,  weather permitting. The images were calibrated using the Pan-STARRS catalog \citep{chambers2019panstarrs1}, and the light curves are publicly available through The Atlas Project Fallingstar photometry service.\footnote{\href{https://atlas.fallingstar.com}{https://atlas.fallingstar.com}} Optical data from ZTF were retrieved from the ZTF second data release \citep{2025A&A...694A...1R} and was obtained using the Palomar 48-inch Schmidt telescope, which scanned the entire northern sky visible from Palomar every two days in the $g$- (3676–5614 \AA) and $r$-band (5498–7394 \AA), while alerts will be distributed in real time to the brokers for the $i$-band (6870–8964 \AA).

\subsection{Host-galaxy template images}\label{sec:template_images}

In cases where the SN occurred in a bright region of the galaxy, it was crucial to remove the underlying host-galaxy light contamination to precisely measure the SN brightness in the image. This was done by checking that the background level around the SN was close to 0.  Ideally, host-galaxy template images should be taken with the same instrument; however, ANDICAM stopped operating right after our program ended in July 2019. As a result, we had to either rely on archival images or obtain new images with other facilities.

The NIR images were obtained through the European Southern Observatory (ESO) archive\footnote{\href{http://archive.eso.org/scienceportal/home}{http://archive.eso.org/scienceportal/home}} and The United Kingdom Infrared Telescope (UKIRT) Hemisphere Survey archive.\footnote{\href{http://wsa.roe.ac.uk/index.html}{http://wsa.roe.ac.uk/index.html}} We also attempted to use 2MASS templates, but the resolution of these images was insufficient to provide a good subtraction. If no archival templates were available, new templates were obtained using the NOTCam camera at the 2.5-meter Nordic Optical Telescope (NOT) in La Palma or SOFI at the NTT in La Silla through the PESSTO collaboration. Furthermore, our project was awarded one night to observe at the 6.5-meter Magellan Baade telescope using the FourStar instrument \citep{2013PASP..125..654P} located on Las Campanas Observatory in Chile. Optical template images on the other hand were obtained from the Dark Energy Survey\footnote{\href{https://www.darkenergysurvey.org}{https://www.darkenergysurvey.org}} (DES), Pan-STARRS\footnote{\href{https://outerspace.stsci.edu/display/PANSTARRS/}{https://outerspace.stsci.edu/display/PANSTARRS/}}, SkyMapper\footnote{\href{https://skymapper.anu.edu.au}{https://skymapper.anu.edu.au}}, or EFOSC at NTT (again through the PESSTO collaboration). In a few cases, when it was not possible to obtain $BVRI$ templates for subtraction, we used $g$-band templates for $B$ and $V$, $r$ for $R$, and $i$ instead of $I$. Since all NIR template images were taken in 2023 or later, the SNe had long since faded, as our SNe were observed in 2018-2019.


\section{Data reduction}\label{sec:reduction}

\subsection{Optical ANDICAM imaging}

The optical ANDICAM images were directly retrieved from the SMARTS FTP archive and had already been reduced using the dedicated ANDICAM pipeline\footnote{\href{http://www.astro.yale.edu/smarts/ANDICAM/data.html}{http://www.astro.yale.edu/smarts/ANDICAM/data.html}}, which applies bias and flat-field corrections to the images.
 
\subsection{NIR ANDICAM imaging}\label{sec:andicam_nir}

NIR images were provided in raw format, so we describe the data reduction process here, including flat-field and dark corrections, sky subtraction, and the combination of dithered images.

\subsubsection{Flat and dark correction}
To reduce our ANDICAM NIR images as well as our standard stars, we used {\sc ccdproc}\footnote{\href{https://github.com/astropy/ccdproc}{https://github.com/astropy/ccdproc}}, an Astropy-affiliated package for the reduction of charge-coupled device (CCD) images. This allowed us to take an image in FITS format and convert it into a {\sc CCDData} object, which is a matrix containing the photon counts at each pixel position in our images. Flats were already combined into master flats and made available in the SMARTS FTP archive. Individual dark frames taken during a night were combined into a master dark. To create the master dark for a given date, all dark images taken on that date were collected and {\sc ccdproc.combine} was used to combine the dark frames entrywise by taking the median. Because flats and darks were not taken every night, we selected the flat and dark images from the FTP archive that were closest in time to when the raw science images were taken (typically 2–3 days apart at worst). Both the master flat and master dark were used for the SN images and standard stars. The master dark was subtracted from each raw image using {\sc ccdproc.subtract\_dark}, which scales the master dark to match the exposure time of the raw image and then subtracts it. In our case, the exposure times were the same, so no scaling was needed. Finally {\sc ccdproc.flat\_correct} was used to normalize the master flat using the median of the matrix (if not already done) and then divided the dark-subtracted images by the flat entrywise.

Figure \ref{fig:dark_and_flat_correction} illustrates the process of dark and flat correction for a single raw image.
We note the presence of an artifact in the middle-right of all our images, which could not be entirely removed during the master flat correction. However, this artifact does not affect the SN magnitude, as the SN is consistently positioned near the center in all our images. For the $Y$-band images, no $Y$-band flats were available in the SMARTS FTP archive, so a $J$-band flat was used instead. By comparing the differences between the $J$- and $H$-band flats, it was found that the variation between them is less than 2\%, resulting in differences in instrumental magnitudes of local sequence stars of 0.02 magnitudes in the worst cases (going in both directions). As a result, an uncertainty to the instrumental magnitude in the $Y$-band of 0.02 magnitudes was added in quadrature.
Because the $Y$-band photometry may have an $\sim 2\%$ offset arising from the use of $J$-band flats during data reduction, we do not recommend using these data for precision cosmology and exclude them from our light-curve fitting.

\subsubsection{Cosmic ray removal}

Cosmic rays were removed using the {\sc cosmicray\_lacosmic} module from {\sc ccdproc}, which detects and removes cosmic rays based on the readout noise, a sigma detection threshold (which was set to the default value of 5), and the science image as input parameters. 
Image F in Figure ~\ref{fig:dark_and_flat_correction} shows the detected cosmic rays, and the image corrected for cosmic rays is seen in image G.

\subsubsection{Quadrant Correction}\label{sec:quadrant_correction}

In some science images, the background levels differ between quadrants, which can be seen in image G in  Figure~\ref{fig:dark_and_flat_correction}.
To address this issue, the image was divided into four quadrants, and bright sources were masked out. The background level in each quadrant was then estimated and subtracted accordingly. The estimated background level for each quadrant can be seen in image H of Figure~\ref{fig:dark_and_flat_correction}, while the science image corrected for quadrant features can be seen in image I. Before subtracting the quadrant backgrounds, the overall background level of the full image was estimated and stored for later use in rescaling the final science frame. Additionally, the background level was used to select images with the similar background levels to construct a sky background (see Sect. \ref{sec:sky_subtraction} for more details).

\subsubsection{Sky subtraction}\label{sec:sky_subtraction}
The next step was to remove the sky background, which can vary significantly in the NIR, even when images were taken just a few minutes apart. To achieve this, the following procedure for a set of dithered images was used: At a given epoch, the median and standard deviation of all images were evaluated. These values were saved to rescale the final science image afterwards, ensuring that reliable uncertainties for photometry were maintained. We estimated the standard deviation of the sky medians and named this value ``$sky\_std$''. Similarly, we estimated the average of the standard deviations and named this value "$sky\_noise$". If $sky\_std < sky\_avg$, some images had background levels that were different from the rest. These were excluded until $sky\_std > sky\_avg$. Then, for a given image in a set of dithered images, {\sc ccdproc.combine} was used on the remaining images to create a sky background. Finally, the sky background was subtracted from each image. In some cases, a large object such as the host galaxy could be bigger than the dithering step size, which could result in a brigther sky image around the location of the host galaxy. To address this issue, bright sources were masked out in each image before combining into a single sky image.

Image J of Figure~\ref{fig:dark_and_flat_correction} shows the sky image with the area around the host galaxy masked out, and the final sky-subtracted image is seen in image K. Here, it was possible to successfully capture and remove the wave-like structure present in the sky background from the dark- and flat-corrected image. 

\subsubsection{World Coordinate System}

Only the NIR ANDICAM images did not have any World Coordinate System (WCS) values in the image header so to obtain these, we used nova.Astronometry.net\footnote{\href{https://nova.astrometry.net}{https://nova.astrometry.net}}, which requires at least three stars that can be detected by the code. This turned out to be difficult, as the stars were faint and the FoV was quite small (2.4$\times$2.4 arcmin$^2$). For SN~2018hhn, SN~2018jag and SN~2019cxx, it was not possible to obtain a WCS. Therefore these images were combined using the relative position of the host galaxy. This turned out to be succesful for SN~2018hhn and SN~2018jag, while it was not possible to detect any stars/galaxies in the images of SN~2019cxx.  

\begin{figure*}[t]
    \centering
    \includegraphics[width=\textwidth]{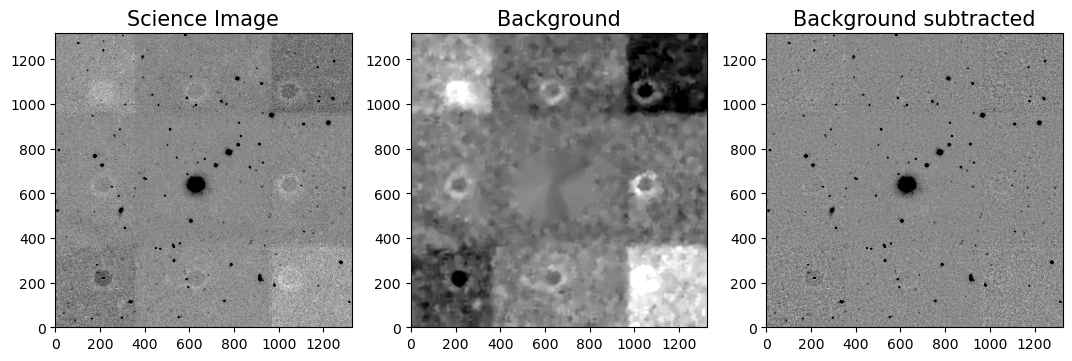}
    \caption{Example of background subtraction for a SOFI image. \textit{Left:} Combined $J$-band science image of SN~2018agk taken with SOFI. \textit{Middle:} Detected image background using  {\sc photutils.Background2D}. \textit{Right:} Background and cosmic ray subtracted image. It is seen that we were able to remove most quadrant structure of the image.}
    \label{fig:cosmic_rays}
\end{figure*}

\subsubsection{Combined final images}
The dithered images can now be combined into a single science image. To do this, each image was projected onto a grid of NaNs based on the relative position of a star. From the list of grids, we used {\sc numpy.nanmedian} to combine the images.
Finally, the edges were cropped to ensure the final image had at least 50 percent of the coadded images throughout. 


\subsection{Complementary imaging}
As aformentioned, we also obtained data from SOFI, ZTF, and ATLAS, which are used in the SN~Ia template light-curve fitting process. ZTF light-curves were retrieved from ZTF DR2 \citep{rigault2024ztfsniadr2}. Below, the data reduction processes for each of the other datasets are described in detail.


\subsubsection{SOFI}
The raw SOFI NIR sciences images, flats, and darks were downloaded from the ESO archive\footnote{\href{http://archive.eso.org/eso/eso\_archive\_main.html}{http://archive.eso.org/eso/eso\_archive\_main.html}} and reduced using the PESSTO pipeline.\footnote{\href{https://github.com/svalenti/pessto}{https://github.com/svalenti/pessto}} This pipeline handles the flat and dark corrections of the raw science images, as well as the sky subtraction and individual frame combination.


\subsubsection{ATLAS}

The ATLAS data were obtained through the ATLAS forced photometry homepage. We requested epochs between $-50$ to $+100$ days of the detection date of the SN to make sure the LC peak(s) were obtained. ATLAS often took four measurements per night so the data was stacked following The Atlas Project's Python plotting software
\citep{Young_plot_atlas_fp}. The fluxes and errors in the output file were given in microJansky, denoted $\mu Jy$ and $d\mu Jy$, respectively. The fluxes were converted into AB magnitudes using
\begin{equation}
    m_{AB} = -2.5\cdot \mathrm{log_{10}}(\mu Jy) + 23.9 \label{eq:ab_magnitude}.
\end{equation}
For the magnitude error, the homepage suggested to do a Taylor expansion of $log_{10}(\mu Jy +/- d\mu Jy)$ when $d\mu Jy/\mu Jy$ was small, but since this was not always the case, we chose to do error progagation instead, where we set
\begin{equation}
    m_{AB,err} = \frac{2.5}{\mathrm{ln(10)}}\cdot\frac{d\mu Jy}{\mu Jy}.
\end{equation}
Non-detections were removed using a $3\sigma$ upper limit
\begin{equation}
    m_{3\sigma} = -2.5\cdot \mathrm{log_{10}}(3 \cdot d\mu Jy) + 23.9.
\end{equation}

In general, baseline correction were not required for ATLAS photometry. However, there are two epochs in which ATLAS changed reference images and therefore the baselines changed. In the cases of SN~2018bie and SN~2018exc, we observed that non-detections both before and after the SN light curve exhibited flux values significantly deviating from zero ($\approx -600$ in flux space). Notably, around 400 days post-explosion, these non-detections returned to zero flux. Without applying a baseline correction, the apparent magnitudes appeared artificially faint. To account for this, a baseline correction was implemented exclusively for these two SNe. The baseline level was determined using a trimmed mean of flux values from epochs earlier than $-30$ days and later than $+400$ days relative to the SN detection date, ensuring that any residual supernova signal was excluded. When performing light-curve fitting (see Sect.~\ref{sec:cal}), the resulting fits were significantly improved with the baseline correction, whereas the fits were poor when no correction was applied.

\subsection{Background subtraction}

\begin{figure*}[t]
    \centering
    \includegraphics[width=\textwidth]{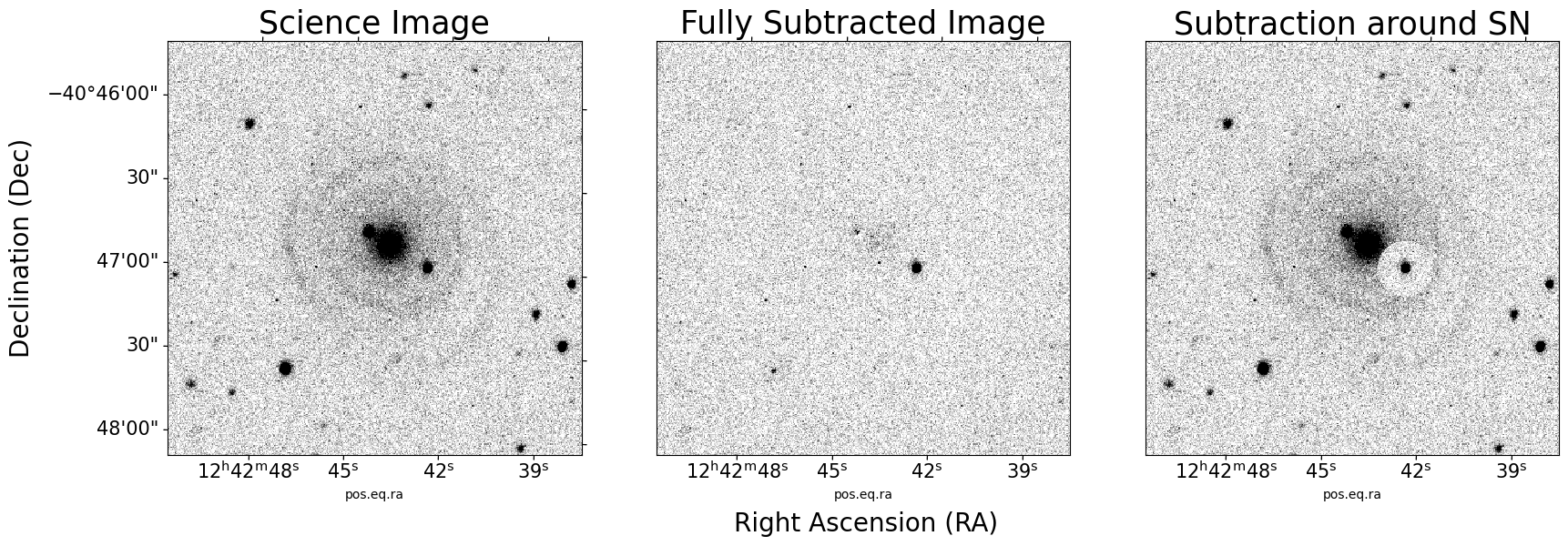}
    \caption{Template subtraction using ImageMatch. \textit{Left:} Science image without template subtraction. \textit{Middle:} Fully subtracted science image. \textit{Right:} Template subtracted image using ImageMatch with the SN in the middle of the subtracted area. ImageMatch does subtraction on the full image, but one can choose to have the subtracted area around the SN inserted back into the original science image, which allowed us to use local sequence stars to determine zeropoints.
    } \label{fig:galaxy_subtraction}
\end{figure*}
\begin{figure}[!t]
    \centering
    \includegraphics[width=\columnwidth]
    {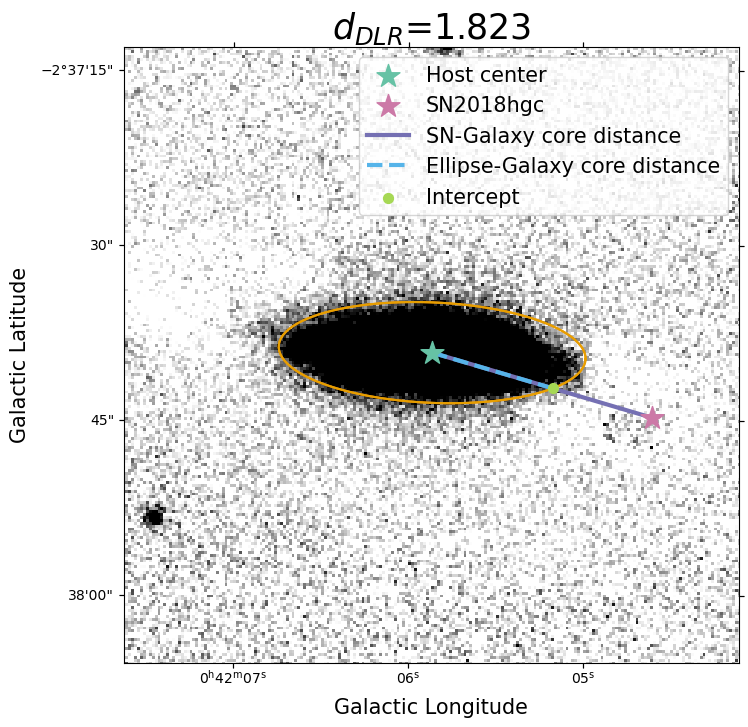}
    \caption{Example of measuring $d_{DLR}$. The shape of the host was measured using $sep$ and plotted in orange. The $d_{DLR}$ is the ratio between distance from the host center to the SN (purple) and the distance from the host center to the edge of the ellipse (dashed blue). }
    \label{fig:dlr_example}
\end{figure}

Lastly, the background was subtracted from the optical ANDICAM images as well as SOFI images (this was already done for the NIR ANDICAM images).
The background was subtracted using {\sc Background2D} from {\sc photutils.background}. Here {\sc MedianBackground} was used, which calculated the background in an array as the sigma-clipped median, with sigma set to 3. Figure~\ref{fig:cosmic_rays} shows an example of background subtraction. All reduced images can be accessed from GitHub.\footnote{\href{http://www.github.com/SN-ICE/ASNOS}{http://www.github.com/SN-ICE/ASNOS}}


\section{Photometry}\label{sec:phot}

Here, we describe the methods used to determine the apparent magnitudes of our SNe and construct light curves. This process includes template subtraction, aperture photometry, calculating zeropoints, and determining color term coefficients. Color term coefficients quantify how a natural system differs from the standard system as a function of the color of an object and allows one to transform magnitudes from one system to the other. A natural system refers to the photometric system native to the telescope and detector setup, or in other words the system response function, while a standard system refers to a widely used photometric system such as the Johnson system. Here the objective is to construct SN light curves in the natural system. 

\begin{figure*}[!t]
    \centering
    \includegraphics[scale = 0.3]
{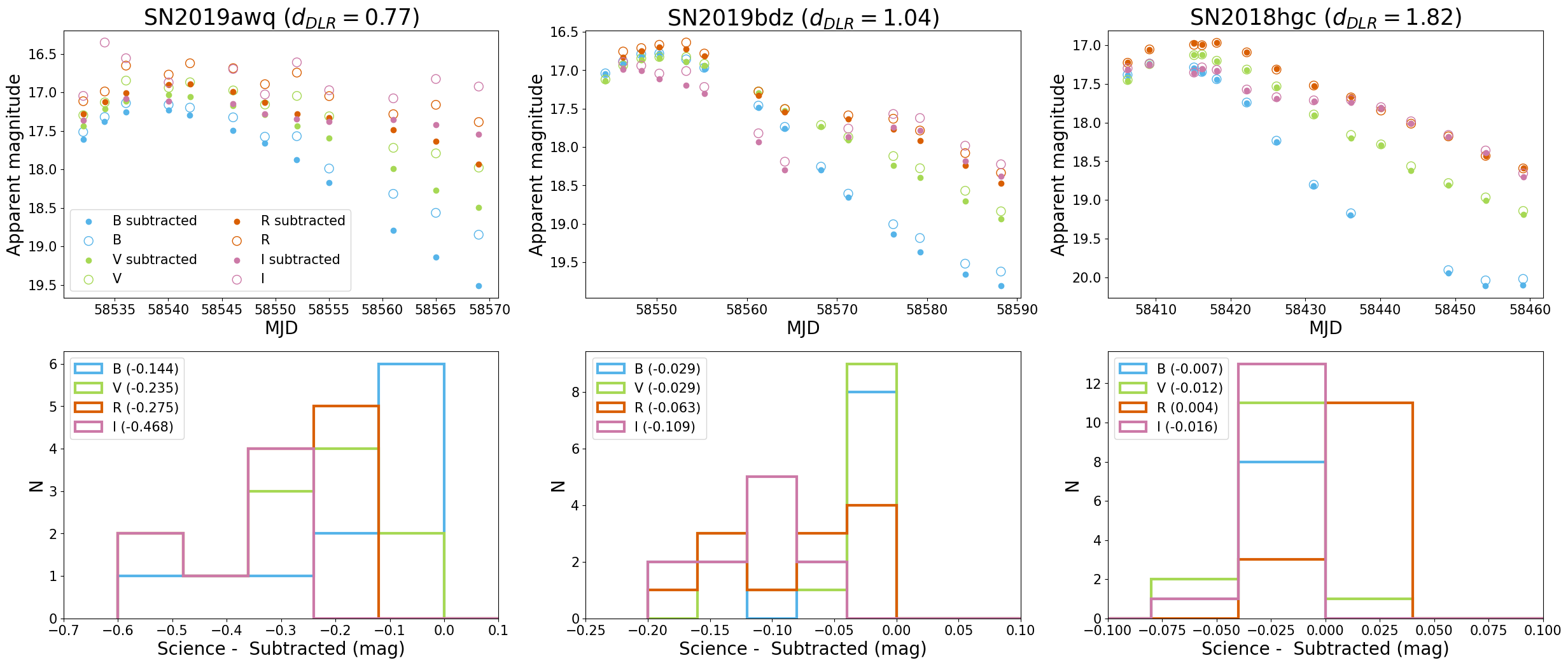 }
    \caption{\textit{Top panels:} Comparison of photometry using subtracted images (filled circles) vs. not using templates (empty circles) for SNe with different $d_{DLR}$. As expected, applying host-galaxy template subtraction to the science images did not have a huge effect on the light curve when the SN was far from its host ($d_{DLR}>1$). On the other hand, the effect was significant when $d_{DLR}<1$, showing why template subtractions were important. \textit{Bottom:} Histogram of the apparent magnitude difference between subtracted and non-subtracted images. The median value is reported in the legend for each filter.}
    \label{fig:template_vs_no_template}
\end{figure*}

\subsection{Template subtraction}

Before our images could be used for photometry, the light from the host galaxy was removed using the software {\sc ImageMatch}.\footnote{\href{https://github.com/obscode/imagematch}{https://github.com/obscode/imagematch}} This software rectifies an image by matching point-sources from the science image to the host-galaxy template and solving for a geometric transformation from one to the other. Afterwards it convolved one image with a ``seeing kernel'' that blurred it so that they matched in resolution, which resulted clean subtractions. Only in a few cases, the seeing in the science image was better than the template, and therefore the science image needed to be blurred so it matched the template in order to do proper subtraction. 
By default, ImageMatch performs subtraction over the entire image. For photometric measurements, a small region centered on the SN from the subtracted image was instead inserted into the original, unsubtracted frame containing all field stars. This procedure allowed the SN’s apparent magnitude to be estimated in a single step: first determining the zeropoint from local sequence stars and then measuring the SN’s instrumental magnitude. To assess the quality of the subtraction, the entire subtracted image was visually inspected to confirm that nearby stars and galaxies were properly removed.

Figure~\ref{fig:galaxy_subtraction} shows an image before and after subtraction. It is clearly seen that the light from the host galaxy had been subtracted around the SN. For a few cases, it was not possible to do proper subtraction when the SN was at the center of the host. This was likely due to the fact that templates and science images were from other instruments/filters. This happended to SNe~2018dda, 2018exb, 2018exc, and 2018feq. Light curve fitting were attempted for these SNe, but since the fits were bad and the SN was not clearly isolated, these images were not used.  

To check the effect of our template subtractions, a comparison between the apparent magnitude of various SNe with and without subtraction was made based on how far away from the host they were. This was accomplished using the angular distance to the SN normalized by the host galaxy’s directional light radius denoted $d_{DLR}$ \citep{Sullivan_2006,Gupta_2016,Sako_2018} and depends on the shape of the galaxy and the SN direction. Generally $d_{DLR}<1$ means that the SN is inside the ellipse of the host galaxy, while $d_{DLR}>1$ means it is outside and therefore would be minimally affected much by the light of its host. Hence SNe with $d_{DLR}>1$ could be used as a consistency check to see if our subtractions were impactful or not. The software {\sc sep}\footnote{\href{https://sep.readthedocs.io}{https://sep.readthedocs.io}} was used to determine the shape of the host galaxy. {\sc sep} is a python implementation of {\sc Source Extractor}\footnote{\href{https://www.astromatic.net/software/sextractor/}{https://www.astromatic.net/software/sextractor/}} and determines the sizes and shapes of stars/galaxies in an image in terms of elliptical parameters such as semimajor and semiminor axis. Figure ~\ref{fig:dlr_example} shows an example of how $d_{DLR}$ was evaluated for a host galaxy with its shape (in orange) determined from {\sc sep} using a $3$-sigma detection threshold. Table \ref{tab:sntable} shows $d_{DLR}$ for all SN hosts in our sample. In each case, $r$-band images were used (mostly from Pan-STARRS, SDSS, DES). The results of estimating $d_{DLR}$ for a few SNe are shown in  Figure~\ref{fig:template_vs_no_template}. The top panels show the apparent magnitude of some SNe with different $d_{DLR}$ with subtraction (filled circles) versus no subtraction (empty circles) for the ANDICAM $BVRI$. The bottom panels show the corresponding histograms of the apparent magnitude difference between non-subtracted and template subtracted images, where the median difference is presented in the legend. Not surprisingly, there was not much difference between subtracted and non-subtracted images when the SN was far away from its host, while there was significant contamination from the host when the SN was close.

\subsection{Absolute Photometry}\label{sec:photometry}

To evaluate the brightness of the local sequence stars and the SN in our images,  the {\sc photutils}\footnote{\href{https://photutils.readthedocs.io/en/stable/index.html}{https://photutils.readthedocs.io/en/stable/index.html}} package was used. With DAOStarFinder, sources in our images that were $3\sigma$ above background level were detected. Each source was then fitted with 2D gaussian to determine their geometric properties such as position, semimajor axis, semiminor axis and full-width-at-half maximum (FWHM), which is the width of the distribution measured between two points where the value of the curve is at half its maximum amplitude. The world coordinate system (WCS) in the image header was used to assign catalog magnitudes to each object if the relative distance between the position obtained from DAOStarFinder and the catalog position was less than 0.5 FWHM. 

\subsubsection{Reference Catalog}

The ATLAS-REFCAT2 catalog \citep{2018ApJ...867..105T}, an all-sky reference catalog containing about one billion stars down to an apparent magnitude of 19, was used to assign catalog magnitudes to our sources.
REFCAT2 consists of a variety of surveys such as PanSTARRS, The AAVSO Photometric All-Sky Survey (APASS), SkyMapper and the Two Micron All Sky Survey (2MASS). The PanSTARRS magnitudes (\textit{gri}) from ATLAS-REFCAT2 were transformed into Johnson \textit{BVRI}, which corresponded to the optical passbands of ANDICAM, using the transformations in \cite{Tonry_2012}
\begin{eqnarray}
 B_{cat,std}  &=&  g_p + 0.213 + 0.587\cdot (g_p-r_p)    \pm 0.034, \nonumber \\
 V_{cat,std}   &=&   r_p + 0.006 + 0.474\cdot (g_p-r_p)    \pm 0.012,  \nonumber \\
 R_{cat,std}   &=&     r_p - 0.138 - 0.131\cdot (g_p-r_p)    \pm 0.015,  \nonumber  \\
 I_{cat,std}  &=&     i_p - 0.367 - 0.149\cdot (g_p-r_p)    \pm 0.016,
\end{eqnarray}
where the ``$p$'' subscript denotes Pan-STARRS passbands, and the ``cat,std" subscript denotes the catalog magnitude in a standard photometric system such as the Johnson system. 
The ANDICAM $Y$~passband was similar to the Dark Energy Camera (DECam) $Y$~passband so our catalog magnitudes had to be transformed into DECam magnitudes. As color term coefficients would be determined later on, we evaluated DECam $z$-band magnitudes as well in order to have a filter that was close to the $Y$~band in terms of wavelength range. This required Pan-STARRS \textit{zy}~bands, but since the ATLAS-REFCAT2 only had PanSTARRS \textit{griz}, we used the 2MASS transformations from \cite{Tonry_2012} to obtain the PanSTARRS $y$~band    
\begin{eqnarray}
 y_p  &=&  J_{2} + 0.531 + 0.916\cdot(J_{2}-H_{2})    \pm 0.061, 
\end{eqnarray}
where the $2$ subscript denotes 2MASS magnitudes. 
Finally the transformations from \citet{Abbott_2021} were used to obtain DECam $zY$ magnitudes
\begin{eqnarray}
 z_{DES}  &=&   y_p - 0.031\cdot(r_p-i_p) - 0.01    \pm 0.015,
\nonumber\\
 Y_{DES}  &=&     y_p - 0.031\cdot(r_p-i_p) + 0.035   \pm 0.017. 
\end{eqnarray}

Before we determined the instrumental magnitude of our sources, a few cuts were applied:
1) Objects that were $3 \cdot FWHM$ from the edges of our images were removed (our images consisted of multiple dithered images, and therefore stars closer to edge had less images compared to the center, which therefore gave less precise measurements).
2) Galaxies were removed. 
3) Saturated objectives were removed. This was achieved by calculating the maximum pixel value for each source and remove objects where the detector limit was exceeded and the detectors would go non-linear. For ANDICAM optical, ANDICAM NIR and SOFI, the limits were 45000, 5000 and 10000 ADU,  respectively. 

\begin{figure}[t]
    \centering
    \includegraphics[width=\columnwidth]{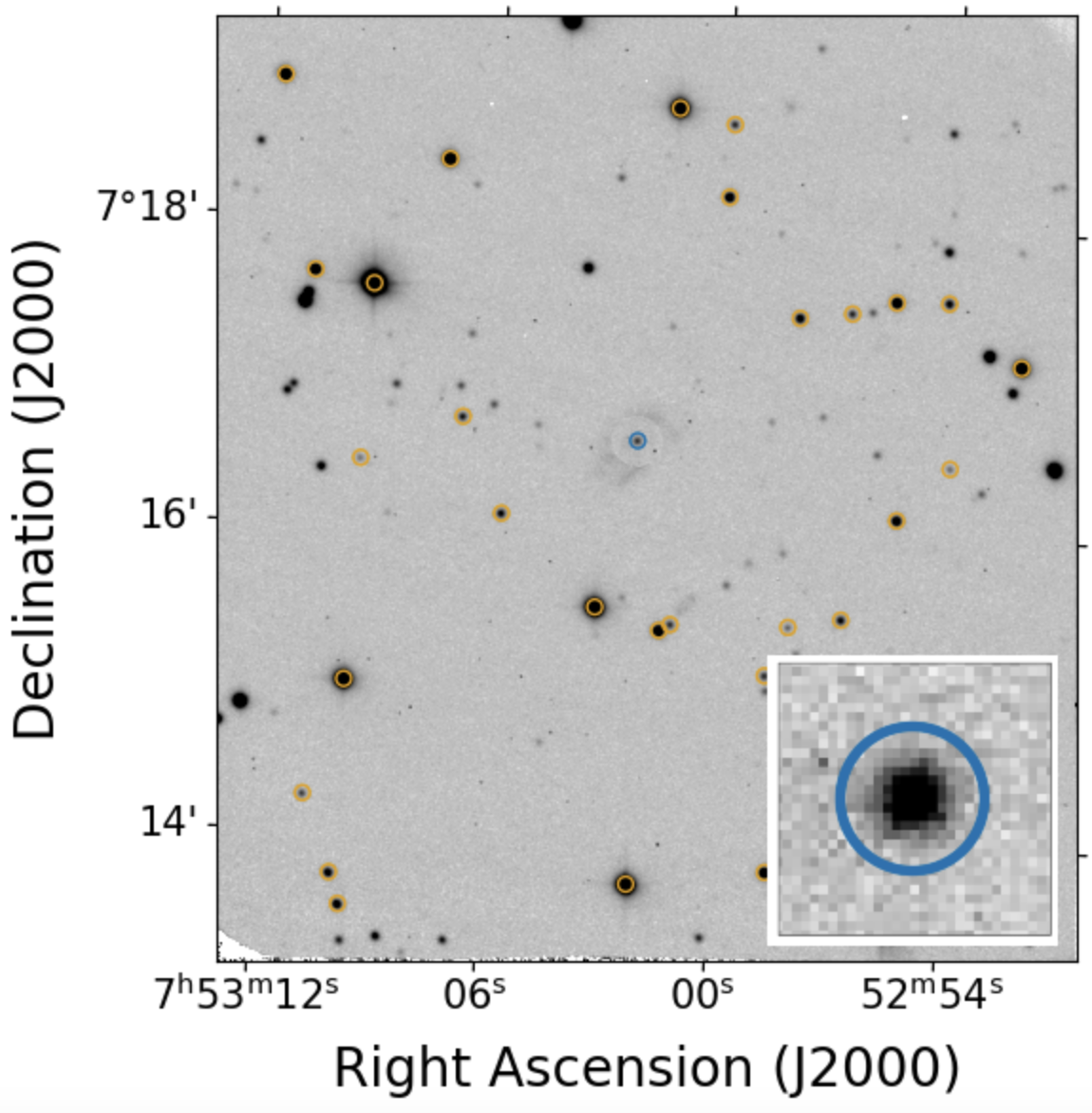}
    \caption{Aperture radii (orange for local sequence and blue for supernova) for one of our science images. To determine the aperture radius (the same aperture is used for all sources), we fitted a 2D gaussian to all local sequence stars and evalauted a 3$\sigma$-clipped median of the FWHM, and multiplied this by $0.6734$ to obtain the aperture radius. The inset at the bottom right is a zoom-in of the SN and its aperture.}
\label{fig:catalog_apertures}
\end{figure}

\subsubsection{Instrumental Magnitude}
To determine the natural instrumental magnitudes of our sources, we used {\sc Circular\_Aperture} from {\sc photutils.aperture}, which evaluated the total counts within a circle of a given radius. For each image, a $3\sigma$-clipped median of the distribution of FWHM of our local sequence stars was evaluated. The aperture radius of all our sources (including the SN) was then set to be $0.6731$ times the median FWHM, which is the optimum aperture size for photometry of sources whose profiles can be approximated as Gaussian.\footnote{https://wise2.ipac.caltech.edu/staff/fmasci/GaussApRadius.pdf}. An example of this procedure can be seen in Figure~\ref{fig:catalog_apertures}.

To convert the counts inside the apertures into instrumental magnitudes in the natural system $m_{inst, nat}$ (i.e., the magnitude system measured by the telescope), we used
\begin{equation}
    m_{inst,nat} = -2.5\cdot \mathrm{log_{10}}(Counts/t_{exposure}),
    \label{inst_mag}
\end{equation}
where $t_{exposure}$ is the total exposure time divided by the number of coadds (for ANDICAM NIR this is just the exposure time of a single image since the coadded images are divided by the mask).
To correct for extinction, say for the $B$-band, we used the equation
\begin{equation}
B_{inst,cor} = B_{inst,nat}-k_B\cdot X.
\label{eq:inst_mag_cor}
\end{equation}
Here $k_B$ is the extinction coefficient, $X$ is the airmass value and was taken from the image header, and $B_{inst,nat}$ is the natural instrumental magnitude from Eq.~\ref{inst_mag}.
Table~1 in \cite{2005PASP..117..810S} was used to correct for extinction the ANDICAM \textit{BVRIY}  passbands from the central wavelength of each filter. The estimated extinction coefficients are shown in Table~\ref{tab:colorterm_table}. In their work, they evaluated extinction coefficients at CTIO out to $11000$~\AA, and the value at this wavelength was $0.003$. Because this value is small, extinction corrections to the ANDICAM $JH$ filters were not applied. Likewise, when we looked at the La Silla Observatory's homepage,\footnote{\href{https://www.eso.org/sci/observing/tools/Extinction.html}{https://www.eso.org/sci/observing/tools/Extinction.html}} there are only extinction coefficients out to $9000$~\AA, and the value at this wavelength was $0.01$.
Therefore, the extinction coefficients in $JHK_s$ were assumed to be zero.

\subsection{Image Zeropoints}\label{sec:zeropoint}

The method used to evaluate the zeropoint for the science images is described here.

\begin{figure}[t]
    \centering
    \includegraphics[width=\columnwidth]{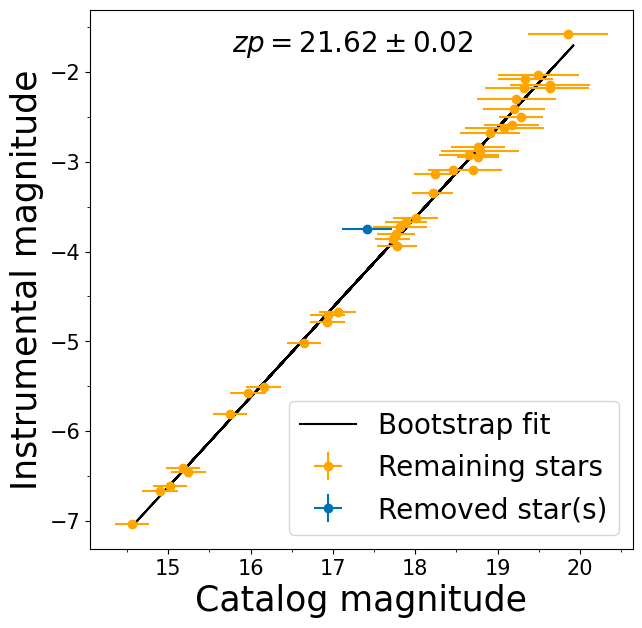}
    \caption{Example of calculating an image zeropoint using bootstrapping with sigma clipping as outlier rejection, where the standard deviation from Chauvenet's criteria was used. At the top of the panel we show the image zeropoint and its error, the black line is the fit using a linear function in the form  of $f(x)=x+b$, and the orange points are the stars that remain after doing sigma clipping, and the blue point(s) are rejected stars. The zeropoint is the intercept value from the fit.} 
    \label{fig:image_zeropoint}
\end{figure}

\subsubsection{ANDICAM optical and SOFI}\label{sec:andicam_optical_zp}

After making cuts described in the previous section, We determined the zeropoint from sigma clipping using Chauvenet’s criterion. A linear function with a slope of unity in the form of $f(x)=x+b$ was fitted, where the zeropoint is the intercept in the instrumental versus catalog magnitude plot, where for a given star the instrumental magnitude is given by Eq.~\ref{eq:inst_mag_cor}. Figure~\ref{fig:image_zeropoint} shows an example of evaluating the image zeropoint, where sigma clipping was used to remove outliers. The error was set to be the weigted mean of the residuals between the fit and the data. Notice that the zeropoints estimated here were not the final zeropoint; they would only be used to determine color term coefficients through an iterative process. 

More generally, the y-axis in Figure~\ref{fig:image_zeropoint} is the $B$-band instrumental magnitude in the ANDICAM natural system $B_{inst,nat}$, while the x-axis is $B_{cat,std}-CT_B\cdot(B-V)_{cat,std}$, where the "cat,std" subscript denotes a catalog magnitude in the Johnson standard system, and $CT_B$ is the $B$-band color term. Initially, the color term was unknown so the x-axis in this case was $B_{cat,std}$. The idea was to use the initial image zeropoint point to estimate an initial color term. With the initial color term, a new image zeropoint could be estimated from the $B_{cat,std}-CT_B\cdot(B-V)_{cat,std}$ versus $B_{inst,nat}$ plot, which in turn would yield a new color term and so forth. This iterative process will be described in greater detail in  Sect.~\ref{sec:colorterms}.

\subsubsection{ANDICAM NIR}

\begin{figure}[!t]
    \centering
    \includegraphics[trim=0cm 0cm 0cm 0cm,clip=True,width=\columnwidth]{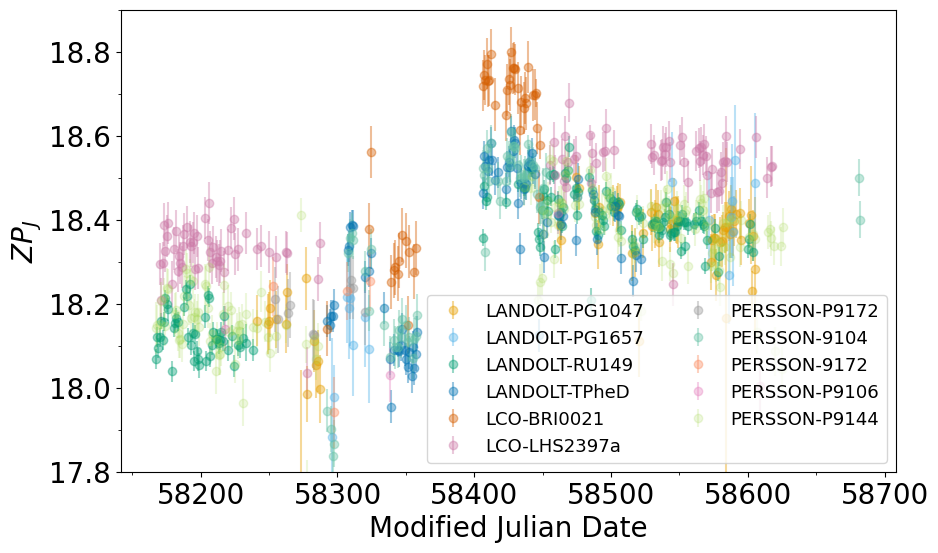}
    \caption{$J$-band zeropoints for different nights. The zeropoints were computed using $J_{inst}-J_{2}$, where $J_{inst}$ is the instrumental magnitude evaluated from aperture photometry and $J_{2}$ is the 2MASS catalog magnitude in the $J$~band. It is seen that standards taken at the same night give different zeropoints, making the standards too uncertain.} 
    \label{fig:standard_zp}
\end{figure}

We evaluated the zeropoints of our standard field observations by plotting them as function of observation date, as shown for the $J$~band in Fig.~\ref{fig:standard_zp}. Even though standards were observed at the same nights, the zeropoints could differ by half a magnitude, which should not be the case. This issue was also seen in \cite{Wang_2020}, who found differences in the nightly zeropoints of NIR standards using ANDICAM, and therefore they used field stars that had magnitudes in the 2MASS catalog. Attempts to apply color-term coefficients did not significantly improve the calibration. Consequently, these standard star fields were not used; instead, local sequence stars with catalog magnitudes from REFCAT2 were adopted to determine the image zeropoint.

\subsection{Color term coefficients}\label{sec:colorterms}

Ideally, color-term coefficients should be determined for each field-of-view on a given night, but the limited number of local sequence stars in most images makes this unfeasible.  Instead, stars observed across all epochs were combined. To determine the color-term coefficients, for example, for the $B$ band, the quantity $B_{cat,std} - B_{inst,cor} - ZP_{B,img}$ was plotted against $(B-V)_{cat,std}$ for all $B$-band images, and the slope of the relation was determined. Here $ZP_{B,img}$ is the image zeropoint, which was determined following the procedure in Sect. \ref{sec:andicam_optical_zp}.
\\
\\
For a given star observed at various epochs, a $3$-sigma clipping was made to remove outliers, and if the residuals were greater than $0.5$ mag for a given star, it was rejected. By doing this, potential variable stars were also excluded. Addtionally, we only selected stars whose uncertainties in instrumental magnitudes were less than $0.05$ mag. The slope of the $B_{cat,std} - B_{inst,cor} - ZP_{B,img}$ versus $(B-V)_{cat}$ plot was determined using Markov Chain Monte Carlo (MCMC) with the Emcee package\footnote{\href{https://emcee.readthedocs.io/en/stable/}{https://emcee.readthedocs.io/en/stable/}}. 

\begin{figure}[!t]
    \centering
    \includegraphics[width=\columnwidth]{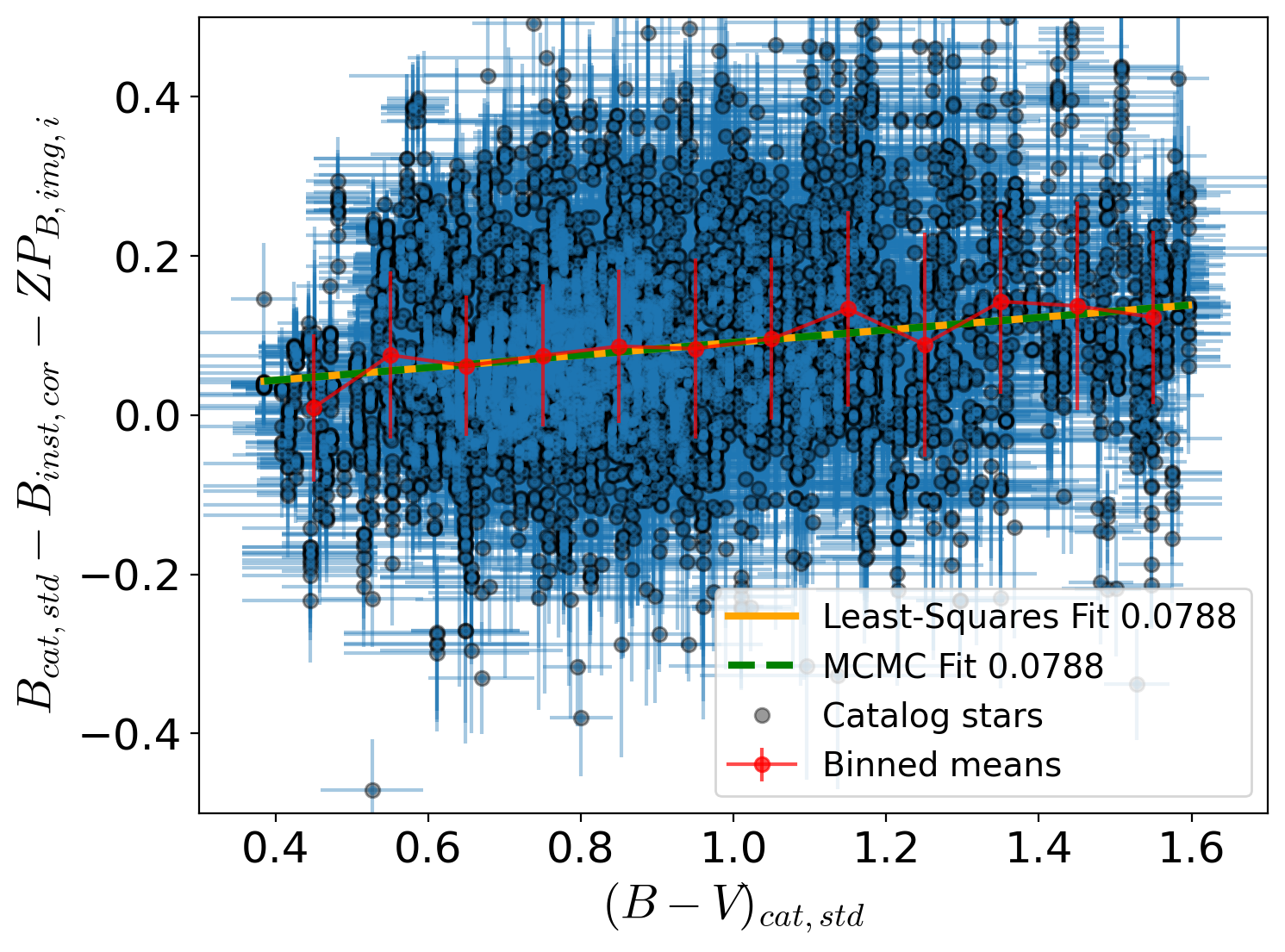}    \caption{Sigma-clipped $B_{cat,std}-B_{inst}-ZP_{B,img,i}$ vs. $(B-V)_{cat,std}$ for all $B$~band images in the sample (black). We estimated binned means (red) and their uncertainties and removed very red/blue stars if the bins did not agree with the fit.  The legend gives the slope using Scipy least squares (orange) and the MCMC fitting code Emcee (green), which shows that the values agree.
    } 
    
    \label{fig:emcee}
\end{figure}

The fitted slope to this data only provided an approximate color term coefficient, as the $ZP_{B,img}$ obtained in the previous step, which was used to ensemble the data from all night into the same scale, did not include any color term correction. Therefore, an iterative approach was taken, where we recomputed the zero-point using the catalog magnitudes corrected by the newly computed color-terms. Subsequently, the new zero point value $ZP_{B,img,i}$ was used in the process of obtaining a new color-term coefficient $CT_{B,i}$. This process was iterated until the color-term value converged to four significant digits. This iterative process was essential because the zero-point and color-term coefficient were interdependent and could only be solved by cyclic computation of the coefficients.
More specifically, $CT_{B,i}$, was used to color correct the catalog magnitude as  $B_{cat,std}-CT_{B,i} \cdot (B-V)_{cat,std}$, and a new image zeropoint, $ZP_{B,img,i}$, was then evaluated in the same manner as in Fig.~\ref{fig:image_zeropoint} from the $B_{inst,nat}$ versus $B_{cat,std}-CT_{B,i} \cdot (B-V)_{cat,std}$ plot. 
All $B$-band images were then used to determine a new slope from the $B_{cat,std}-B_{inst,cor}-ZP_{B,img,i}$ versus $(B-V)_{cat,std}$ plot.
The iterations continued until the slope value converged to four significant figures, which typically occurred after five iterations.
Figure~\ref{fig:emcee} shows the results of the Emcee fit for the final iteration  (green) along with the slope determined from a least squares fit in Scipy\footnote{\href{https://docs.scipy.org/doc/scipy/}{https://docs.scipy.org/doc/scipy/}} (orange), where it is seen that the Scipy and Emcee results agree. 
The mean and standard deviation were estimated for multiple color bins, and stars with extreme colors were excluded if the binned mean value was a $1\sigma$ outlier from the fitted relation.

\begin{table*}
\caption{Color term coefficients for all filters using Emcee along with extinction coefficients.}
\begin{center}
\resizebox{\textwidth}{!}{ 
\begin{tabular}{ccccccccccc}
\hline \hline
Filter &
$B$&
$V$&
$R$&
$I$&
Andicam $Y$&
Andicam $J$&
Andicam $H$&
SOFI $J$&
SOFI $H$ 
&SOFI $K_s$\\
\hline
Color\tablefootmark{a}& $B-V$ & $B-V$ & $R-I$ & $R-I$ & $z-Y$ & $J-H$ & $J-H$ & $J-H$ & $J-H$ & $H-K_s$ \\
\hline
Color term coefficeint & 0.079 & $-0.063$ & 0.126 & 0.001 & $-0.997$ &  $0.113$ & $-0.109$ &  0.095 & $-0.179$ & $-0.464$ \\ 
\hline
Color term error\tablefootmark{b} & 0.003 & 0.001 & 0.001 &  0.001 & 0.147 & 0.009 & 0.011 & 0.015 & 0.018 & 0.037  \\ 
\hline
Extinction coefficient & 0.251 & 0.149 & 0.098 & 0.066 & 0.007 & 0 & 0 & 0 & 0 & 0 \\ 
\hline
\end{tabular}
}
\end{center}
\tablefoot{\\
\tablefoottext{a}{This refers to the catalog magnitude color used to determine the color term for a given filter. For example, $(B-V)_{cat}$ was used to the determine the $B$-band color term coefficient.} 
\\
\tablefoottext{b}{Errors correspond to the 16th and 84th quantile, extinction coefficients were assumed to be 0 for the NIR filters}
}
\label{tab:colorterm_table}
\end{table*}

Table~\ref{tab:colorterm_table} shows the estimated color term coefficients for ANDICAM and SOFI, and Fig. \ref{fig:color_terms} in the Appendix shows the corresponding fit for each filter. We noticed that the color term coefficient for $Y$ was quite large, which was likely due to a small number of local sequence stars and narrow color range. For $K_s$ the color term was large, but the error was only around $5\%$, which suggested that the SOFI $K_s$ filter was different from the 2MASS $K_s$. After the color term coefficients were evaluated, it was now possible to evaluate the SN magnitude in the natural system. Firstly, the catalog magnitudes of the local sequence stars in the standard system were transformed into the natural system. The relation between natural and standard magnitudes for the $B$~band is 
\begin{equation}
    B_{cat,nat} = B_{cat,std} - CT_{B}\cdot(B-V)_{cat,std},
\end{equation}
where $CT_{B}$ is the $B$-band color term coefficient. The newly obtained natural catalog magnitudes were then used to evaluate the natural image zeropoint $ ZP_{img,nat}$ using the same procedure described in Sect.~\ref{sec:andicam_optical_zp}, with the only difference being that the catalog magnitude has been transformed to the ANDICAM natural system, and the instrumental magnitudes are in the natural system and not corrected for extinction. With the natural image zeropoint, the apparent magnitude of the SN in the natural system is obtained following
\begin{equation}
    m_{B,SN, nat} = B_{SN,inst,nat} + ZP_{img, nat},  
\end{equation}
where $ m_{SN,nat}$ is the instrumental magnitude of the SN. Notice that no extinction correction are applied to the local sequence stars/SN, as they were observed in the same small FoV and therefore experience the same amount of airmass and extinction. 

\begin{figure*}[t]
    \centering
    \includegraphics[trim=0cm 0cm 0cm 0cm, width=\textwidth]{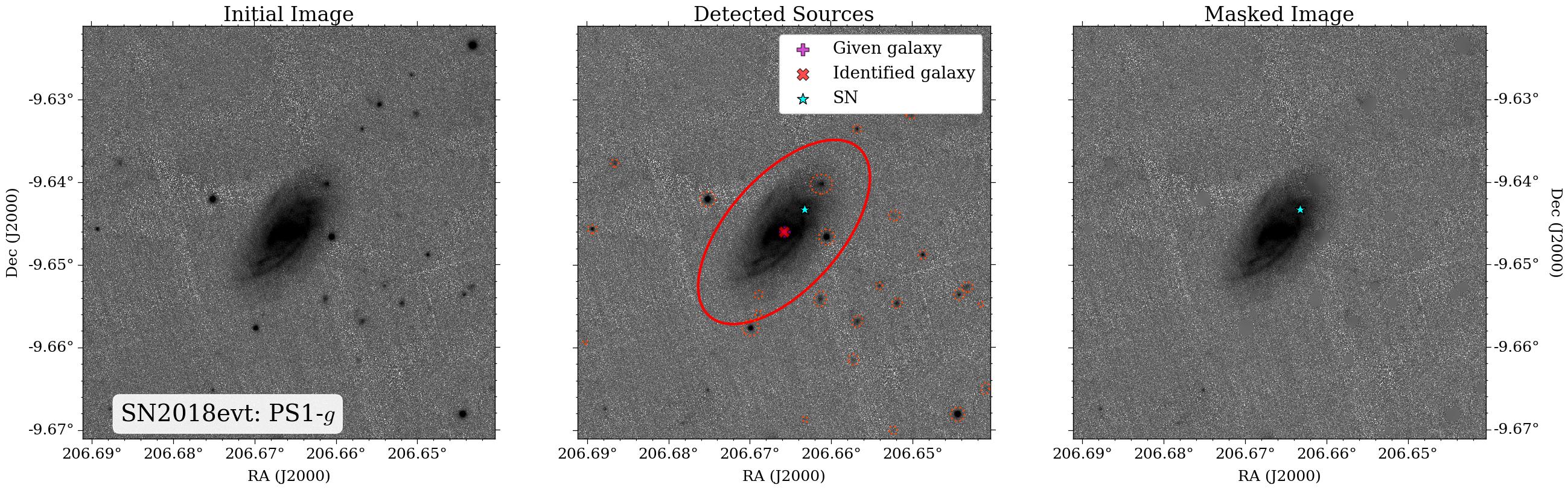}
    \caption{Example of masking out stars using HostPhot. \textit{Left:} Initial image downloaded from Pan-STARRS. \textit{Middle:} Objects detected in HostPhot along with their apertures. \textit{Right:} Image with sources removed.} 
    \label{fig:masking}
\end{figure*}

\begin{figure}[t]
    \centering
    \includegraphics[width=\columnwidth]{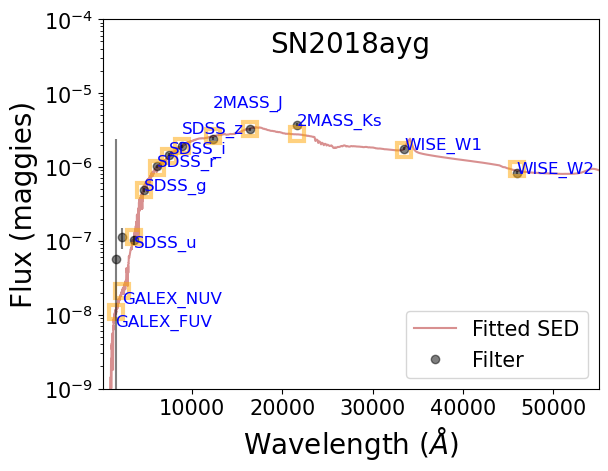}
    \caption{Example of SED fitting in Prospector. Black points are the flux (in maggies) for our survey images. The fitted SED profile is marked in brown, and the yellow boxes indicate the model value for a given survey. Notice this is the initially detected aperture, not the final aperture measured by HostPhot.} 
    \label{fig:prospector_fit}
\end{figure}


\section{Light curve fitting}\label{sec:cal}

We here describe the light curve fitting techniques {\sc SALT3-NIR}, {\sc SNooPy} and {\sc BayeSN} that were used to fit our light curves. For all three cases a flat $\Lambda CDM$ universe with $\Omega_{m0}=0.3$ and $H_0=70 km\,s^{-1}\,Mpc^{-1}$ was used. Furthermore, the dust maps from \cite{Schlegel_1998} was used to obtain the Milky Way color excess and, the reddening laws from \citep{1989ApJ...345..245C} are used to determine Milky Way extinction for SNooPy and SALT3 and \citep{1999PASP..111...63F} for BayeSN (Cardelli would have been used for BayeSN if it were possible, but this was not the case).
\subsection{SALT3-NIR}\label{sec:sncosmo}

{\sc SNCosmo}\footnote{\href{https://sncosmo.readthedocs.io}{https://sncosmo.readthedocs.io}} is a Python library for supernova cosmology analysis. To perform light curve fitting in {\sc SNCosmo}, we used the {\sc SALT3-NIR} model \citep{Pierel_2022}, which is an extension of the SALT3 model \citep{Kenworthy_2021}. This allowed us to do light-curve fitting in the NIR and had been trained on around $\sim 1000$ SN~Ia with data at optical wavelengths, and as well as 166 objects with  NIR light curves. Additionally, SALT3-NIR was also trained with optical and NIR spectra to perform K-corrections. Output parameters include amplitude $x_0$, which is related to the $B$-band peak magnitude ($B_{max} \sim -2.5\cdot log10(x_0)$), time of maximum $t_0$, stretch $x_1$, which is related the shape of the LC, and color $c$, which describes the supernova color (due to dust extinction or intrinsic properties). However, the model only goes out to 20000 \AA~and therefore $K_s$-band images from SOFI could not be included. 

For optical wavelengths, there were issues combining ATLAS data with our optical ANDICAM data; in some cases, fitting ATLAS only or ANDICAM only provided succesful fits, but when combining the two, the $B$-band fit would be overestimated by $0.2-0.3$ magnitudes with respect to the data, which was  not seen in {\sc SNooPy} and {\sc BayeSN}. Also, the same problems were present when using the {\sc SALT2} template instead. We checked the literature to see if some of our objects were observed by other telescopes. SN~2018aoz was studied by \cite{Ni_2022} (see their supplementary Figure~3) and include $B$-band light curves with three different instruments (Korea Microlensing Telescope Network -- KMTNet, Las Cumbres Observatory, and Swift Observatory). The SN peaked with an apparent magnitude of $B_{max} = 12.43 \pm 0.08$ mag (corrected for Milky Way extinction), which is in agreement with our results of $12.47\pm 0.07$, while ATLAS-only data would predict that the SN peaks at $12.7$ mag. Because our data agrees with three other observatories, we therefore chose to exclude the ATLAS data whenever ANDICAM and ATLAS were incompatible. 

Figure~\ref{fig:lc_fits} shows an example of light-curve fitting using SALT3-NIR along with BayeSN and SNooPy. From the output, we used {\sc source\_peakmag} to determine the peak $BJH$-band apparent magnitudes corrected for Milky Way extinction. 

\subsection{SNooPy}\label{sec:snoopy_fitting}

{\sc SNooPy} \citep{2014ApJ...789...32B} is a Python-based tool for modeling and analyzing Type Ia supernova photometry, calibrated using high-quality CSP light curves and templates in the \textit{uBVgriYJH} filters. SNooPy has two different parametrizations for the shape of the light-curves: $\Delta m_{15}$, which is the average decline 15 days after peak brightness when  fitting all bandpasses, and the color-stretch parameter $s_{BV}$ \citep{2015ascl.soft05023B}, which measures the timing of the $B-V$ peak with respect to the average time of 30 days past $B$-band maximum. For the light-curve fitting, we used the {\sc max\_model}, which fits the maximum magnitude for each filter individually and does $K$-corrections as well as a Milky Way exintction correction. The fitted model is of the form 
\begin{equation}
    m_{X} = T_{Y}(t^\prime,\Delta m_{15})+m_{Y}+R_{X}E(B-V)_{gal}+K_{X,Y},
\end{equation}
where $T_Y$ is the light curve template, $m_{X}$ is the observed magnitude in filter $X$, $t^\prime$ is the de-redshifted time relative to $B$-maximum, $\Delta m_{15}$ is the decline rate parameter, $m_{Y}$ is the peak magnitude in filter $Y$, $R_{X}$ is the total-to-selective absorptions for filters $X$ and $K_{XY}$ is the cross-band $K$-correction from rest-frame $X$ to observed filter $Y$. The output from using {\sc max\_model} gives multiple maxima transformed into the CSP filter systems along with time of maxima and $s_{BV}$ or $\Delta m_{15}$.

For the ANDICAM $I$-band photometry, we noticed that {\sc SNooPy} had a tendency to overestimate the secondary peak for most of our SNe while underestimates the first peak. This issue was also seen in \cite{Wee_2018}, who also found this trend in the light curve of SN~2017cbv, a nearby object ( $z\approx 0.004$). Since the $I$-band light curves were still able to reduce the uncertainties in the fitted LC parameters, the $I$-band light curves were still included. 

\subsection{BayeSN}\label{sec:bayesn}

{\sc BayeSN}\footnote{\href{https://github.com/bayesn/bayesn}{https://github.com/bayesn/bayesn}} \citep{Mandel_2021,grayling2024scalablehierarchicalbayesninference} is a hierarchical Bayesian model for SNe Ia SEDs that is continuous over time and wavelength, ranging from the optical to NIR, and the SED is modelled as a combination of distinct host galaxy dust and intrinsic spectral components. We used the \cite{2023ApJ...956..111W} model, which had been trained on the Foundation DR1 compilation \citep{Foley_2017,Jones_2019} and with \cite{Avelino_2019} in the NIR. The training data includes $BgVrizYJH$ filters, ranging from 3000-18500 \AA. Input parameters are the time of observation, flux/magnitude, flux error/magnitude error, a Milky Way color excess, and redshift. The output of {\sc BayeSN} gives a global distance modulus and time of peak maximum along with their errors. From the fits, obtained the peak magnitudes in each band (not corrected for extinction). The fitting results using the three fitters are reported in Table \ref{tab:lc_table} of the Appendix.
Figure~\ref{fig:bmax_cdf} shows the difference in $B_{max}$ for the three light curve fitters. It is seen that BayeSN gives brigther $B_{max}$ compared to SNooPy with most points lying above zero, whereas the SNe for SNooPy versus SALT are more evenly distributed around 0, and the same is  true for BayeSN versus SALT.
Our final SN light-curves are shown in Appendix~\ref{sec:appendix_light_curves} along with ZTF and ATLAS photometry.  

\begin{figure*}[!t]
    \centering
    \includegraphics[width=\textwidth]{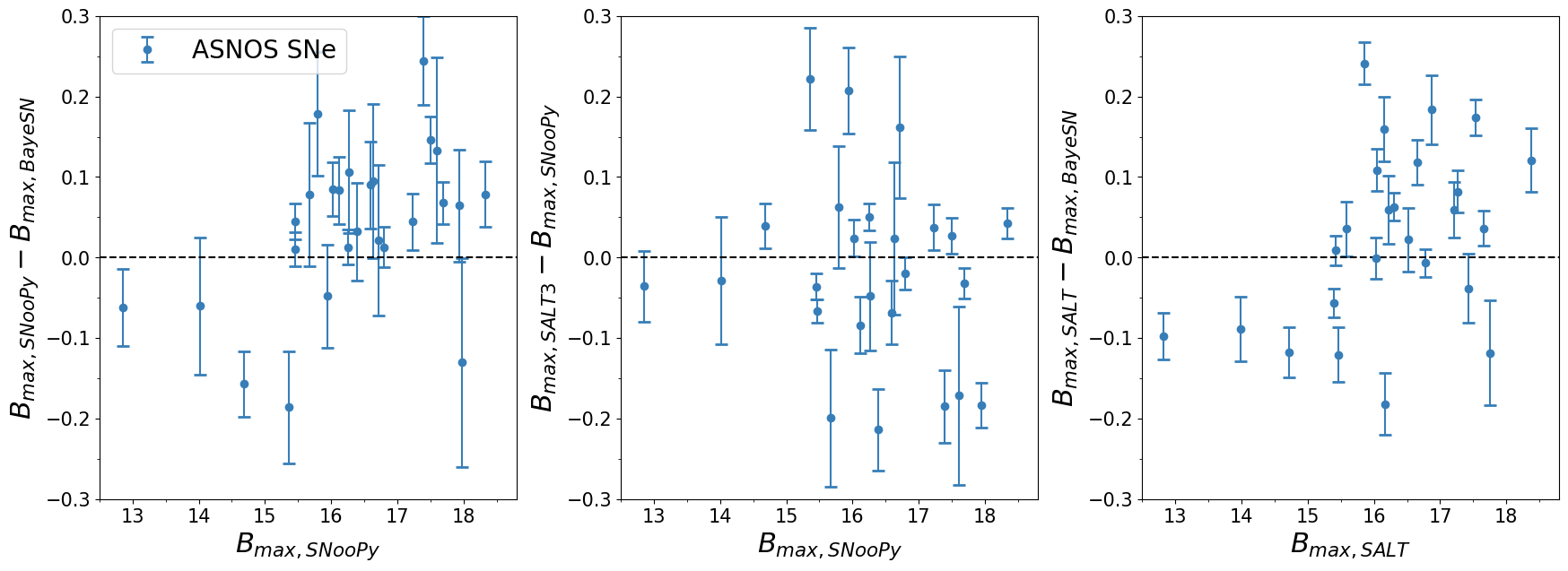}
    \caption{
    Comparison of $B_{\mathrm{max}}$ estimates from BayeSN, SALT3-NIR, and SNooPy. The light-curve fitters generally agree within $\sim0.1$~mag on average. However, BayeSN yields systematically brighter $B_{\mathrm{max}}$ values compared to SNooPy, with most SNe lying above zero. However, the SNe in the SALT3-NIR versus SNooPy appear to be evenly distributed around zero, and the same is true for SALT3-NIR versus BayeSN.  
    }    
    \label{fig:bmax_cdf}
\end{figure*}


\section{Host galaxy masses}\label{sec:host_masses}

It was shown that there is a correlation between the optical brightness of SNe Ia and the host mass of the SN \citep{Kelly_2010,Sullivan_2010,Lampeitl_2010}, where SNe found in massive galaxies were intrinsically brighter after standardization compared to SNe found in galaxies with low masses. To obtain more accurate distance measurements of the SNe, it is therefore important to take this ``mass-step'' into account. In this section we describe how to determine the host galaxy masses for our SNe.

\subsection{Host galaxy photometry}

Low resolution spectral energy distributions (SEDs) of the host galaxies at various wavelengths were constructed using the Python package {\sc HostPhot}\footnote{\href{https://hostphot.readthedocs.io/en/latest/}{https://hostphot.readthedocs.io/en/latest/}} \citep{2022JOSS....7.4508M}. Only using the SN redshift and the host galaxy coordinates, HosPhot downloads images in various filters from several surveys archives, such as GALEX, Pan-STARRS, DES, 2MASS, and unWISE. We then calculated the global apparent magnitude of the host and at the local environment of the SN within certain physical aperture (e.g. 2 kpc diameter using a flat LambdaCDM cosmology with Hubble constant of $H_0 = 70$ km~s$^{-1}$~Mpc$^{-1}$ and matter density parameter $\Omega_{m0} = 0.3$ as default) in all those filters. Before computing photometry, nearby foreground stars were masked out, which would otherwise cause the brightness of the galaxy to be overestimated. An example of this masking is seen in  Figure~\ref{fig:masking}. We defined elliptical apertures and computed global host galaxy photometry for all SNe in our sample, as well as local 1-2-3 kpc circular aperture photometry centered at the SN location.

\subsection{Host galaxy properties}\label{sec:prospector}

Prospector \citep{2017ApJ...837..170L,2021ApJS..254...22J} is a Python module for infer   ring stellar population parameters from photometry and spectroscopy ranging from ultraviolet to infrared wavelengths. It is based on forward modeling the data (i.e. predict observational data by creating a theoretical model based on known physical principles) and Monte Carlo sampling the posterior parameter distribution, which enables complex models and exploration of moderate dimensional parameter spaces. Furthermore, Prospector uses the FSPS code \citep{2009ApJ...699..486C, 2010ApJ...712..833C}, which allowed us derive spectra of stellar populations. Emcee was used to obtain the posterior distributions of our host parameters, and for star formation history (SFH), a non-parametric SFH was used, which divides the galaxy’s lifetime into a series of time bins and estimates the star formation rate (SFR) in each bin independently. More specifically, the ``continuity SFH'' was chosen, which, as described in \citep{2019ApJ...876....3L}, allowed us to obtain a piecewise constant SFH without any sharp transitions in the SFH as function of time. In order to obtain recent values in SFH, the first two bins were fixed between 0-30 Myr and 30-100 Myr, and the oldest time bin was fixed to $0.9t_{univ}$, where $t_{univ}$ is the age of the Universe (we assumed a flat $\Lambda CDM$ universe with $\Omega_{m0}=0.3$ and $H_0=70$ km/s/Mpc, which is the same as for the light curve fitters) at a given observed redshift. The remaining bins were equally spaced in time logarithmically between 100Myr and $0.9t_{univ}$. Finally, the initial mass function from \citep{2002Sci...295...82K} was used, which is a broken power law with an exponent of $-$0.3 for $0.01<m/M_{\odot}<0.08$, $-$1.3 for $0.08<m/M_{\odot}<0.5$, and $-$2.3 at $0.5<m/M_{\odot}<100$. To run Prospector, one needed the galaxy SEDs from HostPhot and the redshift (held fixed). To avoid overfitting and reduce the computing time, only one filter for each wavelength regime was be used (in other words, only one $r$-band was selected if there were multiple $r$-band filters available). So, for the optical regime, we prioritized, based on how deep the surveys went and the number of filters, DES, SDSS, PanSTARRS, Legacy Survey and SkyMapper, in this order. In those cases where the galaxy was too faint and result in negative fluxes (but consistent with 0), the flux was set to zero and the uncertainty to the $1$-sigma upper limit, as suggested in the Prospector documentation\footnote{https://prospect.readthedocs.io/en/latest/faq.html}. In the case of NIR data, the prioritization was UKIDSS, VISTA and 2MASS. Finally, for the mid-infrared data from unWISE, only the $W1$ ($\lambda_{mean}=33526$ Å) and $W2$ ($\lambda_{mean}=46028$ Å) filters were used, as the images in the far-infrared usually had very low resolution. Figure~\ref{fig:prospector_fit} shows an example of SED fitting using Prospector for the host galaxy of SN~2018ayg LEDA 1573011. The global photometry from HostPhot as well as the Prospector parameters can be seen in Table~\ref{tab:host_table}, while the local photometry can be seen in Table~\ref{tab:host_table_local}.


\section{Summary and conclusion}\label{sec:discconc}

In this paper, we presented SMARTS-ANDICAM optical and NIR data for 41 SNe, complemented by NTT-SOFI NIR imaging and described the data reduction and analysis procedures in detail. Given the significantly lower number of publicly available NIR observations of SNe ($\sim$300) compared to optical ($\sim$6000), the ASNOS sample increases by approximately 10\% of the current NIR SN Ia dataset.

The photometry of the ANDICAM NIR images could have been significantly improved if our standard stars had been more consistent with one another when observed at the same night, which, as demonstrated by Figure~\ref{fig:standard_zp}, was not the case. Instead, we had to rely on faint local sequence stars within a limited field of view (FoV) of only 2.4$\times$2.4 arcmin, with only $3-4$ stars available to determine the image zeropoint using the catalog magnitudes from ATLAS-REFCAT2. Ideally, template images for subtraction would have been taken with the same instrument after the SN had faded. However, as the ANDICAM instrument was decommissioned in 2019, this was not possible. Additionally, the SOFI instrument ceased operation around 2023, just as our plan was to acquire template images, so archival data from different instruments had to be used instead. We also observed differences in color-term determinations from broad-band photometry as compared to synthetic photometry computed using the CALSPEC spectrophotometric standards. However, since the color term coefficients remained small in both cases, the impact on the apparent magnitudes of the SNe is minimal.

The resulting light curves were fitted using the three most widely used methods: SNCosmo with the SALT3-NIR template, SNooPy, and BayeSN. Finally, HOSTPHOT was used to derive spectral energy distributions (SEDs) of the host galaxies and local environments, and we employed Prospector to perform stellar population synthesis and estimate key galaxy parameters. After applying standard light-curve selection cuts, a final sample of approximately 20 SNe with NIR observations was retained. This first paper serves as a description of the data reduction process and a data release, reporting all light-curve parameters and host galaxy masses. In a subsequent paper, we will combine literature SNe with our ANDICAM sample to measure extragalactic SN distances, construct Hubble diagrams, analyze differences between optical and NIR observations, and explore correlations between NIR Hubble residuals and host galaxy properties using various techniques such as host galaxy mass and specific star formation rate. 

\subsubsection*{Data Availability}
Tables and data will only be available in electronic form at the CDS via anonymous ftp to cdsarc.u-strasbg.fr (130.79.128.5) or via http://cdsweb.u-strasbg.fr/cgi-bin/qcat?J/A+A/.

\begin{acknowledgements}
The SNICE group acknowledges financial support from the Spanish Ministerio de Ciencia e Innovaci\'on (MCIN) and the Agencia Estatal de Investigaci\'on (AEI) 10.13039/501100011033 under the PID2020-115253GA-I00 HOSTFLOWS and the PID2023-151307NB-I00 SNNEXT projects, from Centro Superior de Investigaciones Cient\'ificas (CSIC) under projects PIE 20215AT016, ILINK23001, and the program Unidad de Excelencia Mar\'ia de Maeztu CEX2020-001058-M, and from the Departament de Recerca i Universitats de la Generalitat de Catalunya through the 2021-SGR-01270 grant.
We acknowledge the financial support from the María de Maeztu Thematic Core at ICE-CSIC.
S. Bose, M.D. Stritzinger, C. Soerensen, and the Aarhus-Barcelona FLOWS project are funded by the Independent Research Fund Denmark (IRFD, grant number  10.46540/2032-00022B) and by an Aarhus University Research Foundation Nova project (AUFF-E-2023-9-28).
\\
\indent This research has used data from the SMARTS 1.3-m telescope, which is operated as part of the SMARTS Consortium.
Based on observations at NSF Cerro Tololo Inter-American Observatory, NSF NOIRLab (NOIRLab Prop. ID NOAO-18A-0047; NOAO-18B-0016; NOAO-19A-0081; PI: L. Galbany; and ID 2025A-754097; PI: K. Phan), which is managed by the Association of Universities for Research in Astronomy (AURA) under a cooperative agreement with the U.S. National Science Foundation.
This work is based (in part) on observations collected at the European Organisation for Astronomical Research in the Southern Hemisphere, Chile as part of PESSTO, (the Public ESO Spectroscopic Survey for Transient Objects Survey) ESO program 188.D-3003, 191.D-0935, 197.D-1075.
This research has made use of the NASA/IPAC Extragalactic Database (NED),
which is operated by the Jet Propulsion Laboratory, California Institute of Technology,
under contract with the National Aeronautics and Space Administration.
This research has made use of the SIMBAD database, operated at CDS, Strasbourg, France.
This research made use of Astropy, a community-developed core Python package for Astronomy \citep{2018AJ....156..123A}, as well as Photutils \citep{larry_bradley_2025_14889440}. 
\end{acknowledgements}
\bibliographystyle{aa}
\bibliography{andi} 

@ARTICLE{Livio2003,
       author = {{Livio}, Mario and {Riess}, Adam G.},
        title = "{Have the Elusive Progenitors of Type Ia Supernovae Been Discovered?}",
      journal = {\apjl},
         year = 2003,
        month = sep,
       volume = {594},
       number = {2},
        pages = {L93-L94},
          doi = {10.1086/378765},
archivePrefix = {arXiv},
       eprint = {astro-ph/0308018},
 primaryClass = {astro-ph},
       adsurl = {https://ui.adsabs.harvard.edu/abs/2003ApJ...594L..93L}
}

@ARTICLE{Kashi2011,
       author = {{Kashi}, Amit and {Soker}, Noam},
        title = "{A circumbinary disc in the final stages of common envelope and the core-degenerate scenario for Type Ia supernovae}",
      journal = {\mnras},
         year = 2011,
        month = oct,
       volume = {417},
       number = {2},
        pages = {1466-1479},
          doi = {10.1111/j.1365-2966.2011.19361.x},
archivePrefix = {arXiv},
       eprint = {1105.5698},
 primaryClass = {astro-ph.SR},
       adsurl = {https://ui.adsabs.harvard.edu/abs/2011MNRAS.417.1466K}
}

@ARTICLE{Kushnir2013,
       author = {{Kushnir}, Doron and {Katz}, Boaz and {Dong}, Subo and {Livne}, Eli and {Fern{\'a}ndez}, Rodrigo},
        title = "{Head-on Collisions of White Dwarfs in Triple Systems Could Explain Type Ia Supernovae}",
      journal = {\apjl},
         year = 2013,
        month = dec,
       volume = {778},
       number = {2},
          eid = {L37},
        pages = {L37},
          doi = {10.1088/2041-8205/778/2/L37},
archivePrefix = {arXiv},
       eprint = {1303.1180},
 primaryClass = {astro-ph.HE},
       adsurl = {https://ui.adsabs.harvard.edu/abs/2013ApJ...778L..37K}
}

@ARTICLE{2022ApJ...934L...7R,
       author = {{Riess}, Adam G. and {Yuan}, Wenlong and {Macri}, Lucas M. and {Scolnic}, Dan and {Brout}, Dillon and {Casertano}, Stefano and {Jones}, David O. and {Murakami}, Yukei and {Anand}, Gagandeep S. and {Breuval}, Louise and {Brink}, Thomas G. and {Filippenko}, Alexei V. and {Hoffmann}, Samantha and {Jha}, Saurabh W. and {D'arcy Kenworthy}, W. and {Mackenty}, John and {Stahl}, Benjamin E. and {Zheng}, WeiKang},
        title = "{A Comprehensive Measurement of the Local Value of the Hubble Constant with 1 km s$^{-1}$ Mpc$^{-1}$ Uncertainty from the Hubble Space Telescope and the SH0ES Team}",
      journal = {\apjl},
         year = 2022,
        month = jul,
       volume = {934},
       number = {1},
          eid = {L7},
        pages = {L7},
          doi = {10.3847/2041-8213/ac5c5b},
archivePrefix = {arXiv},
       eprint = {2112.04510},
 primaryClass = {astro-ph.CO},
       adsurl = {https://ui.adsabs.harvard.edu/abs/2022ApJ...934L...7R}
}

@ARTICLE{1996ApJ...473...88R,
       author = {{Riess}, Adam G. and {Press}, William H. and {Kirshner}, Robert P.},
        title = "{A Precise Distance Indicator: Type IA Supernova Multicolor Light-Curve Shapes}",
      journal = {\apj},
         year = 1996,
        month = dec,
       volume = {473},
        pages = {88},
          doi = {10.1086/178129},
archivePrefix = {arXiv},
       eprint = {astro-ph/9604143},
 primaryClass = {astro-ph},
       adsurl = {https://ui.adsabs.harvard.edu/abs/1996ApJ...473...88R}
}

@ARTICLE{2021ApJ...923..237J,
       author = {{Johansson}, J. and {Cenko}, S.~B. and {Fox}, O.~D. and {Dhawan}, S. and {Goobar}, A. and {Stanishev}, V. and {Butler}, N. and {Lee}, W.~H. and {Watson}, A.~M. and {Fremling}, U.~C. and {Kasliwal}, M.~M. and {Nugent}, P.~E. and {Petrushevska}, T. and {Sollerman}, J. and {Yan}, L. and {Burke}, J. and {Hosseinzadeh}, G. and {Howell}, D.~A. and {McCully}, C. and {Valenti}, S.},
        title = "{Near-infrared Supernova Ia Distances: Host Galaxy Extinction and Mass-step Corrections Revisited}",
      journal = {\apj},
         year = 2021,
        month = dec,
       volume = {923},
       number = {2},
          eid = {237},
        pages = {237},
          doi = {10.3847/1538-4357/ac2f9e},
archivePrefix = {arXiv},
       eprint = {2105.06236},
 primaryClass = {astro-ph.GA},
       adsurl = {https://ui.adsabs.harvard.edu/abs/2021ApJ...923..237J}
}

@ARTICLE{1977SvA....21..675P,
       author = {{Pskovskii}, Iu. P.},
        title = "{Light curves, color curves, and expansion velocity of type I supernovae as functions of the rate of brightness decline}",
      journal = {\sovast},
         year = 1977,
        month = dec,
       volume = {21},
        pages = {675},
       adsurl = {https://ui.adsabs.harvard.edu/abs/1977SvA....21..675P}
}

@ARTICLE{1993ApJ...413L.105P,
       author = {{Phillips}, M.~M.},
        title = "{The Absolute Magnitudes of Type IA Supernovae}",
      journal = {\apjl},
         year = 1993,
        month = aug,
       volume = {413},
        pages = {L105},
          doi = {10.1086/186970},
       adsurl = {https://ui.adsabs.harvard.edu/abs/1993ApJ...413L.105P}
}

@ARTICLE{2018A&A...615A..45S,
       author = {{Stanishev}, V. and {Goobar}, A. and {Amanullah}, R. and {Bassett}, B. and {Fantaye}, Y.~T. and {Garnavich}, P. and {Hlozek}, R. and {Nordin}, J. and {Okouma}, P.~M. and {{\"O}stman}, L. and {Sako}, M. and {Scalzo}, R. and {Smith}, M.},
        title = "{Type Ia supernova Hubble diagram with near-infrared and optical observations}",
      journal = {\aap},
         year = 2018,
        month = jul,
       volume = {615},
          eid = {A45},
        pages = {A45},
          doi = {10.1051/0004-6361/201732357},
       adsurl = {https://ui.adsabs.harvard.edu/abs/2018A&A...615A..45S}
}

@MISC{2015ascl.soft05023B,
       author = {{Burns}, Christopher R. and {Stritzinger}, Maximilian and {Phillips}, M.~M. and {Kattner}, ShiAnne and {Persson}, S.~E. and {Madore}, Barry F. and {Freedman}, Wendy L. and {Boldt}, Luis and {Campillay}, Abdo and {Contreras}, Carlos and {Folatelli}, Gaston and {Gonzalez}, Sergio and {Krzeminski}, Wojtek and {Morrell}, Nidia and {Salgado}, Francisco and {Suntzeff}, Nicholas B.},
        title = "{SNooPy: TypeIa supernovae analysis tools}",
         year = 2015,
        month = may,
          eid = {ascl:1505.023},
        pages = {ascl:1505.023},
archivePrefix = {ascl},
       eprint = {1505.023},
       adsurl = {https://ui.adsabs.harvard.edu/abs/2015ascl.soft05023B}
}

@ARTICLE{2015ApJS..220....9F,
       author = {{Friedman}, Andrew S. and {Wood-Vasey}, W.~M. and {Marion}, G.~H. and {Challis}, Peter and {Mandel}, Kaisey S. and {Bloom}, Joshua S. and {Modjaz}, Maryam and {Narayan}, Gautham and {Hicken}, Malcolm and {Foley}, Ryan J. and {Klein}, Christopher R. and {Starr}, Dan L. and {Morgan}, Adam and {Rest}, Armin and {Blake}, Cullen H. and {Miller}, Adam A. and {Falco}, Emilio E. and {Wyatt}, William F. and {Mink}, Jessica and {Skrutskie}, Michael F. and {Kirshner}, Robert P.},
        title = "{CfAIR2: Near-infrared Light Curves of 94 Type Ia Supernovae}",
      journal = {\apjs},
         year = 2015,
        month = sep,
       volume = {220},
       number = {1},
          eid = {9},
        pages = {9},
          doi = {10.1088/0067-0049/220/1/9},
archivePrefix = {arXiv},
       eprint = {1408.0465},
 primaryClass = {astro-ph.HE},
       adsurl = {https://ui.adsabs.harvard.edu/abs/2015ApJS..220....9F}
}

@MISC{2019hst..prop15889J,
       author = {{Jha}, Saurabh W. and {Avelino}, Arturo and {Burns}, Chris and {Camacho-Neves}, Yssavo and {Dai}, Mi and {Dettman}, Kyle and {Dhawan}, Suhail and {Filippenko}, Alex V. and {Foley}, Ryan and {Friedman}, Andrew and {Galbany}, Lluis and {Garnavich}, Peter M. and {Hlozek}, Renee and {Holoien}, Thomas and {Hounsell}, Rebekah and {Hsiao}, Eric and {Jones}, David Oscar and {Kelly}, Patrick and {Kessler}, Rick and {Kirshner}, Robert P. and {Mandel}, Kaisey and {Matheson}, Thomas and {Narayan}, Gautham and {Phillips}, Mark M. and {Ponder}, Kara and {Rest}, Armin and {Riess}, Adam and {Roberts-Pierel}, Justin and {Rodney}, Steve and {Sand}, David J. and {Scolnic}, Daniel and {Stritzinger}, Maximillian and {Strolger}, Louis-Gregory and {Valenti}, Stefano},
        title = "{Supernovae in the Infrared avec Hubble}",
 howpublished = {HST Proposal. Cycle 27, ID. \#15889},
         year = 2019,
        month = jun,
        pages = {15889},
       adsurl = {https://ui.adsabs.harvard.edu/abs/2019hst..prop15889J}
}

@ARTICLE{2019PASP..131a4002H,
       author = {{Hsiao}, E.~Y. and {Phillips}, M.~M. and {Marion}, G.~H. and {Kirshner}, R.~P. and {Morrell}, N. and {Sand}, D.~J. and {Burns}, C.~R. and {Contreras}, C. and {Hoeflich}, P. and {Stritzinger}, M.~D. and {Valenti}, S. and {Anderson}, J.~P. and {Ashall}, C. and {Baltay}, C. and {Baron}, E. and {Banerjee}, D.~P.~K. and {Davis}, S. and {Diamond}, T.~R. and {Folatelli}, G. and {Freedman}, Wendy L. and {F{\"o}rster}, F. and {Galbany}, L. and {Gall}, C. and {Gonz{\'a}lez-Gait{\'a}n}, S. and {Goobar}, A. and {Hamuy}, M. and {Holmbo}, S. and {Kasliwal}, M.~M. and {Krisciunas}, K. and {Kumar}, S. and {Lidman}, C. and {Lu}, J. and {Nugent}, P.~E. and {Perlmutter}, S. and {Persson}, S.~E. and {Piro}, A.~L. and {Rabinowitz}, D. and {Roth}, M. and {Ryder}, S.~D. and {Schmidt}, B.~P. and {Shahbandeh}, M. and {Suntzeff}, N.~B. and {Taddia}, F. and {Uddin}, S. and {Wang}, L.},
        title = "{Carnegie Supernova Project-II: The Near-infrared Spectroscopy Program}",
      journal = {\pasp},
         year = 2019,
        month = jan,
       volume = {131},
       number = {995},
        pages = {014002},
          doi = {10.1088/1538-3873/aae961},
archivePrefix = {arXiv},
       eprint = {1810.08213},
 primaryClass = {astro-ph.SR},
       adsurl = {https://ui.adsabs.harvard.edu/abs/2019PASP..131a4002H}
}

@ARTICLE{1998A&A...331..815T,
       author = {{Tripp}, Robert},
        title = "{A two-parameter luminosity correction for Type IA supernovae}",
      journal = {\aap},
         year = 1998,
        month = mar,
       volume = {331},
        pages = {815-820},
       adsurl = {https://ui.adsabs.harvard.edu/abs/1998A&A...331..815T}
}

@software{2022zndo....592747B,
       author = {{Barbary}, Kyle and {Bailey}, Stephen and {Barentsen}, Geert and {Barclay}, Tom and {Biswas}, Rahul and {Boone}, Kyle and {Craig}, Matt and {Feindt}, Ulrich and {Friesen}, Brian and {Goldstein}, Danny and {Jha}, Saurabh W. and {Jones}, David O. and {Mondon}, Florian and {Papadogiannakis}, Sem{\'e}li and {Perrefort}, Daniel and {Pierel}, Justin and {Rodney}, Steve and {Rose}, Benjamin and {Saunders}, Clare and {Sip{\H{o}}cz}, Brigitta and {Sofiatti}, Caroline and {Thomas}, Rollin C. and {van Santen}, Jakob and {Vincenzi}, Maria and {Wang}, David and {Wood-Vasey}, Michael},
        title = "{SNCosmo}",
         year = 2025,
        month = jan,
          eid = {10.5281/zenodo.592747},
          doi = {10.5281/zenodo.592747},
      version = {v2.12.0},
    publisher = {Zenodo},
       adsurl = {https://ui.adsabs.harvard.edu/abs/2022zndo....592747B}
}

@INPROCEEDINGS{2020sea..confE..37G,
       author = {{Galbany}, L.},
        title = "{SIRAH - Supernovae in the InfRAred Avec Hubble}",
    booktitle = {XIV.0 Scientific Meeting (virtual) of the Spanish Astronomical Society},
         year = 2020,
        month = jul,
          eid = {37},
        pages = {37},
       adsurl = {https://ui.adsabs.harvard.edu/abs/2020sea..confE..37G}
}

@ARTICLE{2014ApJ...789...32B,
       author = {{Burns}, Christopher R. and {Stritzinger}, Maximilian and {Phillips}, M.~M. and {Hsiao}, E.~Y. and {Contreras}, Carlos and {Persson}, S.~E. and {Folatelli}, Gaston and {Boldt}, Luis and {Campillay}, Abdo and {Castell{\'o}n}, Sergio and {Freedman}, Wendy L. and {Madore}, Barry F. and {Morrell}, Nidia and {Salgado}, Francisco and {Suntzeff}, Nicholas B.},
        title = "{The Carnegie Supernova Project: Intrinsic Colors of Type Ia Supernovae}",
      journal = {\apj},
         year = 2014,
        month = jul,
       volume = {789},
       number = {1},
          eid = {32},
        pages = {32},
          doi = {10.1088/0004-637X/789/1/32},
archivePrefix = {arXiv},
       eprint = {1405.3934},
 primaryClass = {astro-ph.CO},
       adsurl = {https://ui.adsabs.harvard.edu/abs/2014ApJ...789...32B}
}

@ARTICLE{2017AJ....154..211K,
       author = {{Krisciunas}, Kevin and {Contreras}, Carlos and {Burns}, Christopher R. and {Phillips}, M.~M. and {Stritzinger}, Maximilian D. and {Morrell}, Nidia and {Hamuy}, Mario and {Anais}, Jorge and {Boldt}, Luis and {Busta}, Luis and {Campillay}, Abdo and {Castell{\'o}n}, Sergio and {Folatelli}, Gast{\'o}n and {Freedman}, Wendy L. and {Gonz{\'a}lez}, Consuelo and {Hsiao}, Eric Y. and {Krzeminski}, Wojtek and {Persson}, Sven Eric and {Roth}, Miguel and {Salgado}, Francisco and {Ser{\'o}n}, Jacqueline and {Suntzeff}, Nicholas B. and {Torres}, Sim{\'o}n and {Filippenko}, Alexei V. and {Li}, Weidong and {Madore}, Barry F. and {DePoy}, D.~L. and {Marshall}, Jennifer L. and {Rheault}, Jean-Philippe and {Villanueva}, Steven},
        title = "{The Carnegie Supernova Project. I. Third Photometry Data Release of Low-redshift Type Ia Supernovae and Other White Dwarf Explosions}",
      journal = {\aj},
         year = 2017,
        month = nov,
       volume = {154},
       number = {5},
          eid = {211},
        pages = {211},
          doi = {10.3847/1538-3881/aa8df0},
archivePrefix = {arXiv},
       eprint = {1709.05146},
 primaryClass = {astro-ph.IM},
       adsurl = {https://ui.adsabs.harvard.edu/abs/2017AJ....154..211K}
}

@ARTICLE{2012MNRAS.425.1007B,
       author = {{Barone-Nugent}, R.~L. and {Lidman}, C. and {Wyithe}, J.~S.~B. and {Mould}, J. and {Howell}, D.~A. and {Hook}, I.~M. and {Sullivan}, M. and {Nugent}, P.~E. and {Arcavi}, I. and {Cenko}, S.~B. and {Cooke}, J. and {Gal-Yam}, A. and {Hsiao}, E.~Y. and {Kasliwal}, M.~M. and {Maguire}, K. and {Ofek}, E. and {Poznanski}, D. and {Xu}, D.},
        title = "{Near-infrared observations of Type Ia supernovae: the best known standard candle for cosmology}",
      journal = {\mnras},
         year = 2012,
        month = sep,
       volume = {425},
       number = {2},
        pages = {1007-1012},
          doi = {10.1111/j.1365-2966.2012.21412.x},
archivePrefix = {arXiv},
       eprint = {1204.2308},
 primaryClass = {astro-ph.CO},
       adsurl = {https://ui.adsabs.harvard.edu/abs/2012MNRAS.425.1007B}
}

@ARTICLE{1989ApJ...345..245C,
       author = {{Cardelli}, Jason A. and {Clayton}, Geoffrey C. and {Mathis}, John S.},
        title = "{The Relationship between Infrared, Optical, and Ultraviolet Extinction}",
      journal = {\apj},
         year = 1989,
        month = oct,
       volume = {345},
        pages = {245},
          doi = {10.1086/167900},
       adsurl = {https://ui.adsabs.harvard.edu/abs/1989ApJ...345..245C}
}

@ARTICLE{1999PASP..111...63F,
       author = {{Fitzpatrick}, Edward L.},
        title = "{Correcting for the Effects of Interstellar Extinction}",
      journal = {\pasp},
         year = 1999,
        month = jan,
       volume = {111},
       number = {755},
        pages = {63-75},
          doi = {10.1086/316293},
archivePrefix = {arXiv},
       eprint = {astro-ph/9809387},
 primaryClass = {astro-ph},
       adsurl = {https://ui.adsabs.harvard.edu/abs/1999PASP..111...63F}
}

@article{Tonry_2018,
	doi = {10.1088/1538-3873/aabadf},
  
	url = {https://doi.org/10.1088%2F1538-3873%2Faabadf},
  
	year = 2018,
	month = {may},
  
	publisher = {{IOP} Publishing},
  
	volume = {130},
  
	number = {988},
  
	pages = {064505},
  
	author = {J. L. Tonry and L. Denneau and A. N. Heinze and B. Stalder and K. W. Smith and S. J. Smartt and C. W. Stubbs and H. J. Weiland and A. Rest},
  
	title = {{ATLAS}: A High-cadence All-sky Survey System},
  
	journal = {Publications of the Astronomical Society of the Pacific}
}

@misc{chambers2019panstarrs1,
      title={The Pan-STARRS1 Surveys}, 
      author={K. C. Chambers and E. A. Magnier and N. Metcalfe and H. A. Flewelling and M. E. Huber and C. Z. Waters and L. Denneau and P. W. Draper and D. Farrow and D. P. Finkbeiner and C. Holmberg and J. Koppenhoefer and P. A. Price and A. Rest and R. P. Saglia and E. F. Schlafly and S. J. Smartt and W. Sweeney and R. J. Wainscoat and W. S. Burgett and S. Chastel and T. Grav and J. N. Heasley and K. W. Hodapp and R. Jedicke and N. Kaiser and R. -P. Kudritzki and G. A. Luppino and R. H. Lupton and D. G. Monet and J. S. Morgan and P. M. Onaka and B. Shiao and C. W. Stubbs and J. L. Tonry and R. White and E. Bañados and E. F. Bell and R. Bender and E. J. Bernard and M. Boegner and F. Boffi and M. T. Botticella and A. Calamida and S. Casertano and W. -P. Chen and X. Chen and S. Cole and N. Deacon and C. Frenk and A. Fitzsimmons and S. Gezari and V. Gibbs and C. Goessl and T. Goggia and R. Gourgue and B. Goldman and P. Grant and E. K. Grebel and N. C. Hambly and G. Hasinger and A. F. Heavens and T. M. Heckman and R. Henderson and T. Henning and M. Holman and U. Hopp and W. -H. Ip and S. Isani and M. Jackson and C. D. Keyes and A. M. Koekemoer and R. Kotak and D. Le and D. Liska and K. S. Long and J. R. Lucey and M. Liu and N. F. Martin and G. Masci and B. McLean and E. Mindel and P. Misra and E. Morganson and D. N. A. Murphy and A. Obaika and G. Narayan and M. A. Nieto-Santisteban and P. Norberg and J. A. Peacock and E. A. Pier and M. Postman and N. Primak and C. Rae and A. Rai and A. Riess and A. Riffeser and H. W. Rix and S. Röser and R. Russel and L. Rutz and E. Schilbach and A. S. B. Schultz and D. Scolnic and L. Strolger and A. Szalay and S. Seitz and E. Small and K. W. Smith and D. R. Soderblom and P. Taylor and R. Thomson and A. N. Taylor and A. R. Thakar and J. Thiel and D. Thilker and D. Unger and Y. Urata and J. Valenti and J. Wagner and T. Walder and F. Walter and S. P. Watters and S. Werner and W. M. Wood-Vasey and R. Wyse},
      year={2019},
      eprint={1612.05560},
      archivePrefix={arXiv},
      primaryClass={astro-ph.IM}
}

@article{Bellm_2018,
	doi = {10.1088/1538-3873/aaecbe},
  
	url = {https://doi.org/10.1088%2F1538-3873%2Faaecbe},
  
	year = 2018,
	month = {dec},
  
	publisher = {{IOP} Publishing},
  
	volume = {131},
  
	number = {995},
  
	pages = {018002},
  
	author = {Eric C. Bellm and Shrinivas R. Kulkarni and Matthew J. Graham and Richard Dekany and Roger M. Smith and Reed Riddle and Frank J. Masci and George Helou and Thomas A. Prince and Scott M. Adams and C. Barbarino and Tom Barlow and James Bauer and Ron Beck and Justin Belicki and Rahul Biswas and Nadejda Blagorodnova and Dennis Bodewits and Bryce Bolin and Valery Brinnel and Tim Brooke and Brian Bue and Mattia Bulla and Rick Burruss and S. Bradley Cenko and Chan-Kao Chang and Andrew Connolly and Michael Coughlin and John Cromer and Virginia Cunningham and Kishalay De and Alex Delacroix and Vandana Desai and Dmitry A. Duev and Gwendolyn Eadie and Tony L. Farnham and Michael Feeney and Ulrich Feindt and David Flynn and Anna Franckowiak and S. Frederick and C. Fremling and Avishay Gal-Yam and Suvi Gezari and Matteo Giomi and Daniel A. Goldstein and V. Zach Golkhou and Ariel Goobar and Steven Groom and Eugean Hacopians and David Hale and John Henning and Anna Y. Q. Ho and David Hover and Justin Howell and Tiara Hung and Daniela Huppenkothen and David Imel and Wing-Huen Ip and {\v{Z}
}eljko Ivezi{\'{c}} and Edward Jackson and Lynne Jones and Mario Juric and Mansi M. Kasliwal and S. Kaspi and Stephen Kaye and Michael S. P. Kelley and Marek Kowalski and Emily Kramer and Thomas Kupfer and Walter Landry and Russ R. Laher and Chien-De Lee and Hsing Wen Lin and Zhong-Yi Lin and Ragnhild Lunnan and Matteo Giomi and Ashish Mahabal and Peter Mao and Adam A. Miller and Serge Monkewitz and Patrick Murphy and Chow-Choong Ngeow and Jakob Nordin and Peter Nugent and Eran Ofek and Maria T. Patterson and Bryan Penprase and Michael Porter and Ludwig Rauch and Umaa Rebbapragada and Dan Reiley and Mickael Rigault and Hector Rodriguez and Jan van Roestel and Ben Rusholme and Jakob van Santen and S. Schulze and David L. Shupe and Leo P. Singer and Maayane T. Soumagnac and Robert Stein and Jason Surace and Jesper Sollerman and Paula Szkody and F. Taddia and Scott Terek and Angela Van Sistine and Sjoert van Velzen and W. Thomas Vestrand and Richard Walters and Charlotte Ward and Quan-Zhi Ye and Po-Chieh Yu and Lin Yan and Jeffry Zolkower},
  
	title = {The Zwicky Transient Facility: System Overview, Performance, and First Results},
  
	journal = {Publications of the Astronomical Society of the Pacific}
}

@article{Perlmutter_1999,
   title={Measurements of Ω and Λ from 42 High‐Redshift Supernovae},
   volume={517},
   ISSN={1538-4357},
   url={http://dx.doi.org/10.1086/307221},
   DOI={10.1086/307221},
   number={2},
   journal={The Astrophysical Journal},
   publisher={American Astronomical Society},
   author={Perlmutter, S. and Aldering, G. and Goldhaber, G. and Knop, R. A. and Nugent, P. and Castro, P. G. and Deustua, S. and Fabbro, S. and Goobar, A. and Groom, D. E. and Hook, I. M. and Kim, A. G. and Kim, M. Y. and Lee, J. C. and Nunes, N. J. and Pain, R. and Pennypacker, C. R. and Quimby, R. and Lidman, C. and Ellis, R. S. and Irwin, M. and McMahon, R. G. and Ruiz‐Lapuente, P. and Walton, N. and Schaefer, B. and Boyle, B. J. and Filippenko, A. V. and Matheson, T. and Fruchter, A. S. and Panagia, N. and Newberg, H. J. M. and Couch, W. J. and Project, The Supernova Cosmology},
   year={1999},
   month=jun, pages={565–586} }

@article{Riess_1998,
   title={Observational Evidence from Supernovae for an Accelerating Universe and a Cosmological Constant},
   volume={116},
   ISSN={0004-6256},
   url={http://dx.doi.org/10.1086/300499},
   DOI={10.1086/300499},
   number={3},
   journal={The Astronomical Journal},
   publisher={American Astronomical Society},
   author={Riess, Adam G. and Filippenko, Alexei V. and Challis, Peter and Clocchiatti, Alejandro and Diercks, Alan and Garnavich, Peter M. and Gilliland, Ron L. and Hogan, Craig J. and Jha, Saurabh and Kirshner, Robert P. and Leibundgut, B. and Phillips, M. M. and Reiss, David and Schmidt, Brian P. and Schommer, Robert A. and Smith, R. Chris and Spyromilio, J. and Stubbs, Christopher and Suntzeff, Nicholas B. and Tonry, John},
   year={1998},
   month=sep, pages={1009–1038} }

@article{Krisciunas_2004,
   title={Hubble Diagrams of Type Ia Supernovae in the Near-Infrared},
   volume={602},
   ISSN={1538-4357},
   url={http://dx.doi.org/10.1086/382731},
   DOI={10.1086/382731},
   number={2},
   journal={The Astrophysical Journal},
   publisher={American Astronomical Society},
   author={Krisciunas, Kevin and Phillips, Mark M. and Suntzeff, Nicholas B.},
   year={2004},
   month=feb, pages={L81–L84} }

@article{Scolnic_2022,
   title={The Pantheon+ Analysis: The Full Data Set and Light-curve Release},
   volume={938},
   ISSN={1538-4357},
   url={http://dx.doi.org/10.3847/1538-4357/ac8b7a},
   DOI={10.3847/1538-4357/ac8b7a},
   number={2},
   journal={The Astrophysical Journal},
   publisher={American Astronomical Society},
   author={Scolnic, Dan and Brout, Dillon and Carr, Anthony and Riess, Adam G. and Davis, Tamara M. and Dwomoh, Arianna and Jones, David O. and Ali, Noor and Charvu, Pranav and Chen, Rebecca and Peterson, Erik R. and Popovic, Brodie and Rose, Benjamin M. and Wood, Charlotte M. and Brown, Peter J. and Chambers, Ken and Coulter, David A. and Dettman, Kyle G. and Dimitriadis, Georgios and Filippenko, Alexei V. and Foley, Ryan J. and Jha, Saurabh W. and Kilpatrick, Charles D. and Kirshner, Robert P. and Pan, Yen-Chen and Rest, Armin and Rojas-Bravo, Cesar and Siebert, Matthew R. and Stahl, Benjamin E. and Zheng, WeiKang},
   year={2022},
   month=oct, pages={113} }

@article{Peterson_2023,
   title={The DEHVILS survey overview and initial data release: high-quality near-infrared Type Ia supernova light curves at low redshift},
   volume={522},
   ISSN={1365-2966},
   url={http://dx.doi.org/10.1093/mnras/stad1077},
   DOI={10.1093/mnras/stad1077},
   number={2},
   journal={Monthly Notices of the Royal Astronomical Society},
   publisher={Oxford University Press (OUP)},
   author={Peterson, Erik R and Jones, David O and Scolnic, Daniel and Sánchez, Bruno O and Do, Aaron and Riess, Adam G and Ward, Sam M and Dwomoh, Arianna and de Jaeger, Thomas and Jha, Saurabh W and Mandel, Kaisey S and Pierel, Justin D R and Popovic, Brodie and Rose, Benjamin M and Rubin, David and Shappee, Benjamin J and Thorp, Stephen and Tonry, John L and Tully, R Brent and Vincenzi, Maria},
   year={2023},
   month=apr, pages={2478–2494} }

@article{Contreras_2010,
   title={THE CARNEGIE SUPERNOVA PROJECT: FIRST PHOTOMETRY DATA RELEASE OF LOW-REDSHIFT TYPE Ia SUPERNOVAE},
   volume={139},
   ISSN={1538-3881},
   url={http://dx.doi.org/10.1088/0004-6256/139/2/519},
   DOI={10.1088/0004-6256/139/2/519},
   number={2},
   journal={The Astronomical Journal},
   publisher={American Astronomical Society},
   author={Contreras, Carlos and Hamuy, Mario and Phillips, M. M. and Folatelli, Gastón and Suntzeff, Nicholas B. and Persson, S. E. and Stritzinger, Maximilian and Boldt, Luis and González, Sergio and Krzeminski, Wojtek and Morrell, Nidia and Roth, Miguel and Salgado, Francisco and Maureira, María José and Burns, Christopher R. and Freedman, W. L. and Madore, Barry F. and Murphy, David and Wyatt, Pamela and Li, Weidong and Filippenko, Alexei V.},
   year={2010},
   month=jan, pages={519–539} }

@article{Stritzinger_2011,
   title={THE CARNEGIE SUPERNOVA PROJECT: SECOND PHOTOMETRY DATA RELEASE OF LOW-REDSHIFT TYPE Ia SUPERNOVAE},
   volume={142},
   ISSN={1538-3881},
   url={http://dx.doi.org/10.1088/0004-6256/142/5/156},
   DOI={10.1088/0004-6256/142/5/156},
   number={5},
   journal={The Astronomical Journal},
   publisher={American Astronomical Society},
   author={Stritzinger, Maximilian D. and Phillips, M. M. and Boldt, Luis N. and Burns, Chris and Campillay, Abdo and Contreras, Carlos and Gonzalez, Sergio and Folatelli, Gastón and Morrell, Nidia and Krzeminski, Wojtek and Roth, Miguel and Salgado, Francisco and DePoy, D. L. and Hamuy, Mario and Freedman, Wendy L. and Madore, Barry F. and Marshall, J. L. and Persson, Sven E. and Rheault, Jean-Philippe and Suntzeff, Nicholas B. and Villanueva, Steven and Li, Weidong and Filippenko, Alexei V.},
   year={2011},
   month=oct, pages={156} }

@ARTICLE{2019PASP..131a4001P,
       author = {{Phillips}, M.~M. and {Contreras}, Carlos and {Hsiao}, E.~Y. and {Morrell}, Nidia and {Burns}, Christopher R. and {Stritzinger}, Maximilian and {Ashall}, C. and {Freedman}, Wendy L. and {Hoeflich}, P. and {Persson}, S.~E. and {Piro}, Anthony L. and {Suntzeff}, Nicholas B. and {Uddin}, Syed A. and {Anais}, Jorge and {Baron}, E. and {Busta}, Luis and {Campillay}, Abdo and {Castell{\'o}n}, Sergio and {Corco}, Carlos and {Diamond}, T. and {Gall}, Christa and {Gonzalez}, Consuelo and {Holmbo}, Simon and {Krisciunas}, Kevin and {Roth}, Miguel and {Ser{\'o}n}, Jacqueline and {Taddia}, F. and {Torres}, Sim{\'o}n and {Anderson}, J.~P. and {Baltay}, C. and {Folatelli}, Gast{\'o}n and {Galbany}, L. and {Goobar}, A. and {Hadjiyska}, Ellie and {Hamuy}, Mario and {Kasliwal}, Mansi and {Lidman}, C. and {Nugent}, Peter E. and {Perlmutter}, S. and {Rabinowitz}, David and {Ryder}, Stuart D. and {Schmidt}, Brian P. and {Shappee}, B.~J. and {Walker}, Emma S.},
        title = "{Carnegie Supernova Project-II: Extending the Near-infrared Hubble Diagram for Type Ia Supernovae to z {\ensuremath{\sim}} 0.1}",
      journal = {\pasp},
         year = 2019,
        month = jan,
       volume = {131},
       number = {995},
        pages = {014001},
          doi = {10.1088/1538-3873/aae8bd},
archivePrefix = {arXiv},
       eprint = {1810.09252},
 primaryClass = {astro-ph.HE},
       adsurl = {https://ui.adsabs.harvard.edu/abs/2019PASP..131a4001P}
}

@ARTICLE{2018AJ....155..201W,
       author = {{Weyant}, Anja and {Wood-Vasey}, W.~M. and {Joyce}, Richard and {Allen}, Lori and {Garnavich}, Peter and {Jha}, Saurabh W. and {Kroboth}, Jessica R. and {Matheson}, Thomas and {Ponder}, Kara A.},
        title = "{The First Data Release from SweetSpot: 74 Supernovae in 36 Nights on WIYN+WHIRC}",
      journal = {\aj},
         year = 2018,
        month = may,
       volume = {155},
       number = {5},
          eid = {201},
        pages = {201},
          doi = {10.3847/1538-3881/aab901},
archivePrefix = {arXiv},
       eprint = {1703.02402},
 primaryClass = {astro-ph.HE},
       adsurl = {https://ui.adsabs.harvard.edu/abs/2018AJ....155..201W}
}

@article{Jones_2022,
   title={Cosmological Results from the RAISIN Survey: Using Type Ia Supernovae in the Near Infrared as a Novel Path to Measure the Dark Energy Equation of State},
   volume={933},
   ISSN={1538-4357},
   url={http://dx.doi.org/10.3847/1538-4357/ac755b},
   DOI={10.3847/1538-4357/ac755b},
   number={2},
   journal={The Astrophysical Journal},
   publisher={American Astronomical Society},
   author={Jones, D. O. and Mandel, K. S. and Kirshner, R. P. and Thorp, S. and Challis, P. M. and Avelino, A. and Brout, D. and Burns, C. and Foley, R. J. and Pan, Y.-C. and Scolnic, D. M. and Siebert, M. R. and Chornock, R. and Freedman, W. L. and Friedman, A. and Frieman, J. and Galbany, L. and Hsiao, E. and Kelsey, L. and Marion, G. H. and Nichol, R. C. and Nugent, P. E. and Phillips, M. M. and Rest, A. and Riess, A. G. and Sako, M. and Smith, M. and Wiseman, P. and Wood-Vasey, W. M.},
   year={2022},
   month=jul, pages={172} }

@article{Tonry_2012,
   title={THE Pan-STARRS1 PHOTOMETRIC SYSTEM},
   volume={750},
   ISSN={1538-4357},
   url={http://dx.doi.org/10.1088/0004-637X/750/2/99},
   DOI={10.1088/0004-637x/750/2/99},
   number={2},
   journal={The Astrophysical Journal},
   publisher={American Astronomical Society},
   author={Tonry, J. L. and Stubbs, C. W. and Lykke, K. R. and Doherty, P. and Shivvers, I. S. and Burgett, W. S. and Chambers, K. C. and Hodapp, K. W. and Kaiser, N. and Kudritzki, R.-P. and Magnier, E. A. and Morgan, J. S. and Price, P. A. and Wainscoat, R. J.},
   year={2012},
   month=apr, pages={99} }

@ARTICLE{2005PASP..117..810S,
       author = {{Stritzinger}, Maximilian and {Suntzeff}, Nicholas B. and {Hamuy}, Mario and {Challis}, Peter and {Demarco}, Ricardo and {Germany}, Lisa and {Soderberg}, A.~M.},
        title = "{An Atlas of Spectrophotometric Landolt Standard Stars}",
      journal = {\pasp},
         year = 2005,
        month = aug,
       volume = {117},
       number = {834},
        pages = {810-822},
          doi = {10.1086/431468},
archivePrefix = {arXiv},
       eprint = {astro-ph/0504244},
 primaryClass = {astro-ph},
       adsurl = {https://ui.adsabs.harvard.edu/abs/2005PASP..117..810S}
}

@article{Wang_2020,
   title={Optical and Near-infrared Observations of the Nearby SN Ia 2017cbv},
   volume={904},
   ISSN={1538-4357},
   url={http://dx.doi.org/10.3847/1538-4357/abba82},
   DOI={10.3847/1538-4357/abba82},
   number={1},
   journal={The Astrophysical Journal},
   publisher={American Astronomical Society},
   author={Wang, Lingzhi and Contreras, Carlos and Hu, Maokai and Hamuy, Mario A. and Hsiao, Eric Y. and Sand, David J. and Anderson, Joseph P. and Ashall, Chris and Burns, Christopher R. and Chen, Juncheng and Diamond, Tiara R. and Davis, Scott and Förster, Francisco and Galbany, Lluís and González-Gaitán, Santiago and Gromadzki, Mariusz and Hoeflich, Peter and Li, Wenxiong and Marion, G. H. and Morrell, Nidia and Pignata, Giuliano and Prieto, Jose L. and Phillips, Mark M. and Shahbandeh, Melissa and Suntzeff, Nicholas B. and Valenti, Stefano and Wang, Lifan and Wang, Xiaofeng and Young, D. R. and Yu, Lixin and Zhang, Jujia},
   year={2020},
   month=nov, pages={14} }

@article{Wee_2018,
   title={Optical and Infrared Photometry of the nearby SN 2017cbv},
   volume={863},
   ISSN={1538-4357},
   url={http://dx.doi.org/10.3847/1538-4357/aacd4e},
   DOI={10.3847/1538-4357/aacd4e},
   number={1},
   journal={The Astrophysical Journal},
   publisher={American Astronomical Society},
   author={Wee, Jerrick and Chakraborty, Nilotpal and Wang, Jiayun and Penprase, Bryan Edward},
   year={2018},
   month=aug, pages={90} }

@article{Pierel_2022,
   title={SALT3-NIR: Taking the Open-source Type Ia Supernova Model to Longer Wavelengths for Next-generation Cosmological Measurements},
   volume={939},
   ISSN={1538-4357},
   url={http://dx.doi.org/10.3847/1538-4357/ac93f9},
   DOI={10.3847/1538-4357/ac93f9},
   number={1},
   journal={The Astrophysical Journal},
   publisher={American Astronomical Society},
   author={Pierel, J. D. R. and Jones, D. O. and Kenworthy, W. D. and Dai, M. and Kessler, R. and Ashall, C. and Do, A. and Peterson, E. R. and Shappee, B. J. and Siebert, M. R. and Barna, T. and Brink, T. G. and Burke, J. and Calamida, A. and Camacho-Neves, Y. and Jaeger, T. de and Filippenko, A. V. and Foley, R. J. and Galbany, L. and Fox, O. D. and Gomez, S. and Hiramatsu, D. and Hounsell, R. and Howell, D. A. and Jha, S. W. and Kwok, L. A. and Pérez-Fournon, I. and Poidevin, F. and Rest, A. and Rubin, D. and Scolnic, D. M. and Shirley, R. and Strolger, L. G. and Tinyanont, S. and Wang, Q.},
   year={2022},
   month=oct, pages={11} }

@misc{descollaboration2024darkenergysurveycosmology,
      title={The Dark Energy Survey: Cosmology Results With ~1500 New High-redshift Type Ia Supernovae Using The Full 5-year Dataset}, 
      author={{DES Collaboration} and T. M. C. Abbott and M. Acevedo and M. Aguena and A. Alarcon and S. Allam and O. Alves and A. Amon and F. Andrade-Oliveira and J. Annis and P. Armstrong and J. Asorey and S. Avila and D. Bacon and B. A. Bassett and K. Bechtol and P. H. Bernardinelli and G. M. Bernstein and E. Bertin and J. Blazek and S. Bocquet and D. Brooks and D. Brout and E. Buckley-Geer and D. L. Burke and H. Camacho and R. Camilleri and A. Campos and A. Carnero Rosell and D. Carollo and A. Carr and J. Carretero and F. J. Castander and R. Cawthon and C. Chang and R. Chen and A. Choi and C. Conselice and M. Costanzi and L. N. da Costa and M. Crocce and T. M. Davis and D. L. DePoy and S. Desai and H. T. Diehl and M. Dixon and S. Dodelson and P. Doel and C. Doux and A. Drlica-Wagner and J. Elvin-Poole and S. Everett and I. Ferrero and A. Ferté and B. Flaugher and R. J. Foley and P. Fosalba and D. Friedel and J. Frieman and C. Frohmaier and L. Galbany and J. García-Bellido and M. Gatti and E. Gaztanaga and G. Giannini and K. Glazebrook and O. Graur and D. Gruen and R. A. Gruendl and G. Gutierrez and W. G. Hartley and K. Herner and S. R. Hinton and D. L. Hollowood and K. Honscheid and D. Huterer and B. Jain and D. J. James and N. Jeffrey and E. Kasai and L. Kelsey and S. Kent and R. Kessler and A. G. Kim and R. P. Kirshner and E. Kovacs and K. Kuehn and O. Lahav and J. Lee and S. Lee and G. F. Lewis and T. S. Li and C. Lidman and H. Lin and U. Malik and J. L. Marshall and P. Martini and J. Mena-Fernández and F. Menanteau and R. Miquel and J. J. Mohr and J. Mould and J. Muir and A. Möller and E. Neilsen and R. C. Nichol and P. Nugent and R. L. C. Ogando and A. Palmese and Y. -C. Pan and M. Paterno and W. J. Percival and M. E. S. Pereira and A. Pieres and A. A. Plazas Malagón and B. Popovic and A. Porredon and J. Prat and H. Qu and M. Raveri and M. Rodríguez-Monroy and A. K. Romer and A. Roodman and B. Rose and M. Sako and E. Sanchez and D. Sanchez Cid and M. Schubnell and D. Scolnic and I. Sevilla-Noarbe and P. Shah and J. Allyn. Smith and M. Smith and M. Soares-Santos and E. Suchyta and M. Sullivan and N. Suntzeff and M. E. C. Swanson and B. O. Sánchez and G. Tarle and G. Taylor and D. Thomas and C. To and M. Toy and M. A. Troxel and B. E. Tucker and D. L. Tucker and S. A. Uddin and M. Vincenzi and A. R. Walker and N. Weaverdyck and R. H. Wechsler and J. Weller and W. Wester and P. Wiseman and M. Yamamoto and F. Yuan and B. Zhang and Y. Zhang},
      year={2024},
      eprint={2401.02929},
      archivePrefix={arXiv},
      primaryClass={astro-ph.CO},
      url={https://arxiv.org/abs/2401.02929}, 
}

@article{Johnson_2021,
   title={Stellar Population Inference with Prospector},
   volume={254},
   ISSN={1538-4365},
   url={http://dx.doi.org/10.3847/1538-4365/abef67},
   DOI={10.3847/1538-4365/abef67},
   number={2},
   journal={The Astrophysical Journal Supplement Series},
   publisher={American Astronomical Society},
   author={Johnson, Benjamin D. and Leja, Joel and Conroy, Charlie and Speagle, Joshua S.},
   year={2021},
   month=may, pages={22} }

@article{Kelly_2010,
   title={HUBBLE RESIDUALS OF NEARBY TYPE Ia SUPERNOVAE ARE CORRELATED WITH HOST GALAXY MASSES},
   volume={715},
   ISSN={1538-4357},
   url={http://dx.doi.org/10.1088/0004-637X/715/2/743},
   DOI={10.1088/0004-637x/715/2/743},
   number={2},
   journal={The Astrophysical Journal},
   publisher={American Astronomical Society},
   author={Kelly, Patrick L. and Hicken, Malcolm and Burke, David L. and Mandel, Kaisey S. and Kirshner, Robert P.},
   year={2010},
   month=may, pages={743–756} }

@article{Sullivan_2010,
   title={The dependence of Type Ia Supernovae luminosities on their host galaxies: SN Ia host galaxies},
   ISSN={1365-2966},
   url={http://dx.doi.org/10.1111/j.1365-2966.2010.16731.x},
   DOI={10.1111/j.1365-2966.2010.16731.x},
   journal={Monthly Notices of the Royal Astronomical Society},
   publisher={Oxford University Press (OUP)},
   author={Sullivan, M. and Conley, A. and Howell, D. A. and Neill, J. D. and Astier, P. and Balland, C. and Basa, S. and Carlberg, R. G. and Fouchez, D. and Guy, J. and Hardin, D. and Hook, I. M. and Pain, R. and Palanque-Delabrouille, N. and Perrett, K. M. and Pritchet, C. J. and Regnault, N. and Rich, J. and Ruhlmann-Kleider, V. and Baumont, S. and Hsiao, E. and Kronborg, T. and Lidman, C. and Perlmutter, S. and Walker, E. S.},
   year={2010},
   month=may, pages={no-no} }

@article{Lampeitl_2010,
   title={First-year Sloan Digital Sky Survey-II supernova results: consistency and constraints with other intermediate-redshift data sets},
   volume={401},
   ISSN={1365-2966},
   url={http://dx.doi.org/10.1111/j.1365-2966.2009.15851.x},
   DOI={10.1111/j.1365-2966.2009.15851.x},
   number={4},
   journal={Monthly Notices of the Royal Astronomical Society},
   publisher={Oxford University Press (OUP)},
   author={Lampeitl, H. and Nichol, R. C. and Seo, H.-J. and Giannantonio, T. and Shapiro, C. and Bassett, B. and Percival, W. J. and Davis, T. M. and Dilday, B. and Frieman, J. and Garnavich, P. and Sako, M. and Smith, M. and Sollerman, J. and Becker, A.C. and Cinabro, D. and Filippenko, A. V. and Foley, R. J. and Hogan, C. J. and Holtzman, J. A. and Jha, S. W. and Konishi, K. and Marriner, J. and Richmond, M. W. and Riess, A. G. and Schneider, D. P. and Stritzinger, M. and van der Heyden, K. J. and VanderPlas, J. T. and Wheeler, J. C. and Zheng, C.},
   year={2010},
   month=feb, pages={2331–2342} }

@ARTICLE{2015A&A...579A..40S,
       author = {{Smartt}, S.~J. and {Valenti}, S. and {Fraser}, M. and {Inserra}, C. and {Young}, D.~R. and {Sullivan}, M. and {Pastorello}, A. and {Benetti}, S. and {Gal-Yam}, A. and {Knapic}, C. and {Molinaro}, M. and {Smareglia}, R. and {Smith}, K.~W. and {Taubenberger}, S. and {Yaron}, O. and {Anderson}, J.~P. and {Ashall}, C. and {Balland}, C. and {Baltay}, C. and {Barbarino}, C. and {Bauer}, F.~E. and {Baumont}, S. and {Bersier}, D. and {Blagorodnova}, N. and {Bongard}, S. and {Botticella}, M.~T. and {Bufano}, F. and {Bulla}, M. and {Cappellaro}, E. and {Campbell}, H. and {Cellier-Holzem}, F. and {Chen}, T. -W. and {Childress}, M.~J. and {Clocchiatti}, A. and {Contreras}, C. and {Dall'Ora}, M. and {Danziger}, J. and {de Jaeger}, T. and {De Cia}, A. and {Della Valle}, M. and {Dennefeld}, M. and {Elias-Rosa}, N. and {Elman}, N. and {Feindt}, U. and {Fleury}, M. and {Gall}, E. and {Gonzalez-Gaitan}, S. and {Galbany}, L. and {Morales Garoffolo}, A. and {Greggio}, L. and {Guillou}, L.~L. and {Hachinger}, S. and {Hadjiyska}, E. and {Hage}, P.~E. and {Hillebrandt}, W. and {Hodgkin}, S. and {Hsiao}, E.~Y. and {James}, P.~A. and {Jerkstrand}, A. and {Kangas}, T. and {Kankare}, E. and {Kotak}, R. and {Kromer}, M. and {Kuncarayakti}, H. and {Leloudas}, G. and {Lundqvist}, P. and {Lyman}, J.~D. and {Hook}, I.~M. and {Maguire}, K. and {Manulis}, I. and {Margheim}, S.~J. and {Mattila}, S. and {Maund}, J.~R. and {Mazzali}, P.~A. and {McCrum}, M. and {McKinnon}, R. and {Moreno-Raya}, M.~E. and {Nicholl}, M. and {Nugent}, P. and {Pain}, R. and {Pignata}, G. and {Phillips}, M.~M. and {Polshaw}, J. and {Pumo}, M.~L. and {Rabinowitz}, D. and {Reilly}, E. and {Romero-Ca{\~n}izales}, C. and {Scalzo}, R. and {Schmidt}, B. and {Schulze}, S. and {Sim}, S. and {Sollerman}, J. and {Taddia}, F. and {Tartaglia}, L. and {Terreran}, G. and {Tomasella}, L. and {Turatto}, M. and {Walker}, E. and {Walton}, N.~A. and {Wyrzykowski}, L. and {Yuan}, F. and {Zampieri}, L.},
        title = "{PESSTO: survey description and products from the first data release by the Public ESO Spectroscopic Survey of Transient Objects}",
      journal = {\aap},
         year = 2015,
        month = jul,
       volume = {579},
          eid = {A40},
        pages = {A40},
          doi = {10.1051/0004-6361/201425237},
archivePrefix = {arXiv},
       eprint = {1411.0299},
 primaryClass = {astro-ph.SR},
       adsurl = {https://ui.adsabs.harvard.edu/abs/2015A&A...579A..40S}
}

@ARTICLE{2023ApJ...956..111W,
       author = {{Ward}, Sam M. and {Thorp}, Stephen and {Mandel}, Kaisey S. and {Dhawan}, Suhail and {Jones}, David O. and {Taggart}, Kirsty and {Foley}, Ryan J. and {Narayan}, Gautham and {Chambers}, Kenneth C. and {Coulter}, David A. and {Davis}, Kyle W. and {de Boer}, Thomas and {de Soto}, Kaylee and {Earl}, Nicholas and {Gagliano}, Alex and {Gao}, Hua and {Hjorth}, Jens and {Huber}, Mark E. and {Izzo}, Luca and {Langeroodi}, Danial and {Magnier}, Eugene A. and {McGill}, Peter and {Rest}, Armin and {Rojas-Bravo}, C{\'e}sar and {Wojtak}, Rados{\l}aw and {Young Supernova Experiment}},
        title = "{Relative Intrinsic Scatter in Hierarchical Type Ia Supernova Sibling Analyses: Application to SNe 2021hpr, 1997bq, and 2008fv in NGC 3147}",
      journal = {\apj},
         year = 2023,
        month = oct,
       volume = {956},
       number = {2},
          eid = {111},
        pages = {111},
          doi = {10.3847/1538-4357/acf7bb},
archivePrefix = {arXiv},
       eprint = {2209.10558},
 primaryClass = {astro-ph.CO},
       adsurl = {https://ui.adsabs.harvard.edu/abs/2023ApJ...956..111W}
}

@misc{grayling2024scalablehierarchicalbayesninference,
      title={Scalable hierarchical BayeSN inference: Investigating dependence of SN Ia host galaxy dust properties on stellar mass and redshift}, 
      author={Matthew Grayling and Stephen Thorp and Kaisey S. Mandel and Suhail Dhawan and Ana Sofia M. Uzsoy and Benjamin M. Boyd and Erin E. Hayes and Sam M. Ward},
      year={2024},
      eprint={2401.08755},
      archivePrefix={arXiv},
      primaryClass={astro-ph.CO},
      url={https://arxiv.org/abs/2401.08755}, 
}

@article{Foley_2017,
   title={The Foundation Supernova Survey: motivation, design, implementation, and first data release},
   volume={475},
   ISSN={1365-2966},
   url={http://dx.doi.org/10.1093/mnras/stx3136},
   DOI={10.1093/mnras/stx3136},
   number={1},
   journal={Monthly Notices of the Royal Astronomical Society},
   publisher={Oxford University Press (OUP)},
   author={Foley, Ryan J and Scolnic, Daniel and Rest, Armin and Jha, S W and Pan, Y-C and Riess, A G and Challis, P and Chambers, K C and Coulter, D A and Dettman, K G and Foley, M M and Fox, O D and Huber, M E and Jones, D O and Kilpatrick, C D and Kirshner, R P and Schultz, A S B and Siebert, M R and Flewelling, H A and Gibson, B and Magnier, E A and Miller, J A and Primak, N and Smartt, S J and Smith, K W and Wainscoat, R J and Waters, C and Willman, M},
   year={2017},
   month=dec, pages={193–219} }

@article{Jones_2019,
   title={The Foundation Supernova Survey: Measuring Cosmological Parameters with Supernovae from a Single Telescope},
   volume={881},
   ISSN={1538-4357},
   url={http://dx.doi.org/10.3847/1538-4357/ab2bec},
   DOI={10.3847/1538-4357/ab2bec},
   number={1},
   journal={The Astrophysical Journal},
   publisher={American Astronomical Society},
   author={Jones, D. O. and Scolnic, D. M. and Foley, R. J. and Rest, A. and Kessler, R. and Challis, P. M. and Chambers, K. C. and Coulter, D. A. and Dettman, K. G. and Foley, M. M. and Huber, M. E. and Jha, S. W. and Johnson, E. and Kilpatrick, C. D. and Kirshner, R. P. and Manuel, J. and Narayan, G. and Pan, Y.-C. and Riess, A. G. and Schultz, A. S. B. and Siebert, M. R. and Berger, E. and Chornock, R. and Flewelling, H. and Magnier, E. A. and Smartt, S. J. and Smith, K. W. and Wainscoat, R. J. and Waters, C. and Willman, M.},
   year={2019},
   month=aug, pages={19} }

@article{Avelino_2019,
   title={Type Ia Supernovae Are Excellent Standard Candles in the Near-infrared},
   volume={887},
   ISSN={1538-4357},
   url={http://dx.doi.org/10.3847/1538-4357/ab2a16},
   DOI={10.3847/1538-4357/ab2a16},
   number={1},
   journal={The Astrophysical Journal},
   publisher={American Astronomical Society},
   author={Avelino, Arturo and Friedman, Andrew S. and Mandel, Kaisey S. and Jones, David O. and Challis, Peter J. and Kirshner, Robert P.},
   year={2019},
   month=dec, pages={106} }

@article{Schlegel_1998,
   title={Maps of Dust Infrared Emission for Use in Estimation of Reddening and Cosmic Microwave Background Radiation Foregrounds},
   volume={500},
   ISSN={1538-4357},
   url={http://dx.doi.org/10.1086/305772},
   DOI={10.1086/305772},
   number={2},
   journal={The Astrophysical Journal},
   publisher={American Astronomical Society},
   author={Schlegel, David J. and Finkbeiner, Douglas P. and Davis, Marc},
   year={1998},
   month=jun, pages={525–553} }

@article{Ni_2022,
   title={Infant-phase reddening by surface Fe-peak elements in a normal type Ia supernova},
   volume={6},
   ISSN={2397-3366},
   url={http://dx.doi.org/10.1038/s41550-022-01603-4},
   DOI={10.1038/s41550-022-01603-4},
   number={5},
   journal={Nature Astronomy},
   publisher={Springer Science and Business Media LLC},
   author={Ni, Yuan Qi and Moon, Dae-Sik and Drout, Maria R. and Polin, Abigail and Sand, David J. and González-Gaitán, Santiago and Kim, Sang Chul and Lee, Youngdae and Park, Hong Soo and Howell, D. Andrew and Nugent, Peter E. and Piro, Anthony L. and Brown, Peter J. and Galbany, Lluís and Burke, Jamison and Hiramatsu, Daichi and Hosseinzadeh, Griffin and Valenti, Stefano and Afsariardchi, Niloufar and Andrews, Jennifer E. and Antoniadis, John and Arcavi, Iair and Beaton, Rachael L. and Bostroem, K. Azalee and Carlberg, Raymond G. and Cenko, S. Bradley and Cha, Sang-Mok and Dong, Yize and Gal-Yam, Avishay and Haislip, Joshua and Holoien, Thomas W.-S. and Johnson, Sean D. and Kouprianov, Vladimir and Lee, Yongseok and Matzner, Christopher D. and Morrell, Nidia and McCully, Curtis and Pignata, Giuliano and Reichart, Daniel E. and Rich, Jeffrey and Ryder, Stuart D. and Smith, Nathan and Wyatt, Samuel and Yang, Sheng},
   year={2022},
   month=feb, pages={568–576} }

@misc{rigault2024ztfsniadr2,
      title={ZTF SN Ia DR2: Overview}, 
      author={Mickael Rigault and Mathew Smith and Ariel Goobar and Kate Maguire and Georgios Dimitriadis and Umut Burgaz and Suhail Dhawan and Jesper Sollerman and Nicolas Regnault and Marek Kowalski and Melissa Amenouche and Marie Aubert and Chloé Barjou-Delayre and Julian Bautista and Josh S. Bloom and Bastien Carreres and Tracy X. Chen and Yannick Copin and Maxime Deckers and Dominique Fouchez and Christoffer Fremling and Lluis Galbany and Madeleine Ginolin and Matthew Graham and Mancy M. Kasliwal and W. D'Arcy Kenworthy and Young-Lo Kim and Dylan Kuhn and Frank F. Masci and Tomas Müller-Bravo and Adam Miller and Joel Johansson and Jakob Nordin and Peter Nugent and Igor Andreoni and Eric Bellm and Marc Betoule and Mahmoud Osman and Dan Perley and Brodie Popovic and Philippe Rosnet and Damiano Rosselli and Florian Ruppin and Robert Senzel and Ben Rusholme and Tassilo Schweyer and Jacco H. Terwel and Alice Townsend and Andy Tzanidakis and Avery Wold and Josiah Purdum and Yu-Jing Qin and Benjamin Racine and Simeon Reusch and Reed Riddle and Lin Yan},
      year={2024},
      eprint={2409.04346},
      archivePrefix={arXiv},
      primaryClass={astro-ph.CO},
      url={https://arxiv.org/abs/2409.04346}, 
}

@ARTICLE{2023A&A...679A..95G,
       author = {{Galbany}, L. and {de Jaeger}, T. and {Riess}, A.~G. and {M{\"u}ller-Bravo}, T.~E. and {Dhawan}, S. and {Phan}, K. and {Stritzinger}, M.~D. and {Karamehmetoglu}, E. and {Leibundgut}, B. and {Burns}, C. and {Peterson}, E. and {D'Arcy Kenworthy}, W. and {Johansson}, J. and {Maguire}, K. and {Jha}, S.~W.},
        title = "{An updated measurement of the Hubble constant from near-infrared observations of Type Ia supernovae}",
      journal = {\aap},
         year = 2023,
        month = nov,
       volume = {679},
          eid = {A95},
        pages = {A95},
          doi = {10.1051/0004-6361/202244893},
archivePrefix = {arXiv},
       eprint = {2209.02546},
 primaryClass = {astro-ph.CO},
       adsurl = {https://ui.adsabs.harvard.edu/abs/2023A&A...679A..95G}
}

@INPROCEEDINGS{2003SPIE.4841..827D,
       author = {{DePoy}, Darren L. and {Atwood}, Bruce and {Belville}, Stanley R. and {Brewer}, David F. and {Byard}, Paul L. and {Gould}, Andrew and {Mason}, Jerry A. and {O'Brien}, Thomas P. and {Pappalardo}, Daniel P. and {Pogge}, Richard W. and {Steinbrecher}, David P. and {Teiga}, Edward J.},
        title = "{A Novel Double Imaging Camera (ANDICAM)}",
    booktitle = {Instrument Design and Performance for Optical/Infrared Ground-based Telescopes},
         year = 2003,
       editor = {{Iye}, Masanori and {Moorwood}, Alan F.~M.},
       series = {Society of Photo-Optical Instrumentation Engineers (SPIE) Conference Series},
       volume = {4841},
        month = mar,
        pages = {827-838},
          doi = {10.1117/12.459907},
       adsurl = {https://ui.adsabs.harvard.edu/abs/2003SPIE.4841..827D}
}

@ARTICLE{2025A&A...694A...1R,
       author = {{Rigault}, M. and {Smith}, M. and {Goobar}, A. and {Maguire}, K. and {Dimitriadis}, G. and {Johansson}, J. and {Nordin}, J. and {Burgaz}, U. and {Dhawan}, S. and {Sollerman}, J. and {Regnault}, N. and {Kowalski}, M. and {Nugent}, P. and {Andreoni}, I. and {Amenouche}, M. and {Aubert}, M. and {Barjou-Delayre}, C. and {Bautista}, J. and {Bellm}, E. and {Betoule}, M. and {Bloom}, J.~S. and {Carreres}, B. and {Chen}, T.~X. and {Copin}, Y. and {Deckers}, M. and {de Jaeger}, T. and {Feinstein}, F. and {Fouchez}, D. and {Fremling}, C. and {Galbany}, L. and {Ginolin}, M. and {Graham}, M. and {Groom}, S.~L. and {Harvey}, L. and {Kasliwal}, M.~M. and {Kenworthy}, W.~D. and {Kim}, Y. -L. and {Kuhn}, D. and {Kulkarni}, S.~R. and {Lacroix}, L. and {Laher}, R.~R. and {Masci}, F.~J. and {M{\"u}ller-Bravo}, T.~E. and {Miller}, A. and {Osman}, M. and {Perley}, D. and {Popovic}, B. and {Purdum}, J. and {Qin}, Y. -J. and {Racine}, B. and {Reusch}, S. and {Riddle}, R. and {Rosnet}, P. and {Rosselli}, D. and {Ruppin}, F. and {Senzel}, R. and {Rusholme}, B. and {Schweyer}, T. and {Terwel}, J.~H. and {Townsend}, A. and {Tzanidakis}, A. and {Wold}, A. and {Yan}, L.},
        title = "{ZTF SN Ia DR2: Overview}",
      journal = {\aap},
         year = 2025,
        month = feb,
       volume = {694},
          eid = {A1},
        pages = {A1},
          doi = {10.1051/0004-6361/202450388},
archivePrefix = {arXiv},
       eprint = {2409.04346},
 primaryClass = {astro-ph.CO},
       adsurl = {https://ui.adsabs.harvard.edu/abs/2025A&A...694A...1R}
}

@ARTICLE{2000ARA&A..38..191H,
       author = {{Hillebrandt}, Wolfgang and {Niemeyer}, Jens C.},
        title = "{Type IA Supernova Explosion Models}",
      journal = {\araa},
         year = 2000,
        month = jan,
       volume = {38},
        pages = {191-230},
          doi = {10.1146/annurev.astro.38.1.191},
archivePrefix = {arXiv},
       eprint = {astro-ph/0006305},
 primaryClass = {astro-ph},
       adsurl = {https://ui.adsabs.harvard.edu/abs/2000ARA&A..38..191H}
}

@ARTICLE{Young_plot_atlas_fp,
    author = {Young, David R.},
    doi = {10.5281/zenodo.10978968},
    license = {GPL-3.0-only},
    title = {{plot_atlas_fp.py}},
    url = {https://zenodo.org/doi/10.5281/zenodo.10978968},
    year = {2020}
}

@article{Do_2024,
   title={Hawai‘i Supernova Flows: a peculiar velocity survey using over a Thousand Supernovae in the near-infrared},
   volume={536},
   ISSN={1365-2966},
   url={http://dx.doi.org/10.1093/mnras/stae2501},
   DOI={10.1093/mnras/stae2501},
   number={1},
   journal={Monthly Notices of the Royal Astronomical Society},
   publisher={Oxford University Press (OUP)},
   author={Do, Aaron and Shappee, Benjamin J and Tonry, John L and Tully, R Brent and de Jaeger, Thomas and Rubin, David and Ashall, Chris and Burns, Christopher R and Desai, Dhvanil D and Hinkle, Jason T and Hoogendam, Willem B and Huber, Mark E and Jones, David O and Mandel, Kaisey S and Payne, Anna V and Peterson, Erik R and Scolnic, Dan and Tucker, Michael A},
   year={2024},
   month=nov, pages={624–663} }

@article{Gupta_2016,
   title={HOST GALAXY IDENTIFICATION FOR SUPERNOVA SURVEYS},
   volume={152},
   ISSN={1538-3881},
   url={http://dx.doi.org/10.3847/0004-6256/152/6/154},
   DOI={10.3847/0004-6256/152/6/154},
   number={6},
   journal={The Astronomical Journal},
   publisher={American Astronomical Society},
   author={Gupta, Ravi R. and Kuhlmann, Steve and Kovacs, Eve and Spinka, Harold and Kessler, Richard and Goldstein, Daniel A. and Liotine, Camille and Pomian, Katarzyna and D’Andrea, Chris B. and Sullivan, Mark and Carretero, Jorge and Castander, Francisco J. and Nichol, Robert C. and Finley, David A. and Fischer, John A. and Foley, Ryan J. and Kim, Alex G. and Papadopoulos, Andreas and Sako, Masao and Scolnic, Daniel M. and Smith, Mathew and Tucker, Brad E. and Uddin, Syed and Wolf, Rachel C. and Yuan, Fang and Abbott, Tim M. C. and Abdalla, Filipe B. and Benoit-Lévy, Aurélien and Bertin, Emmanuel and Brooks, David and Rosell, Aurelio Carnero and Kind, Matias Carrasco and Cunha, Carlos E. and Costa, Luiz N. da and Desai, Shantanu and Doel, Peter and Eifler, Tim F. and Evrard, August E. and Flaugher, Brenna and Fosalba, Pablo and Gaztañaga, Enrique and Gruen, Daniel and Gruendl, Robert and James, David J. and Kuehn, Kyler and Kuropatkin, Nikolay and Maia, Marcio A. G. and Marshall, Jennifer L. and Miquel, Ramon and Plazas, Andrés A. and Romer, A. Kathy and Sánchez, Eusebio and Schubnell, Michael and Sevilla-Noarbe, Ignacio and Sobreira, Flávia and Suchyta, Eric and Swanson, Molly E. C. and Tarle, Gregory and Walker, Alistair R. and Wester, William},
   year={2016},
   month=nov, pages={154} }

@article{Sako_2018,
   title={The Data Release of the Sloan Digital Sky Survey-II Supernova Survey},
   volume={130},
   ISSN={1538-3873},
   url={http://dx.doi.org/10.1088/1538-3873/aab4e0},
   DOI={10.1088/1538-3873/aab4e0},
   number={988},
   journal={Publications of the Astronomical Society of the Pacific},
   publisher={IOP Publishing},
   author={Sako, Masao and Bassett, Bruce and Becker, Andrew C. and Brown, Peter J. and Campbell, Heather and Wolf, Rachel and Cinabro, David and D’Andrea, Chris B. and Dawson, Kyle S. and DeJongh, Fritz and Depoy, Darren L. and Dilday, Ben and Doi, Mamoru and Filippenko, Alexei V. and Fischer, John A. and Foley, Ryan J. and Frieman, Joshua A. and Galbany, Lluis and Garnavich, Peter M. and Goobar, Ariel and Gupta, Ravi R. and Hill, Gary J. and Hayden, Brian T. and Hlozek, Renée and Holtzman, Jon A. and Hopp, Ulrich and Jha, Saurabh W. and Kessler, Richard and Kollatschny, Wolfram and Leloudas, Giorgos and Marriner, John and Marshall, Jennifer L. and Miquel, Ramon and Morokuma, Tomoki and Mosher, Jennifer and Nichol, Robert C. and Nordin, Jakob and Olmstead, Matthew D. and Östman, Linda and Prieto, Jose L. and Richmond, Michael and Romani, Roger W. and Sollerman, Jesper and Stritzinger, Max and Schneider, Donald P. and Smith, Mathew and Wheeler, J. Craig and Yasuda, Naoki and Zheng, Chen},
   year={2018},
   month=may, pages={064002} }

@article{Kenworthy_2021,
   title={SALT3: An Improved Type Ia Supernova Model for Measuring Cosmic Distances},
   volume={923},
   ISSN={1538-4357},
   url={http://dx.doi.org/10.3847/1538-4357/ac30d8},
   DOI={10.3847/1538-4357/ac30d8},
   number={2},
   journal={The Astrophysical Journal},
   publisher={American Astronomical Society},
   author={Kenworthy, W. D. and Jones, D. O. and Dai, M. and Kessler, R. and Scolnic, D. and Brout, D. and Siebert, M. R. and Pierel, J. D. R. and Dettman, K. G. and Dimitriadis, G. and Foley, R. J. and Jha, S. W. and Pan, Y.-C. and Riess, A. and Rodney, S. and Rojas-Bravo, C.},
   year={2021},
   month=dec, pages={265} }

@article{Mandel_2021,
   title={A hierarchical Bayesian SED model for Type Ia supernovae in the optical to near-infrared},
   volume={510},
   ISSN={1365-2966},
   url={http://dx.doi.org/10.1093/mnras/stab3496},
   DOI={10.1093/mnras/stab3496},
   number={3},
   journal={Monthly Notices of the Royal Astronomical Society},
   publisher={Oxford University Press (OUP)},
   author={Mandel, Kaisey S and Thorp, Stephen and Narayan, Gautham and Friedman, Andrew S and Avelino, Arturo},
   year={2021},
   month=dec, pages={3939–3966} }

@ARTICLE{2021ApJS..254...22J,
    author = {{Johnson}, Benjamin D. and {Leja}, Joel and {Conroy}, Charlie and {Speagle}, Joshua S.},
        title = "{Stellar Population Inference with Prospector}",
    journal = {\apjs},
        year = 2021,
        month = jun,
    volume = {254},
    number = {2},
        eid = {22},
        pages = {22},
        doi = {10.3847/1538-4365/abef67},
archivePrefix = {arXiv},
    eprint = {2012.01426},
primaryClass = {astro-ph.GA},
    adsurl = {https://ui.adsabs.harvard.edu/abs/2021ApJS..254...22J}
}

@ARTICLE{2017ApJ...837..170L,
       author = {{Leja}, Joel and {Johnson}, Benjamin D. and {Conroy}, Charlie and {van Dokkum}, Pieter G. and {Byler}, Nell},
        title = "{Deriving Physical Properties from Broadband Photometry with Prospector: Description of the Model and a Demonstration of its Accuracy Using 129 Galaxies in the Local Universe}",
      journal = {\apj},
         year = 2017,
        month = mar,
       volume = {837},
       number = {2},
          eid = {170},
        pages = {170},
          doi = {10.3847/1538-4357/aa5ffe},
archivePrefix = {arXiv},
       eprint = {1609.09073},
 primaryClass = {astro-ph.GA},
       adsurl = {https://ui.adsabs.harvard.edu/abs/2017ApJ...837..170L}
}

@ARTICLE{2009ApJ...699..486C,
       author = {{Conroy}, Charlie and {Gunn}, James E. and {White}, Martin},
        title = "{The Propagation of Uncertainties in Stellar Population Synthesis Modeling. I. The Relevance of Uncertain Aspects of Stellar Evolution and the Initial Mass Function to the Derived Physical Properties of Galaxies}",
      journal = {\apj},
         year = 2009,
        month = jul,
       volume = {699},
       number = {1},
        pages = {486-506},
          doi = {10.1088/0004-637X/699/1/486},
archivePrefix = {arXiv},
       eprint = {0809.4261},
 primaryClass = {astro-ph},
       adsurl = {https://ui.adsabs.harvard.edu/abs/2009ApJ...699..486C}
}

@ARTICLE{2019ApJ...876....3L,
       author = {{Leja}, Joel and {Carnall}, Adam C. and {Johnson}, Benjamin D. and {Conroy}, Charlie and {Speagle}, Joshua S.},
        title = "{How to Measure Galaxy Star Formation Histories. II. Nonparametric Models}",
      journal = {\apj},
         year = 2019,
        month = may,
       volume = {876},
       number = {1},
          eid = {3},
        pages = {3},
          doi = {10.3847/1538-4357/ab133c},
archivePrefix = {arXiv},
       eprint = {1811.03637},
 primaryClass = {astro-ph.GA},
       adsurl = {https://ui.adsabs.harvard.edu/abs/2019ApJ...876....3L}
}

@ARTICLE{2010ApJ...712..833C,
       author = {{Conroy}, Charlie and {Gunn}, James E.},
        title = "{The Propagation of Uncertainties in Stellar Population Synthesis Modeling. III. Model Calibration, Comparison, and Evaluation}",
      journal = {\apj},
         year = 2010,
        month = apr,
       volume = {712},
       number = {2},
        pages = {833-857},
          doi = {10.1088/0004-637X/712/2/833},
archivePrefix = {arXiv},
       eprint = {0911.3151},
 primaryClass = {astro-ph.CO},
       adsurl = {https://ui.adsabs.harvard.edu/abs/2010ApJ...712..833C}
}

@ARTICLE{2002Sci...295...82K,
       author = {{Kroupa}, Pavel},
        title = "{The Initial Mass Function of Stars: Evidence for Uniformity in Variable Systems}",
      journal = {Science},
         year = 2002,
        month = jan,
       volume = {295},
       number = {5552},
        pages = {82-91},
          doi = {10.1126/science.1067524},
archivePrefix = {arXiv},
       eprint = {astro-ph/0201098},
 primaryClass = {astro-ph},
       adsurl = {https://ui.adsabs.harvard.edu/abs/2002Sci...295...82K}
}

@ARTICLE{2018ApJ...867..105T,
       author = {{Tonry}, J.~L. and {Denneau}, L. and {Flewelling}, H. and {Heinze}, A.~N. and {Onken}, C.~A. and {Smartt}, S.~J. and {Stalder}, B. and {Weiland}, H.~J. and {Wolf}, C.},
        title = "{The ATLAS All-Sky Stellar Reference Catalog}",
      journal = {\apj},
         year = 2018,
        month = nov,
       volume = {867},
       number = {2},
          eid = {105},
        pages = {105},
          doi = {10.3847/1538-4357/aae386},
archivePrefix = {arXiv},
       eprint = {1809.09157},
 primaryClass = {astro-ph.IM},
       adsurl = {https://ui.adsabs.harvard.edu/abs/2018ApJ...867..105T}
}

@article{Abbott_2021,
   title={The Dark Energy Survey Data Release 2},
   volume={255},
   ISSN={1538-4365},
   url={http://dx.doi.org/10.3847/1538-4365/ac00b3},
   DOI={10.3847/1538-4365/ac00b3},
   number={2},
   journal={The Astrophysical Journal Supplement Series},
   publisher={American Astronomical Society},
   author={Abbott, T. M. C. and Adamów, M. and Aguena, M. and Allam, S. and Amon, A. and Annis, J. and Avila, S. and Bacon, D. and Banerji, M. and Bechtol, K. and Becker, M. R. and Bernstein, G. M. and Bertin, E. and Bhargava, S. and Bridle, S. L. and Brooks, D. and Burke, D. L. and Carnero Rosell, A. and Carrasco Kind, M. and Carretero, J. and Castander, F. J. and Cawthon, R. and Chang, C. and Choi, A. and Conselice, C. and Costanzi, M. and Crocce, M. and da Costa, L. N. and Davis, T. M. and De Vicente, J. and DeRose, J. and Desai, S. and Diehl, H. T. and Dietrich, J. P. and Drlica-Wagner, A. and Eckert, K. and Elvin-Poole, J. and Everett, S. and Evrard, A. E. and Ferrero, I. and Ferté, A. and Flaugher, B. and Fosalba, P. and Friedel, D. and Frieman, J. and García-Bellido, J. and Gaztanaga, E. and Gelman, L. and Gerdes, D. W. and Giannantonio, T. and Gill, M. S. S. and Gruen, D. and Gruendl, R. A. and Gschwend, J. and Gutierrez, G. and Hartley, W. G. and Hinton, S. R. and Hollowood, D. L. and Honscheid, K. and Huterer, D. and James, D. J. and Jeltema, T. and Johnson, M. D. and Kent, S. and Kron, R. and Kuehn, K. and Kuropatkin, N. and Lahav, O. and Li, T. S. and Lidman, C. and Lin, H. and MacCrann, N. and Maia, M. A. G. and Manning, T. A. and Maloney, J. D. and March, M. and Marshall, J. L. and Martini, P. and Melchior, P. and Menanteau, F. and Miquel, R. and Morgan, R. and Myles, J. and Neilsen, E. and Ogando, R. L. C. and Palmese, A. and Paz-Chinchón, F. and Petravick, D. and Pieres, A. and Plazas, A. A. and Pond, C. and Rodriguez-Monroy, M. and Romer, A. K. and Roodman, A. and Rykoff, E. S. and Sako, M. and Sanchez, E. and Santiago, B. and Scarpine, V. and Serrano, S. and Sevilla-Noarbe, I. and Smith, J. Allyn and Smith, M. and Soares-Santos, M. and Suchyta, E. and Swanson, M. E. C. and Tarle, G. and Thomas, D. and To, C. and Tremblay, P. E. and Troxel, M. A. and Tucker, D. L. and Turner, D. J. and Varga, T. N. and Walker, A. R. and Wechsler, R. H. and Weller, J. and Wester, W. and Wilkinson, R. D. and Yanny, B. and Zhang, Y. and Nikutta, R. and Fitzpatrick, M. and Jacques, A. and Scott, A. and Olsen, K. and Huang, L. and Herrera, D. and Juneau, S. and Nidever, D. and Weaver, B. A. and Adean, C. and Correia, V. and de Freitas, M. and Freitas, F. N. and Singulani, C. and Vila-Verde, G.},
   year={2021},
   month=jul, pages={20} }

@ARTICLE{2013PASP..125..654P,
       author = {{Persson}, S.~E. and {Murphy}, D.~C. and {Smee}, S. and {Birk}, C. and {Monson}, A.~J. and {Uomoto}, A. and {Koch}, E. and {Shectman}, S. and {Barkhouser}, R. and {Orndorff}, J. and {Hammond}, R. and {Harding}, A. and {Scharfstein}, G. and {Kelson}, D. and {Marshall}, J. and {McCarthy}, P.~J.},
        title = "{FourStar: The Near-Infrared Imager for the 6.5 m Baade Telescope at Las Campanas Observatory}",
      journal = {\pasp},
         year = 2013,
        month = jun,
       volume = {125},
       number = {928},
        pages = {654},
          doi = {10.1086/671164},
       adsurl = {https://ui.adsabs.harvard.edu/abs/2013PASP..125..654P}
}

@ARTICLE{1973ApJ...186.1007W,
       author = {{Whelan}, John and {Iben}, Jr., Icko},
        title = "{Binaries and Supernovae of Type I}",
      journal = {\apj},
         year = 1973,
        month = dec,
       volume = {186},
        pages = {1007-1014},
          doi = {10.1086/152565},
       adsurl = {https://ui.adsabs.harvard.edu/abs/1973ApJ...186.1007W}
}

@ARTICLE{1984ApJ...277..355W,
       author = {{Webbink}, R.~F.},
        title = "{Double white dwarfs as progenitors of R Coronae Borealis stars and type I supernovae.}",
      journal = {\apj},
         year = 1984,
        month = feb,
       volume = {277},
        pages = {355-360},
          doi = {10.1086/161701},
       adsurl = {https://ui.adsabs.harvard.edu/abs/1984ApJ...277..355W}
}

@article{Thompson_2011,
   title={ACCELERATING COMPACT OBJECT MERGERS IN TRIPLE SYSTEMS WITH THE KOZAI RESONANCE: A MECHANISM FOR “PROMPT” TYPE Ia SUPERNOVAE, GAMMA-RAY BURSTS, AND OTHER EXOTICA},
   volume={741},
   ISSN={1538-4357},
   url={http://dx.doi.org/10.1088/0004-637X/741/2/82},
   DOI={10.1088/0004-637x/741/2/82},
   number={2},
   journal={The Astrophysical Journal},
   publisher={American Astronomical Society},
   author={Thompson, Todd A.},
   year={2011},
   month=oct, pages={82} }

@ARTICLE{2022A&A...665A.123M,
       author = {{M{\"u}ller-Bravo}, T.~E. and {Galbany}, L. and {Karamehmetoglu}, E. and {Stritzinger}, M. and {Burns}, C. and {Phan}, K. and {I{\'a}{\~n}ez Ferres}, A. and {Anderson}, J.~P. and {Ashall}, C. and {Baron}, E. and {Hoeflich}, P. and {Hsiao}, E.~Y. and {de Jaeger}, T. and {Kumar}, S. and {Lu}, J. and {Phillips}, M.~M. and {Shahbandeh}, M. and {Suntzeff}, N. and {Uddin}, S.~A.},
        title = "{Testing the homogeneity of type Ia Supernovae in near-infrared for accurate distance estimations}",
      journal = {\aap},
         year = 2022,
        month = sep,
       volume = {665},
          eid = {A123},
        pages = {A123},
          doi = {10.1051/0004-6361/202243845},
archivePrefix = {arXiv},
       eprint = {2207.04780},
 primaryClass = {astro-ph.CO},
       adsurl = {https://ui.adsabs.harvard.edu/abs/2022A&A...665A.123M}
}

@article{Sullivan_2006,
   title={Rates and Properties of Type Ia Supernovae as a Function of Mass and Star Formation in Their Host Galaxies},
   volume={648},
   ISSN={1538-4357},
   url={http://dx.doi.org/10.1086/506137},
   DOI={10.1086/506137},
   number={2},
   journal={The Astrophysical Journal},
   publisher={American Astronomical Society},
   author={Sullivan, M. and Le Borgne, D. and Pritchet, C. J. and Hodsman, A. and Neill, J. D. and Howell, D. A. and Carlberg, R. G. and Astier, P. and Aubourg, E. and Balam, D. and Basa, S. and Conley, A. and Fabbro, S. and Fouchez, D. and Guy, J. and Hook, I. and Pain, R. and Palanque‐Delabrouille, N. and Perrett, K. and Regnault, N. and Rich, J. and Taillet, R. and Baumont, S. and Bronder, J. and Ellis, R. S. and Filiol, M. and Lusset, V. and Perlmutter, S. and Ripoche, P. and Tao, C.},
   year={2006},
   month=sep, pages={868–883} }

@ARTICLE{2022JOSS....7.4508M,
       author = {{M{\"u}ller-Bravo}, Tom{\'a}s and {Galbany}, Llu{\'\i}s},
        title = "{HostPhot: global and local photometry of galaxies hosting supernovae or other transients}",
      journal = {The Journal of Open Source Software},
         year = 2022,
        month = aug,
       volume = {7},
       number = {76},
          eid = {4508},
        pages = {4508},
          doi = {10.21105/joss.04508},
archivePrefix = {arXiv},
       eprint = {2208.08117},
 primaryClass = {astro-ph.CO},
       adsurl = {https://ui.adsabs.harvard.edu/abs/2022JOSS....7.4508M}
}

@ARTICLE{2018AJ....156..123A,
       author = {{Astropy Collaboration} and {Price-Whelan}, A.~M. and {Sip{\H{o}}cz}, B.~M. and {G{\"u}nther}, H.~M. and {Lim}, P.~L. and {Crawford}, S.~M. and {Conseil}, S. and {Shupe}, D.~L. and {Craig}, M.~W. and {Dencheva}, N. and {Ginsburg}, A. and {VanderPlas}, J.~T. and {Bradley}, L.~D. and {P{\'e}rez-Su{\'a}rez}, D. and {de Val-Borro}, M. and {Aldcroft}, T.~L. and {Cruz}, K.~L. and {Robitaille}, T.~P. and {Tollerud}, E.~J. and {Ardelean}, C. and {Babej}, T. and {Bach}, Y.~P. and {Bachetti}, M. and {Bakanov}, A.~V. and {Bamford}, S.~P. and {Barentsen}, G. and {Barmby}, P. and {Baumbach}, A. and {Berry}, K.~L. and {Biscani}, F. and {Boquien}, M. and {Bostroem}, K.~A. and {Bouma}, L.~G. and {Brammer}, G.~B. and {Bray}, E.~M. and {Breytenbach}, H. and {Buddelmeijer}, H. and {Burke}, D.~J. and {Calderone}, G. and {Cano Rodr{\'\i}guez}, J.~L. and {Cara}, M. and {Cardoso}, J.~V.~M. and {Cheedella}, S. and {Copin}, Y. and {Corrales}, L. and {Crichton}, D. and {D'Avella}, D. and {Deil}, C. and {Depagne}, {\'E}. and {Dietrich}, J.~P. and {Donath}, A. and {Droettboom}, M. and {Earl}, N. and {Erben}, T. and {Fabbro}, S. and {Ferreira}, L.~A. and {Finethy}, T. and {Fox}, R.~T. and {Garrison}, L.~H. and {Gibbons}, S.~L.~J. and {Goldstein}, D.~A. and {Gommers}, R. and {Greco}, J.~P. and {Greenfield}, P. and {Groener}, A.~M. and {Grollier}, F. and {Hagen}, A. and {Hirst}, P. and {Homeier}, D. and {Horton}, A.~J. and {Hosseinzadeh}, G. and {Hu}, L. and {Hunkeler}, J.~S. and {Ivezi{\'c}}, {\v{Z}}. and {Jain}, A. and {Jenness}, T. and {Kanarek}, G. and {Kendrew}, S. and {Kern}, N.~S. and {Kerzendorf}, W.~E. and {Khvalko}, A. and {King}, J. and {Kirkby}, D. and {Kulkarni}, A.~M. and {Kumar}, A. and {Lee}, A. and {Lenz}, D. and {Littlefair}, S.~P. and {Ma}, Z. and {Macleod}, D.~M. and {Mastropietro}, M. and {McCully}, C. and {Montagnac}, S. and {Morris}, B.~M. and {Mueller}, M. and {Mumford}, S.~J. and {Muna}, D. and {Murphy}, N.~A. and {Nelson}, S. and {Nguyen}, G.~H. and {Ninan}, J.~P. and {N{\"o}the}, M. and {Ogaz}, S. and {Oh}, S. and {Parejko}, J.~K. and {Parley}, N. and {Pascual}, S. and {Patil}, R. and {Patil}, A.~A. and {Plunkett}, A.~L. and {Prochaska}, J.~X. and {Rastogi}, T. and {Reddy Janga}, V. and {Sabater}, J. and {Sakurikar}, P. and {Seifert}, M. and {Sherbert}, L.~E. and {Sherwood-Taylor}, H. and {Shih}, A.~Y. and {Sick}, J. and {Silbiger}, M.~T. and {Singanamalla}, S. and {Singer}, L.~P. and {Sladen}, P.~H. and {Sooley}, K.~A. and {Sornarajah}, S. and {Streicher}, O. and {Teuben}, P. and {Thomas}, S.~W. and {Tremblay}, G.~R. and {Turner}, J.~E.~H. and {Terr{\'o}n}, V. and {van Kerkwijk}, M.~H. and {de la Vega}, A. and {Watkins}, L.~L. and {Weaver}, B.~A. and {Whitmore}, J.~B. and {Woillez}, J. and {Zabalza}, V. and {Astropy Contributors}},
        title = "{The Astropy Project: Building an Open-science Project and Status of the v2.0 Core Package}",
      journal = {\aj},
         year = 2018,
        month = sep,
       volume = {156},
       number = {3},
          eid = {123},
        pages = {123},
          doi = {10.3847/1538-3881/aabc4f},
archivePrefix = {arXiv},
       eprint = {1801.02634},
 primaryClass = {astro-ph.IM},
       adsurl = {https://ui.adsabs.harvard.edu/abs/2018AJ....156..123A}
}

@software{larry_bradley_2025_14889440,
  author       = {Larry Bradley and
                  Brigitta Sip{\H o}cz and
                  Thomas Robitaille and
                  Erik Tollerud and
                  Z\`e Vin{\'{\i}}cius and
                  Christoph Deil and
                  Kyle Barbary and
                  Tom J Wilson and
                  Ivo Busko and
                  Axel Donath and
                  Hans Moritz G{\"u}nther and
                  Mihai Cara and
                  P. L. Lim and
                  Sebastian Me{\ss}linger and
                  Zach Burnett and
                  Simon Conseil and
                  Michael Droettboom and
                  Azalee Bostroem and
                  E. M. Bray and
                  Lars Andersen Bratholm and
                  William Jamieson and
                  Adam Ginsburg and
                  Geert Barentsen and
                  Matt Craig and
                  Sergio Pascual and
                  Shivangee Rathi and
                  Marshall Perrin and
                  Brett M. Morris},
  title        = {astropy/photutils: 2.2.0},
  month        = feb,
  year         = 2025,
  publisher    = {Zenodo},
  version      = {2.2.0},
  doi          = {10.5281/zenodo.14889440},
  url          = {https://doi.org/10.5281/zenodo.14889440},
  swhid        = {swh:1:dir:11159107f27a28985192ed1118b1f2055709d093
                   ;origin=https://doi.org/10.5281/zenodo.596036;visi
                   t=swh:1:snp:ae8c4a55d349d43e53cfe9ce92e678fcfe840f
                   3b;anchor=swh:1:rel:0117f67e8888adcdfc85308287dd9c
                   854b466389;path=astropy-photutils-ffb96c5
                  },
}

\clearpage

\appendix
\onecolumn
\section{Supernova sample}
\FloatBarrier

\begin{table*}[!h]
\caption{SN types, coordinates, redshifts, host names, discovery and classfication groups and number of observations in the NIR.}
\begin{center}
\resizebox{\textwidth}{!}{ 
\begin{tabular}{llccllllccccccc}
\hline \hline
 & & & & & & & & \multicolumn{7}{c}{Epochs\tablefootmark{a}}\\
SN&Type&RA&Dec&z&Host name&Disc. &Class. & &\multicolumn{3}{c}{ANDICAM}& \multicolumn{3}{c}{SOFI} \\
 & & & & & & & & Total&$Y$&$J$&$H$&$J_s$&$H_s$&$K_s$\\
\hline
2018rw  & Ia      & 06:09:39.200 & $-$33:35:23.24 & 0.037052 & FRL 1138                  & ASAS-SN    & ePESSTO      &  7 & 0& 6& 1&0&0&0\\ 
2018yu  & Ia      & 05:22:32.371 & $-$11:29:13.86 & 0.009150 & NGC 1888                  & DLT40      & Zhang+      & 32 & 0&16&16&0&0&0\\ 
2018agk & Ia      & 13:10:36.373 & $-$04:29:08.67 & 0.026131 & IC 0855                   & Rest+      & ASAS-SN     & 20 & 0& 4& 4&4&4&4\\ 
2018aoz & Ia      & 11:51:01.810 & $-$28:44:38.69 & 0.005767 & NGC 3923                  & DLT40      & GSP         & 40 & 0&20&20&0&0&0\\ 
2018ayg & Ia      & 14:43:39.091 & +18:52:12.51 & 0.030955 & LEDA 1573011              & ASAS-SN    & Asiago      &  4 & 0& 2& 2&0&0&0\\ 
2018bie & Ia-91T  & 12:35:44.330 & $-$00:13:16.43 & 0.023133 & CGCG 014-073              & ATLAS      & ePESSTO      & 30 & 0&15&15&0&0&0\\ 
2018bta & Ia      & 16:57:58.670 & $-$62:43:54.05 & 0.019691 & ESO 101- G 020            & BOSS       & ePESSTO      & 28 & 0&14&14&0&0&0\\ 
2018cow & Ic-BL   & 16:16:00.220 & +22:16:04.91 & 0.014060 & 2MASX J16160054+2216080   & ATLAS      & ZTF         & 22 & 0&11&11&0&0&0\\ 
2018dda & Ia      & 22:08:14.148 & $-$25:03:41.21 & 0.018405 & ESO 532- G 021            & ASAS-SN    & USC         & 39 &13&13&13&0&0&0\\ 
2018ebu & Ia      & 21:57:19.870 & $-$45:34:39.29 & 0.061    & LEDA 525236               & ASAS-SN    & ATEL11894   & 17 & 0& 7& 7&1&1&1\\ 
2018enc & Ia      & 15:19:28.620 & $-$09:52:49.94 & 0.023890 & LEDA 985222               & ATLAS      & ePESSTO      & 30 & 7& 7& 7&3&3&3\\ 
2018eov & Ia      & 16:15:17.420 & $-$61:07:53.54 & 0.016508 & 2MFGC 13057               & GaiaAlerts & ePESSTO      & 30 & 7& 7& 7&3&3&3\\ 
2018evt & Ia-CSM  & 13:46:39.181 & $-$09:38:36.00 & 0.025352 & MCG -01-35-011            & ASAS-SN    & PESSTO      & 51 &17&17&17&0&0&0\\ 
2018exb & Ia-91T  & 21:13:08.580 & $-$20:42:38.77 & 0.047406 & 2MASX J21130828-2042398   & ATLAS      & Dong+       & 21 & 3& 3& 3&4&4&4\\ 
2018exc & Ia      & 21:00:08.018 & $-$40:21:30.94 & 0.050575 & 2MFGC 15903               & ATLAS      & ePESSTO      & 30 & 5& 5& 5&5&5&5\\ 
2018feq & Ia-91T  & 01:15:18.506 & $-$44:55:09.95 & 0.031505 & ESO 244- G 007            & ASAS-SN    & ePESSTO      & 18 & 3& 3& 3&3&3&3\\ 
2018hfp & Ia      & 20:59:47.870 & $-$16:38:12.52 & 0.029037 & MCG -03-53-015            & ASAS-SN    & USC         & 42 &12&12&12&2&2&2\\ 
2018hgc & Ia      & 00:42:04.605 & $-$02:37:44.75 & 0.051782 & LEDA 1086813              & ASAS-SN    & USC         & 45 &13&13&13&2&2&2\\ 
2018hhn & Ia      & 22:52:32.100 & +11:40:26.62 & 0.029313 & UGC 12222                 & PSH        & QUB         & 42 &12&12&12&2&2&2\\ 
2018hjw & Ia-91T  & 07:53:02.180 & +07:16:35.04 & 0.039316 & CGCG 030-025              & ASAS-SN    & ePESSTO      & 39 &12&12&12&1&1&1\\ 
2018ilu & Ia      & 23:33:20.969 & +04:48:34.74 & 0.018070 & SDSS J233320.80+044839.0  & ATLAS      & PESSTO      & 39 &12&12&12&1&1&1\\ 
2018jag & Ia-91bg & 01:03:48.310 & +10:35:32.38 & 0.040657 & MCG +02-03-030            & TNTS       & SCAT        & 42 &13&13&13&1&1&1\\ 
2018jky & Ia      & 03:26:02.140 & $-$17:33:46.44 & 0.014503 & NGC 1329                  & ASAS-SN    & GSP         & 39 &12&12&12&1&1&1\\ 
2019gf  & Ia-91T  & 08:05:35.263 & $-$09:34:56.06 & 0.068469     & -                         & ATLAS      & ePESSTO      & 39 &12&12&12&1&1&1\\ 
2019jf  & Ia      & 08:36:14.793 & $-$05:21:02.02 & 0.041382 & 2MASX J08361402-0521041   & ATLAS      & ePESSTO      & 39 &12&12&12&1&1&1\\ 
2019rm  & Ia      & 05:53:13.270 & $-$73:06:56.74 & 0.023    & 2MASX J05531375-7306576   & ASAS-SN    & ePESSTO      & 36 &11&11&11&1&1&1\\ 
2019so  & Ia-91bg & 12:42:36.420 & $-$40:44:46.79 & 0.013747 & NGC 4622                  & ATLAS      & ePESSTO      & 47 &15&15&15&1&1&0\\ 
2019ahi & Ia      & 13:51:36.599 & $-$08:50:24.29 & 0.028763 & LEDA 170305               & ATLAS      & QUB         &  3 & 1& 1& 1&0&0&0\\ 
2019akg & Ia-91T  & 10:45:03.732 & +00:06:16.37 & 0.039798 & UGC 05867                 & ATLAS      & SCAT        & 39 &12&12&12&1&1&1\\ 
2019awq & Ia      & 04:23:29.390 & $-$15:46:00.55 & 0.036709 & LEDA 146227               & ATLAS      & Cartier+    & 48 &14&14&14&2&2&2\\ 
2019bdz & Ia      & 14:48:36.894 & +06:48:51.93 & 0.034632 & CGCG 048-018              & ATLAS      & Asiago      & 45 &14&14&14&1&1&1\\ 
2019bus & Ia-91T  & 14:29:25.706 & +16:40:55.42 & 0.082326 & LEDA 1514443              & ATLAS      & ePESSTO      & 27 & 9& 9& 9&0&0&0\\ 
2019cvl & Ia      & 19:32:22.129 & $-$29:23:08.31 & 0.023833 & MCG -05-46-002            & ASAS-SN    & SCAT        & 39 &13&13&13&0&0&0\\ 
2019cxu & Ia-02cx & 13:05:30.082 & $-$08:45:52.68 & 0.044014  & -                         & ZTF        & GSP         & 24 & 8& 8& 8&0&0&0\\ 
2019cxx & Ia      & 11:17:48.187 & +13:43:41.55 & 0.024784 & IC 2695                   & Tanaka+    & Tanaka+     & 33 &10&10&10&1&1&1\\ 
2019dks & Ia-91T  & 11:44:05.602 & $-$04:40:25.21 & 0.060682   & WISEA J114405.79-044026.4 & ZTF        & GSP         & 36 &12&12&12&0&0&0\\ 
2019eim & Ia      & 23:49:42.938 & $-$69:42:02.38 & 0.039    & LEDA 277632               & ASAS-SN    & ePESSTO      & 33 &11&11&11&0&0&0\\ 
2019fcf & Ia      & 17:40:17.630 & +17:41:39.16 & 0.058425    & LEDA 1539132              & ATLAS      & ePESSTO      & 33 &11&11&11&0&0&0\\
2019fmr & Ia-91T  & 23:17:38.900 & $-$59:15:42.77 & 0.026915 & LEDA 372765               & ASAS-SN    & Stritzinger & 30 &10&10&10&0&0&0\\ 
2019gbx & Ia      & 12:50:02.812 & $-$14:45:59.96 & 0.013146 & MCG-02-33-017             & ATLAS      & Zhang+      & 57 &19&19&19&0&0&0\\ 
2019gwa & Ia-91T  & 15:58:41.190 & +11:14:25.04 & 0.054953 & SDSS J155841.10+111425.5  & ZTF        & SCAT        & 45 &15&15&15&0&0&0\\
\hline
\end{tabular}
}
\end{center}
\label{tab:sntable}
\tablefoot{\\
\tablefoottext{a}{The number outside the parenthesis indicates the total number of NIR images, and the ``s'' subscript denotes SOFI filters.}
}
\end{table*}

\FloatBarrier
\begin{sidewaystable}
\section{Host galaxy parameters}
\caption{Global photometry for our SN hosts using HostPhot along with host galaxy properties from Prospector.}

\resizebox{\textwidth}{!}{ 
\begin{tabular}{c|cc|cccccccc|ccccc|cc|ccccc|c}
\hline \hline
 & \multicolumn{2}{c|}{GALEX} & \multicolumn{8}{c|}{Optical\tablefootmark{a}} & \multicolumn{5}{c|}{NIR}& \multicolumn{2}{c|}{unWISE}&\multicolumn{5}{c|}{Prospector output\tablefootmark{b}} &    \\
SN&\textit{FUV}&\textit{NUV}&\textit{u}&\textit{v}&\textit{g}&\textit{r}&\textit{i}&\textit{z}&\textit{Y} & Survey&\textit{Y}&\textit{J}&\textit{H}&$K_s$& Survey &\textit{W1}&\textit{W2}&$log(M/M_{\odot})$& $log(z_{\odot})$  &$log(SFR_{100})$&$log(SFR_{250})$&$log(SFR_{400})$& $d_{DLR}$ \\
\hline
2018rw & - & - & 16.22 (0.16) & 18.25 (0.80) & 15.33 (0.01) & 14.65 (0.01) & 14.36 (0.01) & 14.13 (0.13) & 13.96 (0.01) & SkyMapper,DES & - & 13.42 (0.02) &  12.49 (0.02) & 12.11 (0.03) & 2MASS & 11.20 (0.04) & 11.12 (0.09) & 10.72$^{+0.08}_{-0.09}$ &  -0.61$^{+0.26}_{-0.27}$ & 1.51$^{+0.15}_{-0.14}$ & 1.11$^{+0.16}_{-0.14}$ & 0.91$^{+0.16}_{-0.14}$ & 0.50  \\ 
2018yu & - & - & 15.47 (0.75) & - & 12.67 (0.04) & 11.87 (0.03) & 11.11 (0.01) & 10.86 (0.01) & - & SkyMapper & - & 9.86 (0.00) & 9.14 (0.00) & 8.85 (0.01) & 2MASS & 7.90 (0.03) & 7.79 (0.04) & 10.52$^{+0.06}_{-0.09}$ & -0.72$^{+0.18}_{-0.16}$ & 1.76$^{+0.25}_{-0.10}$ & 1.38$^{+0.28}_{-0.10}$& 1.18$^{+0.28}_{-0.10}$ & 0.67\\ 
2018agk & 15.83 (13.92) &  14.66 (0.10) & - & - & 14.70 (0.01) & 14.31 (0.01) & 14.16 (0.02) & 14.06 (0.02) & 14.00 (0.02) & Pan-STARRS & - & 12.98 (0.01) & 12.30 (0.01)& 12.12 (0.02) &VISTA & 11.01 (0.09) & 10.63 (0.18) & 9.94$^{+0.15}_{-0.18}$ & -0.28$^{+0.09}_{-0.15}$ & 2.23$^{+0.45}_{-0.26}$ & 1.84$^{+0.45}_{-0.26}$& 1.64$^{+0.45}_{-0.25}$ & 0.46  \\ 
2018aoz & 14.39 (0.19) & 13.39 (0.07) & 13.95 (0.35) & 13.18 (0.23) & 10.47 (0.01) & 9.74 (0.01) & 9.38 (0.02) & 9.10 (0.02) & 8.90 (0.02) & SkyMapper,Pan-STARRS & - & 7.90 (0.00) & 8.20 (0.00) & 7.00 (0.00) & 2MASS & 6.13 (0.02) & 6.14 (0.03) & 11.07$^{+0.07}_{-0.08}$ & -0.68$^{+0.23}_{-0.24}$  & 1.54$^{+0.22}_{-0.16}$ & 1.16$^{+0.21}_{-0.13}$ & 0.98$^{+0.19}_{-0.11}$& 2.48 \\ 
2018ayg  & 18.11 (43.95) & 17.36 (0.35) & 17.48 (0.16) & - & 15.79 (0.06) &14.98 (0.04) & 14.59 (0.03) & 14.30 (0.03) & -  & SDSS & - & 13.14 (0.01) & 12.33 (0.01) & 11.72 (0.01) & 2MASS & 11.68 (0.02) & 11.85 (0.05) & 10.51$^{+0.04}_{-0.05}$ & -0.31$^{+0.15}_{-0.21}$  & 1.19$^{+0.24}_{-0.23}$ & 0.80$^{+0.23}_{-0.21}$ & 0.61$^{+0.23}_{-0.20}$ & 0.21  \\ 
2018bie & 17.60 (0.67) & 15.92 (0.17) & 17.24 (0.14) & - & 16.33 (0.01) & 15.88 (0.01) & 15.63 (0.02) & 15.45 (0.02) & 15.38 (0.02) & SDSS,Pan-STARRS & - & 13.90 (0.02) & 13.94 (0.04) & 13.92 (0.08) & 2MASS & 12.97 (0.09) & 12.52 (0.17) & 8.91$^{+0.10}_{-0.08}$ & -0.94$^{+0.10}_{-0.04}$  & 1.42$^{+0.23}_{-0.09}$ & 1.52$^{+0.41}_{-0.24}$ &1.52$^{+0.43}_{-0.31}$ & 1.09 \\ 
2018bta  & - & - & 15.10 (0.13) & 14.63 (0.10) & 13.01 (0.02) & 12.58 (0.02) & 12.30 (0.01) &12.10 (0.02)  & - & SkyMapper & -  & 10.86 (0.01) & - & 9.93 (0.01) & VISTA & 8.52 (0.03) & 8.68 (0.06) & 11.17$^{+0.08}_{-0.12}$ & -0.77$^{+0.19}_{-0.15}$  & 1.73$^{+0.23}_{-0.11}$ & 1.34$^{+0.25}_{-0.12}$ & 1.14$^{+0.25}_{-0.12}$ &  0.98 \\ 
2018cow  & 16.43 (17.84)  & 15.90 (0.21) & 16.34 (0.10) & -  & 15.33 (0.05)  & 14.85 (0.04) & 14.62 (0.03) & 14.45 (0.03) & - & SDSS & 13.84 (0.01) & 13.74 (0.01) & 12.98 (0.02) & 12.80 (0.03) &  UKIDSS & 11.40 (0.03) & 11.54 (0.08) & 9.18$^{+0.13}_{-0.15}$ & -0.86$^{+0.18}_{-0.10}$  & 1.60$^{+0.34}_{-0.12}$ & 1.52$^{+0.34}_{-0.18}$& 1.47$^{+0.37}_{-0.22}$ & 0.44\\ 
2018dda  & 16.19 (14.72)  & 15.31 (0.14) & - & - & 14.27 (0.01) & 13.70 (0.01)& 13.44 (0.02) & 13.27 (0.02) & 13.02 (0.02) & Pan-STARRS & - & 12.00 (0.01) & - & 11.20 (0.01) & VISTA&  10.65 (0.06) & 10.61 (0.14) & 10.23$^{+0.05}_{-0.05}$ & 0.29$^{+0.11}_{-0.09}$  & 1.67$^{+0.27}_{-0.17}$ & 1.29$^{+0.27}_{-0.16}$ & 1.09$^{+0.26}_{-0.15}$ & 0.21\\ 
2018ebu  & 17.50 (29.40) & 16.46 (0.22) & - & - & 16.77 (0.01) & 16.29 (0.01) &16.10 (0.01) & 15.92 (0.01) &15.82 (0.01) & DES & - & 14.77 (0.01) &  14.21 (0.01) & 13.95 (0.02) & VISTA & 13.51 (0.19) & - & 10.26$^{+0.05}_{-0.07}$ & 0.22$^{+0.20}_{-0.14}$  & 1.25$^{+0.22}_{-0.22}$ & 0.92$^{+0.19}_{-0.17}$ & 0.77$^{+0.18}_{-0.15}$ & 0.25 \\ 
2018enc & - & - & - & - & 16.75 (0.02) & 16.58 (0.02) & 16.50 (0.02) &16.47 (0.02) & 16.36 (0.04) & Pan-STARRS & - & 15.40 (0.04) & 15.27 (0.12) & 14.66 (0.08) & VISTA & 14.38 (0.14) & 14.18 (0.36) & 9.12$^{+0.04}_{-0.06}$ & -0.85$^{+0.19}_{-0.11}$  & 1.66$^{+0.18}_{-0.10}$ & 1.27$^{+0.19}_{-0.10}$ & 1.07$^{+0.19}_{-0.10}$ & 1.30 \\ 
2018eov  & - & - & 17.10 (0.90) & 15.34 (0.61)& 13.44 (0.02) & 12.95 (0.02) & 12.46 (0.02) & 12.41 (0.02) & - & SkyMapper  & - & 11.40 (0.05) & -  & 10.55 (0.16) & VISTA & 9.12 (0.06) & 9.11 (0.07) & 10.81$^{+0.06}_{-0.06}$ & -0.82$^{+0.09}_{-0.11}$  & 1.47$^{+0.14}_{-0.11}$ & 1.07$^{+0.14}_{-0.11}$ & 0.87$^{+0.14}_{-0.11}$ &  0.95 \\ 
2018evt  & 15.75 (11.03)  & 15.15 (0.14) & - & - & 14.77 (0.01) & 14.24 (0.01) & 14.00 (0.02) & 13.84 (0.02) &13.67 (0.02) & Pan-STARRS & - & 12.72 (0.33) & 12.12 (0.17) & 11.95 (0.48) & VISTA & 11.07 (0.05) & 11.15 (0.12) & 10.37$^{+0.08}_{-0.10}$ & -0.77$^{+0.16}_{-0.11}$  & 1.48$^{+0.26}_{-0.19}$ & 1.16$^{+0.22}_{-0.13}$ & 1.03$^{+0.20}_{-0.12}$ & 0.48 \\ 
2018exb  & 17.85 (39.51) & 16.60 (0.25) & 16.35 (0.17) & - & 16.05 (0.01) & 15.37 (0.01) & 15.01 (0.02) & 14.77 (0.02) & 14.61 (0.02) & SkyMapper, Pan-STARRS & - & 13.70 (0.02) &12.40 (0.02) & 12.17 (0.03) & 2MASS  & 11.63 (0.05) & - & 10.71$^{+0.05}_{-0.06}$ & -0.26$^{+0.14}_{-0.18}$  & 1.58$^{+0.33}_{-0.21}$ & 1.20$^{+0.32}_{-0.19}$ & 1.01$^{+0.31}_{-0.17}$  &  0.82\\ 
2018exc  & 17.77 (34.83) & 16.75 (0.27) & 16.34 (0.28) & 15.94 (0.26) & 15.86 (0.13) & 15.24 (0.08) & 15.09 (0.07) &  14.87 (0.07)& - & SkyMapper & - & 13.31 (0.01) & -  & 12.26 (0.01) & VISTA & 11.53 (0.04) & 11.31 (0.08) & 10.75$^{+0.06}_{-0.08}$ & 0.17$^{+0.02}_{-0.04}$  & 1.79$^{+0.29}_{-0.19}$ & 1.40$^{+0.29}_{-0.18}$ & 1.21$^{+0.29}_{-0.17}$ &  0.80\\ 
2018feq  & 17.22 (24.43) & 16.12 (0.18) & - & - & 14.97 (0.01) & 14.42 (0.00) & 14.23 (0.01) & 14.05 (0.01) & 13.90 (0.01) & DES & - & 12.77 (0.01) & 12.13 (0.01) & 11.79 (0.01) & VISTA & 11.09 (0.04) & 10.80 (0.07) & 10.38$^{+0.08}_{-0.11}$ & -0.07$^{+0.18}_{-0.22}$  & 1.79$^{+0.37}_{-0.20}$ & 1.40$^{+0.36}_{-0.18}$ & 1.21$^{+0.35}_{-0.16}$ & 0.30\\ 
2018hfp  & 14.67 (14.07) & 14.93 (0.13) & 14.67 (0.13) & - & 14.11 (0.01) & 13.49 (0.01) & 13.18 (0.02) & 12.98 (0.02) & 12.65 (0.02) & SkyMapper,Pan-STARRS & - & 11.71 (0.01) & - & 10.81 (0.01) & VISTA & 9.91 (0.04) & 9.86 (0.09) & 11.21$^{+0.04}_{-0.07}$ & 0.16$^{+0.02}_{-0.04}$  & 2.16$^{+0.56}_{-0.30}$ & 1.77$^{+0.55}_{-0.30}$ & 1.57$^{+0.55}_{-0.30}$ &  0.50 \\ 
2018hgc  & 18.28 (0.93) & 17.33 (0.34) & 17.27 (0.14) & -  & 16.14 (0.07) & 15.69 (0.05) & 15.34 (0.04) & 15.15 (0.04) & -  & SDSS & - & 13.64 (0.01) & 13.02 (0.02) & 13.03 (0.03) & 2MASS & 12.26 (0.04) & 12.00 (0.07)& 10.51$^{+0.09}_{-0.09}$ & -0.46$^{+0.31}_{-0.33}$  & 1.57$^{+0.29}_{-0.20}$ & 1.19$^{+0.28}_{-0.16}$ & 1.01$^{+0.26}_{-0.14}$ & 1.82 \\ 
2018hhn & 15.84 (13.17) & 15.03 (0.14) & 15.10 (0.06) & - & 13.90 (0.03) & 13.27 (0.02) & 12.95 (0.02) & 12.70 (0.01) & -  & SDSS  & 12.29 (0.02) & 12.48 (0.02) & 11.22 (0.01) & 1.94 (0.02) & UKIDSS & 10.21 (0.02) & 10.20 (0.04) & 10.46$^{+0.18}_{-0.13}$ & -0.57$^{+0.12}_{-0.11}$  & 1.44$^{+0.20}_{-0.11}$ & 1.47$^{+0.43}_{-0.23}$ & 1.46$^{+0.49}_{-0.31}$ &  1.21\\ 
2018hjw  & 16.96 (17.56) & 16.10 (0.19) & 16.73 (0.11) & - & 15.83 (0.06) & 15.56 (0.05) & 15.23 (0.04) &15.33 (9.05) & - & SDSS & - & 14.12 (0.01) & 13.37 (0.01) & 12.51 (0.01) & 2MASS  & 11.86 (0.03) & 11.56 (0.07) & 9.62$^{+0.18}_{-0.20}$ & -0.27$^{+0.16}_{-0.16}$  & 2.08$^{+0.43}_{-0.23}$ & 1.69$^{+0.42}_{-0.22}$ & 1.49$^{+0.41}_{-0.20}$ & 0.07 \\ 
2018ilu  & 19.16 (63.06) & 18.50 (0.64) & 19.96 (0.65) & - & 19.39 (0.33) & 19.17 (0.29) & 18.91 (0.26) &18.78 (0.30) & - & SDSS &18.29 (0.22) & 18.95 (0.41) & 17.50 (0.28) & 17.52 (0.53) & UKIDSS & - & - & 7.72$^{+0.19}_{-0.36}$ & -0.75$^{+0.32}_{-0.17}$  & 1.98$^{+0.62}_{-0.44}$ & 1.60$^{+0.63}_{-0.44}$ & 1.40$^{+0.63}_{-0.44}$ & 1.36\\ 
2018jag  & 17.59 (43.75) & 16.24 (0.2) & 16.75 (0.12) & - & 15.06 (0.04) & 14.21 (0.03) & 13.77 (0.02) & 13.48 (0.02) & - & SDSS & 12.84 (0.01) & 12.62 (0.02) & - & 11.38 (0.01) & UKIDSS & 10.76 (0.02) & 10.72 (0.05) & 10.79$^{+0.11}_{-0.16}$ & -0.55$^{+0.23}_{-0.24}$  & 1.48$^{+0.24}_{-0.15}$ & 1.17$^{+0.21}_{-0.09}$ &1.02$^{+0.22}_{-0.09}$ &  2.14\\ 
2018jky  & 15.87 (0.33) & 15.13 (0.13) & - & - & 13.04 (0.01) & 12.32 (0.00) & 12.00 (0.01) &11.75 (0.01) & - & DES & - & 10.54 (0.01) & 9.93 (0.01) & 9.65 (0.01) & 2MASS & 9.30 (0.03) & 9.30 (0.05) & 10.75$^{+0.04}_{-0.03}$ & -0.18$^{+0.19}_{-0.21}$  & 1.40$^{+0.22}_{-0.16}$ & 1.02$^{+0.21}_{-0.15}$ & 0.83$^{+0.20}_{-0.14}$ & 3.26 \\ 
2019gf  & - & - & 21.06 (1.41) & - & 20.64 (0.04) & 20.16 (0.03) & 20.00 (0.02) & 20.06 (0.05) & - & SDSS,Pan-STARRS & -  & 16.10 (0.04) & 17.26 (0.23) & - & 2MASS & - & - & 8.73$^{+0.06}_{-0.06}$ & -0.05$^{+0.21}_{-0.75}$  & 1.05$^{+0.35}_{-0.34}$ & 0.65$^{+0.35}_{-0.34}$ & 0.45$^{+0.35}_{-0.34}$ &  1.60\\ 
2019jf  & 18.74 (142.02) & 17.14 (0.32) & - & - & 15.68 (0.01) & 14.94 (0.01) & 14.56 (0.02) & 14.40 (0.02) & - & Pan-STARRS & - & 13.11 (0.01) & - & 12.17 (0.01) & VISTA & 11.16 (0.04) & 10.98 (0.06) & 10.84$^{+0.07}_{-0.08}$ & -0.62$^{+0.32}_{-0.25}$  & 1.52$^{+0.38}_{-0.21}$ & 1.14$^{+0.36}_{-0.18}$ & 0.95$^{+0.35}_{-0.16}$ & 0.75\\ 
2019rm  & 15.50 (8.84) & 14.84 (124.77) & - & - & - & - & - & 14.79 (0.05) & - & LegacySurvey & - & 13.53 (0.03) & 13.18 (0.05) & 12.31 (0.04) & 2MASS & 10.81 (0.03) & 10.64 (0.04) & 9.39$^{+0.16}_{-0.34}$ & -0.82$^{+0.15}_{-0.12}$  & 2.76$^{+0.44}_{-0.19}$ & 2.36$^{+0.44}_{-0.20}$ & 2.16$^{+0.44}_{-0.20}$  &  0.24 \\ 
2019so  & 14.46 (5.85)  & 14.01 (0.11) & - & - & 13.13 (0.08) & 13.21 (0.08) & 12.07 (0.03) &11.59 (0.02) & -  & SkyMapper & - & 10.41 (0.01) & - & 9.49 (0.01) & VISTA & 8.71 (0.04) & 8.72 (0.07) & 11.01$^{+0.02}_{-0.02}$ & -0.12$^{+0.20}_{-0.32}$  & 0.93$^{+0.21}_{-0.28}$ & 0.54$^{+0.21}_{-0.28}$ & 0.33$^{+0.21}_{-0.28}$ & 0.31 \\ 
2019ahi  & 15.75 (9.83) & 14.88 (0.12) & - & - & 14.20 (0.01) & 13.90 (0.01) & 13.72 (0.02) & 13.62 (0.02) & 13.46 (0.02) & Pan-STARRS & - & 12.74 (0.13) & 11.93 (0.10) & 11.64 (0.06) & VISTA & 10.90 (0.06) & 10.63 (0.11) & 10.50$^{+0.13}_{-0.15}$ & -0.44$^{+0.28}_{-0.17}$  & 1.35$^{+0.20}_{-0.15}$ & 1.13$^{+0.19}_{-0.11}$ & 1.02$^{+0.26}_{-0.12}$ &  0.65\\ 
2019akg  &  16.26 (0.39) & 15.35 (0.15) & 15.91 (0.08) & - & 14.68 (0.04) & 14.14 (0.03) & 13.82 (0.02) & 13.61 (0.02) & - &  SDSS & 13.18 (0.01) & 13.17 (0.01) & 12.19 (0.02) & 11.87 (0.02) & UKIDSS & 11.05 (0.03) & 11.25 (0.08) & 10.63$^{+0.10}_{-0.13}$ & -0.78$^{+0.11}_{-0.08}$  & 1.50$^{+0.23}_{-0.14}$ & 1.19$^{+0.17}_{-0.09}$ & 1.05$^{+0.17}_{-0.09}$ & 0.40 \\
2019awq  & 15.77 (28.61) & 16.98 (0.30) & - & - & 15.93 (0.01) &15.27 (0.01) & 14.92 (0.02) & 14.72 (0.02) & 14.53 (0.02) & Pan-STARRS & - & 13.37 (0.01) & - & 12.37 (0.01) & VISTA & 11.63 (0.04) & 11.90 (0.11) & 10.52$^{+0.09}_{-0.13}$ & -0.72$^{+0.25}_{-0.19}$  & 1.39$^{+0.25}_{-0.20}$ & 1.08$^{+0.20}_{-0.15}$ & 0.94$^{+0.18}_{-0.13}$ & 0.77 \\ 
2019bdz  & 16.85 (21.61) & 15.87 (0.18) &  16.21 (0.08) & - & 15.18 (0.05) & 14.71 (0.03) & 14.42 (0.03) & 14.24 (0.03) & - & SDSS & 13.65 (0.01) & 13.51 (0.01) & 13.14 (0.01) & 12.80 (0.02) & UKIDSS & 11.58 (0.04) & 11.54 (0.07) & 9.94$^{+0.12}_{-0.10}$ & -0.96$^{+0.07}_{-0.03}$  &1.54$^{+0.23}_{-0.10}$ & 1.34$^{+0.20}_{-0.10}$ & 1.26$^{+0.23}_{-0.14}$ & 0.86 \\ 
2019bus  & 19.44 (69.53) & 18.71 (0.63) & 17.69 (0.28) & - & 17.77 (0.14) & 17.22 (0.11) & 16.93 (0.10) & 16.73 (0.09) & - &  SDSS & - & 15.50 (0.03) & - & 14.70 (0.05) & 2MASS & 13.33 (0.03) & 13.58 (0.10) & 9.92$^{+0.14}_{-0.21}$ & -0.72$^{+0.23}_{-0.20}$  & 1.58$^{+0.28}_{-0.15}$ & 1.89$^{+1.08}_{-0.43}$ & 1.90$^{+1.16}_{-0.50}$ &  0.70\\ 
2019cvl  & 14.64 (0.25) & 13.95 (0.10) & - & - & 13.55 (0.01) &13.02 (0.01) & 12.82 (0.02) &12.69 (0.02) & 12.37 (0.02) & Pan-STARRS & - & 11.36 (0.01) & - & 10.49 (0.01) & VISTA & 9.27 (0.06) &9.14 (0.07) & 10.92$^{+0.07}_{-0.10}$ & -0.84$^{+0.15}_{-0.12}$  & 1.59$^{+0.24}_{-0.17}$ & 1.22$^{+0.22}_{-0.12}$ & 1.06$^{+0.20}_{-0.10}$ &  1.20\\ 
2019cxu  & 19.13 (61.85) & 18.16 (0.50) & - & - & 19.65 (0.03) & 19.46 (0.03) & 19.32 (0.04) & 19.10 (0.05) & 19.27 (0.11) & Pan-STARRS & - & 19.19 (0.38) & - & 16.71 (0.17) & VISTA & 13.75 (0.07) & 17.63 (6.65) & 8.25$^{+0.16}_{-0.14}$ & -0.96$^{+0.06}_{-0.03}$  & 3.37$^{+0.65}_{-0.56}$ & 2.97$^{+0.65}_{-0.55}$ & 2.76$^{+0.65}_{-0.55}$  & 0.68 \\
2019cxx  & 21.42 (734.05) & 19.87 (1.05) & 17.34 (0.14) & - & 16.35 (0.07) &15.95 (0.06) & 15.73 (0.05) &15.61 (0.05) & -  & SDSS & 15.00 (0.02) & 15.40 (0.02) & 14.16 (0.03) & 13.91 (0.04) & UKIDSS & 14.84 (0.27) &20.20 (111.20) & 9.05$^{+0.04}_{-0.92}$ & -0.24$^{+0.13}_{-0.14}$  & 0.81$^{+0.38}_{-0.43}$ & 2.51$^{+1.92}_{-0.82}$ & 2.60$^{+1.93}_{-0.83}$ &  0.46\\ 
2019dks  & 19.89 (112.63) & 18.49 (0.56) & - & - & 19.71 (0.03) & 19.44 (0.03) & 19.31 (0.03) &19.06 (0.05) & 19.16 (0.11) & Pan-STARRS  & - & 18.03 (0.13) & 18.22 (0.30) & - & VISTA & 13.61 (0.08) & 13.66 (0.22) & 8.55$^{+0.16}_{-0.15}$ & -0.78$^{+0.07}_{-0.09}$  & 3.19$^{+0.65}_{-0.52}$ & 2.79$^{+0.65}_{-0.52}$ & 2.59$^{+0.64}_{-0.51}$ & 0.97\\ 
2019eim  & 17.70 (38.68) & 16.75 (0.26) & - & - & 17.20 (0.18) & 16.47 (0.12) & - & 15.97 (0.09)& - & LegacySurvey  & 15.14 (0.02) & 14.80 (0.02) & - & 13.75 (0.06) & VISTA & 12.90 (0.09) &14.38 (0.80) & 9.91$^{+0.08}_{-0.010}$ & -0.27$^{+0.30}_{-0.38}$  & 1.57$^{+0.18}_{-0.17}$ & 1.17$^{+0.18}_{-0.17}$ & 0.97$^{+0.18}_{-0.17}$ &  0.19\\ 
2019fcf  & 16.86 (21.88) & 16.19 (0.23) & - & - & 16.71 (0.01) & 16.00 (0.01) & 15.59 (0.02) & 15.36 (0.02) & 15.03 (0.02) & Pan-STARRS & - & 13.75 (0.02) & 13.03 (0.02) & 12.87 (0.03) & 2MASS & 11.67 (0.03) & 12.12 (0.10) & 10.66$^{+0.11}_{-0.17}$ & -0.66$^{+0.25}_{-0.23}$ & 1.58$^{+0.27}_{-0.18}$ & 1.22$^{+0.24}_{-0.13}$ & 1.06$^{+0.24}_{-0.12}$ &  0.73\\
2019fmr  & 16.93 (0.48) & 15.93 (0.17) & - & - & 16.01 (0.01) & 15.63 (0.01)& 15.53 (0.01) &15.39 (0.01) & 15.14 (0.01) & DES & - & 14.58 (0.02) & 14.31 (0.03) & 13.77 (0.03) & DES & 12.71 (0.18) & - & 9.54$^{+0.07}_{-0.12}$ & -0.94$^{+0.14}_{-0.05}$  & 1.59$^{+0.26}_{-0.20}$ & 1.24$^{+0.24}_{-0.14}$ & 1.08$^{+0.22}_{-0.13}$ &  0.57\\ 
2019gbx  & 15.59 (9.27) & 14.24 (0.09) & 15.01 (0.35) & 13.81 (0.15) & 13.27 (0.04) & 12.61 (0.03) & 11.86 (0.02) &11.46 (0.02) & - & SkyMapper & 10.89 (0.02) & 10.43 (0.01) & 9.77 (0.01) & 9.52 (0.01) & VISTA & 8.38 (0.04) & 8.37 (0.05)& 101.02$^{+0.04}_{-0.05}$ & -0.24$^{+0.23}_{-0.18}$  & 1.65$^{+0.35}_{-0.21}$ & 1.26$^{+0.35}_{-0.20}$ & 1.06$^{+0.35}_{-0.19}$ & 2.62 \\ 
2019gwa & - & 17.90 (0.48) & 18.25 (0.24) & - & 17.11 (0.11) & 16.68 (0.08) & 16.38 (0.07) &16.23 (0.07) & - & SDSS & - & 15.03 (0.05) & 14.13 (0.04) & 14.32 (0.09) & 2MASS & 14.09 (0.09) & 13.98 (0.22) & 10.12$^{+0.02}_{-0.01}$ & -0.97$^{+0.04}_{-0.02}$  & 0.84$^{+0.30}_{-0.37}$ & 0.53$^{+0.44}_{-0.36}$ & 0.36$^{+0.49}_{-0.36}$ & 0.33 \\ 
\hline
\end{tabular}
}
\label{tab:host_table}

\tablefoot{\\
\tablefoottext{a}{We only include the filters used for Prospector. If multiple surveys are indicated in the optical survey column, only the $u$\,band is either from SDSS or SkyMapper, and the remaining filters from the other survey. Furthermore, $v$\,band is always from SkyMapper.}
}
\\
\tablefoottext{b}{Host stellar mass is given in units of solar masses ($M_\odot$), metallicity in units of solar metallicity ($Z_\odot$), and star-formation rate (SFR) in solar masses per year ($M_\odot\, yr^{-1}$). The subscripts (100, 250, 400) denote SFRs averaged over the past 100, 250, and 400 Myr, respectively.}
\end{sidewaystable}

\begin{sidewaystable}
\caption{Same as Table \ref{tab:host_table} but with local photometry. For each filter we indicate the apparent magnitude inside apertures of 1, 2 and 3 kpc centered around the SN position. Note that in some cases it was not possible to measure any flux because there is no host light within the aperture. 
}
\resizebox{\textwidth}{!}{ 
\begin{tabular}{c|cc|cccccccc|ccccc|cc}
\hline \hline
 & \multicolumn{2}{c|}{GALEX} & \multicolumn{8}{c|}{Optical} & \multicolumn{5}{c|}{NIR}& \multicolumn{2}{c}{unWISE} \\
SN&FUV&NUV&u&v&g&r&i&z&Y & Survey&Y&J&H&K\_s& Survey &W1&W2 \\
\hline
2018rw  & - & - & 22.64 (1.64), 20.65 (0.55), 20.23 (0.55)& 22.73 (1.72), 20.50 (0.45), 19.84 (0.37) & 20.71 (0.02), 19.03 (0.01), 18.18 (0.01) & 20.08 (0.01), 18.40 (0.01), 17.51 (0.01) & 19.81 (0.01), 18.14 (0.01), 17.24 & 19.60 (0.01), 17.93 (0.01), 17.03 (0.01) &  19.45 (0.03), 17.80 (0.01), 16.90 (0.01) & SkyMapper,DES & - & 18.07 (0.65), 16.38 (0.28), 15.46 (0.18) &  17.04 (0.52), 15.51 (0.25), 14.72 (0.19) & 17.16 (1.32), 15.35 (0.50), 14.40 (0.31) & 2MASS & 16.17 (0.12), 14.61 (0.11), 13.73 (0.11) & 16.00 (0.13), 14.43 (0.11), 13.57 (0.11) \\
2018yu  & - & - & -,17.26 (0.44), 17.15 (0.60) & - & 18.48 (0.44), 16.88 (0.20), 16.07 (0.15) & 17.97 (0.33), 16.88 (0.24), 16.30 (0.21) & 16.87 (0.02) 15.49 (0.02), 14.66 (0.01) & 16.74 (0.04), 15.28 (0.03), 14.47 (0.02)& - & SkyMapper & - & 15.47 (0.28), 13.96 (0.14), 13.17 (0.11) & 14.73 (0.33), 13.29 (0.18), 12.49 (0.13) & 14.52 (0.51), 13.17 (0.30), 12.40 (0.23) & 2MASS & 14.36 (0.24), 12.87 (0.24), 12.05 (0.26) & 14.34 (0.35), 12.86 (0.36), 12.01 (0.37) \\
2018agk & 23.28 (1.53), 20.29 (0.34), 19.27 (0.22) & 21.24 (0.20), 19.59 (0.09), 20.90 (0.49) & - & - & 19.03 (0.02), 17.73 (0.01), 17.00 (0.01) & 18.68 (0.02), 17.35 (0.01), 16.60 (0.01) & 18.60 (0.02), 17.25 (0.02), 16.47 (0.02) & 18.49 (0.02), 17.14 (0.02), 16.35 (0.02) & 18.41 (0.03), 17.07 (0.02), 16.26 (0.02) & Pan-STARRS & - & 17.49 (0.07), 16.10 (0.04), 15.31 (0.03) & 16.95 (0.07), 15.53 (0.04), 14.71 (0.03) & 16.59 (0.11), 15.31 (0.07), 14.46 (0.05) &VISTA & 16.87 (0.25), 15.32 (0.23), 14.43 (0.23) & 16.74 (0.37), 15.19 (0.34), 14.30 (0.34) \\
2018aoz & 21.22 (0.14), 19.90 (0.09), 19.00 (0.85) & 20.52 (0.06), 19.03 (0.04), 18.17 (0.03) & -,-,18.54 (2.32) &  21.21 (15.00),-,19.02 (6.00) & - & - & 21.04 (0.32), 19.34 (0.14), 18.65 (0.11) & - & - & SkyMapper,Pan-STARRS & - & 18.48 (8.31), -, - & 16.05 (1.92), 15.04 (1.62), 15.29 (3.34) & - & 2MASS & 15.03 (1.57), 13.49 (1.52), 12.65 (1.58) & 15.02 (1.89), 13.51 (1.87), 12.64 (1.89) \\
2018ayg & -, 24.25 (2.92), 23.24 (1.78) & 24.63 (1.20), 22.37 (0.41), 21.56 (0.29) & 19.67 (0.52), 18.49 (0.26), 18.13 (0.22) & - & 17.85 (0.15), 16.72 (0.09), 16.37 (0.08) & 17.06 (0.10), 15.90 (0.06), 15.55 (0.06) & 16.69 (0.09), 15.51 (0.05), 15.16 (0.04) & 16.38 (0.07), 15.22 (0.04), 14.88 (0.04) & -  & SDSS & - &  16.05 (0.13), 14.58 (0.07), 13.86 (0.05) & 15.35 (0.18), 13.91 (0.10), 13.18 (0.08) & 15.02 (0.22), 13.58 (0.12), 12.84 (0.09) & 2MASS & 16.09 (0.13), 14.48 (0.12), 13.62 (0.12) & 16.01 (0.15), 14.54 (0.14), 13.68 (0.14) \\
2018bie & 24.89 (0.66), 23.59 (0.01), 22.76 (0.03) & 23.65 (0.18), 22.20 (0.10), 21.26 (0.07) & 24.53 (24.86), 21.77 (2.48), 20.73 (1.21) & - & 22.44 (1.84), 20.54 (0.59), 19.50 (0.35) & 21.53 (0.98), 19.92 (0.41), 18.95 (0.25)  &21.31 (1.11), 19.66 (0.39), 18.67 (0.23) & 20.83 (0.84), 18.06 (0.33), 18.21 (0.22) & 21.00 (0.20), 19.43 (0.09), 18.53 (0.06) & SDSS,Pan-STARRS & - & -,-,17.48 (2.14) & - & 17.90 (4.99), 17.06 (4.61), 15.37 (1.48) & 2MASS & 19.73 (2.14), 18.06 (1.82), 16.91 (1.45) & 18.68 (2.16), 17.12 (2.00), 16.47 (2.45) \\
2018bta  & - & - & -,-,19.57 (0.99) & -, 20.78 (1.55), 18.81 (0.39) & 19.94 (0.08), 18.28 (0.04), 17.08 (0.02) & 19.52 (0.06), 17.89 (0.03), 16.66 (0.02) & 19.39 (0.12)., 17.83 (0.06), 16.43 (0.03) & 19.23 (0.21), 17.60 (0.09), 16.27 (0.04)  & - & SkyMapper & -  & 18.53 (0.23), 17.18 (0.13), 15.60 (0.05) & - & 17.62 (0.36), 16.10 (0.17), 14.74 (0.08) & VISTA & 15.85 (0.30), 14.24 (0.27), 13.26 (0.25) & 15.74 (0.35), 14.13 (0.31), 13.13 (0.28) \\
2018cow  & 20.50 (0.45), 18.88 (0.22), 18.27 (0.17) & 19.83 (0.12), 18.46 (0.07), 17.88 (0.06) &  18.38 (0.27), 17.26 (0.15), 16.79 (0.12) & -  & 17.47 (0.15), 16.27 (0.08), 15.77 (0.06)  & 17.00 (0.10), 15.77 (0.06), 15.27 (0.05) & 16.77 (0.09), 15.53 (0.05), 15.02 (0.04) & 16.60 (0.08), 15.34 (0.05), 14.84 (0.04) & - & SDSS & 16.14 (0.05), 14.81 (0.03), 14.22 (0.03) & 15.71 (0.03), 14.43 (0.02), 13.91 (0.02) & 15.21 (0.03), 13.89 (0.02), 13.31 (0.02) & 15.01 (0.06), 13.70 (0.04), 13.13 (0.04) &  UKIDSS &  15.25 (0.20), 13.77 (0.20), 13.05 (0.21) & 15.17 (0.25), 13.70 (0.25), 12.97 (0.28) \\
2018dda  & 21.25 (0.38), 19.65 (0.19), 18.92 (0.14) & 20.93 (0.18), 19.23 (0.09), 18.34 (0.06) & - & - & 17.82 (0.01), 16.21 (0.01), 15.50 (0.01) & 17.14 (0.01), 15.51 (0.01), 14.83 (0.01) & 16.80 (0.02), 15.19 (0.02), 14.52 (0.02) & 16.58 (0.02), 14.96 (0.02), 14.31 (0.02) & 16.34 (0.02), 14.73 (0.02), 14.08 (0.02) &  Pan-STARRS & - & 15.16 (0.01), 13.52 (0.11), 12.88 (0.01) & - & 14.28 (0.02), 12.62 (0.01), 12.00 (0.01) & VISTA&  13.85 (0.08), 12.58 (0.08), 11.94 (0.09) & 13.86 (0.09), 12.56 (0.10), 11.90 (0.11) \\
2018ebu  & 23.27 (0.78), 22.02 (0.46), 21.35 (0.34) & 22.97 (0.39), 21.60 (0.21), 20.81 (0.15) & - & - & 20.76 (0.01), 19.27 (0.01), 18.38 (0.01) & 20.31 (0.01), 18.76 (0.01), 17.82 (0.01) & 20.13 (0.01), 18.56 (0.01), 17.60 (0.01) & 19.96 (0.01), 18.37 (0.01), 17.41 (0.01) & 19.90 (0.02), 18.27 (0.01), 17.35 (0.01) & DES & - & 18.09 (0.05), 16.62 (0.03), 15.95 (0.02) & 17.42 (0.05), 15.95 (0.03), 15.30 (0.02) & 17.15 (0.08), 15.65 (0.04), 14.98 (0.03) & VISTA & 18.89 (0.31), 17.30 (0.26), 16.40 (0.25) & 18.70 (0.43), 17.17 (0.36), 16.28 (0.34) \\
2018enc  & - & - & -  &  - &21.00 (0.09), 19.21 (0.04), 18.34 (0.02) & 20.81 (0.06), 19.11 (0.03), 18.17 (0.02) & 20.83 (0.05), 19.07 (0.02), 18.11 (0.02) & 20.69 (0.10), 19.04 (0.05), 18.08 (0.03) & 20.28 (0.15), 18.87 (0.08), 17.85 (0.05) & Pan-STARRS & - & 19.61 (0.51), 17.84 (0.20), 17.09 (0.15) & 19.42 (0.83), 17.44 (0.27), 16.37 (0.15) & 18.89 (0.95), 17.56 (0.56), 16.28 (0.26) & VISTA & - & - \\
2018eov  & - & - & 19.89 (0.63), 18.45 (0.38), 18.18 (0.44) &19.43 (0.34), 18.19 (0.23), 18.29 (0.17) & 18.13 (0.03), 16.48 (0.02), 15.65 (0.02) & 17.73 (0.02), 16.07 (0.02), 15.22 (0.02) & 17.36 (0.03), 15.78 (0.02), 14.83 (0.02) & 17.18 (0.04), 15.56 (0.03), 14.64 (0.02) & - & SkyMapper  & - & 15.84 (0.06), 14.18 (0.06), 13.21 (0.05) & -  & 15.00 (0.17), 13.31 (0.16), 12.35 (0.16) & VISTA & 14.92 (0.42), 13.31 (0.38), 12.34 (0.35) & 14.78 (0.40), 13.19 (0.37), 12.27 (0.35) \\
2018evt  & 21.92 (0.67), 20.72 (0.40), 20.02 (0.29) & 21.99 (0.40), 20.39 (0.20), 19.65 (0.15) & - & - & 19.48 (0.02), 18.08 (0.02), 17.29 (0.02) & 19.02 (0.02), 17.58 (0.02), 16.77 (0.01) & 18.78 (0.02), 17.36 (0.02), 16.55 (0.02)& 18.63 (0.02), 17.20 (0.02), 16.40 (0.02) & 18.46 (0.03), 17.05 (0.02), 16.24 (0.02) & Pan-STARRS & - & 17.39 (0.39), 15.93 (0.35), 15.11 (0.34) & 16.77 (0.26), 15.32 (0.20), 14.60 (0.19) & 16.63 (0.60), 15.10 (0.51), 14.33 (0.49) & VISTA & 16.88 (0.25), 15.42 (0.24), 14.57 (0.24) & 16.87 (0.36), 15.36 (0.35), 14.49 (0.35) \\
2018exb  & -, 25.66 (6.35), 23.32 (1.39) & 25.29 (1.45), 22.99 (0.51), 21.85 (0.31) & 20.41 (0.39), 19.36 (0.30), 18.69 (0.25) & - & 19.87 (0.02), 18.42 (0.02), 17.58 (0.02), & 19.18 (0.02), 17.73 (0.02), 16.89 (0.02) & 18.80 (0.02), 17.37 (0.02), 16.51 (0.02) & 18.58 (0.02), 17.14 (0.02), 16.28 (0.02) & 18.40 (0.02), 16.99 (0.02), 16.13 (0.02) & SkyMapper, Pan-STARRS & - & 17.14 (0.27), 15.64 (0.13), 14.78 (0.09) & 16.32 (0.29), 14.79 (0.14), 13.93 (0.10) & 16.00 (0.39), 14.48 (0.19), 13.63 (0.13) & 2MASS  & 16.36 (0.12), 14.85 (0.10), 14.05 (0.10) & 16.32 (0.15), 14.83 (0.13), 14.03 (0.14) \\
2018exc  & 23.54 (1.00), 22.50 (0.67), 22.01 (0.56) & 23.58 (0.55), 21.93 (0.27), 20.94 (0.17) & 23.29 (7.79), 20.91 (1.74), 19.69 (0.86) & -, 19.14 (0.50), 18.56 (0.44) & 19.96 (0.29), 18.38 (0.14), 17.63 (0.11) & 19.11 (0.15), 17.69 (0.08), 16.98 (0.07) & 18.56 (0.10), 17.26 (0.06), 16.61 (0.05) & 18.20 (0.08), 16.86 (0.05), 16.23 (0.04) & - & SkyMapper & - & 16.47 (0.02), 15.21 (0.01), 14.62 (0.01) & -  & 15.28 (0.02), 14.05 (0.01), 13.49 (0.01) & VISTA & 16.69 (0.13), 15.19 (0.11), 14.34 (0.11) & 16.59 (0.14), 15.09 (0.12), 14.23 (0.12) \\
2018feq  & 21.39 (0.37), 20.31 (0.23), 19.32 (0.15) & 20.70 (0.15), 19.19 (0.08), 18.30 (0.06) & - & - & 17.88 (0.01), 16.21 (0.01), 15.52 (0.01) & 17.34 (0.01), 15.65 (0.00), 14.98 (0.00) & 17.22 (0.01), 15.48 (0.01), 14.80 (0.01) & 17.03 (0.01), 15.28 (0.01), 14.61 (0.01) & 16.83 (0.01), 15.17 (0.01), 14.50 (0.01) & DES & - & 16.01 (0.02), 14.25 (0.01), 13.42 (0.01) & 15.40 (0.01), 13.63 (0.01), 12.80 (0.01) & 15.05 (0.02), 13.25 (0.01), 12.42 (0.02) & VISTA & 15.97 (0.13), 14.31 (0.11), 13.38 (0.11) & 15.64 (0.13), 14.00 (0.11), 13.09 (0.11) \\
2018hfp  & 25.49 (3.55), 21.81 (0.49), 20.20 (0.23) & 21.60 (0.25), 20.05 (0.10), 19.03 (0.07) & 21.26 (1.24), 18.98 (0.33), 17.74 (0.19) & -  & 19.45 (0.02), 17.74 (0.01), 16.43 (0.01) & 18.74 (0.02), 17.02 (0.01), 15.73 (0.01) & 18.41 (0.02), 16.67 (0.02), 15.39 (0.02) & 18.21 (0.02), 16.47 (0.02), 15.18 (0.02) &17.88 (0.02), 16.17 (0.02), 14.91 (0.02) & SkyMapper,Pan-STARRS & - & 16.63 (0.03), 14.83 (0.01), 13.43 (0.01) & - & 15.80 (0.05), 13.92 (0.02), 12.42 (0.01) & VISTA & 15.14 (0.10), 13.55 (0.09) , 12.53 (0.08)& 14.88 (0.11), 13.33 (0.10), 12.34 (0.09) \\
2018hgc  & 27.78 (2.10), 25.43 (0.70), 24.30 (0.42) & 25.37 (0.42), 24.11 (0.25), 23.30 (0.17) & 24.83 (37.80), 23.75 (14.64), 22.04 (3.33) & -  & 27.02 (95.81), 24.23 (8.60), 22.69 (2.60) & 24.86 (11.99), 23.31 (3.77), 24.42 (2.15) & 24.36 (16.30), 22.72 (3.97), 21.60 (1.63) & 22.87 (2.87), -, 21.80 (2.53) & -  & SDSS & - & 19.79 (2.11), 18.71 (1.56), 18.53 (1.98) & 19.33 (3.25), 17.83 (1.63), 16.97 (1.11) & 19.17 (5.66), 17.88 (6.38), 17.55 (3.84) & 2MASS & 22.13 (4.52), 20.50 (3.64), 19.57 (3.41) & - \\
2018hhn & 23.84 (1.21), 22.00 (0.55), 21.52 (0.45) & 23.32 (0.24), 21.95 (0.13), 21.20 (0.10) & 24.16 (25.05), 21.56 (2.81), 20.87 (1.83) & - & 21.84 (1.44), 20.45 (0.60), 19.74 (0.42) & 21.45 (0.98), 20.01 (0.44), 19.26 (0.31) & 21.56 (1.67), 20.03 (0.56), 19.14 (0.33) & 21.13 (1.66), 19.59 (0.62), 19.07 (0.52) & -  & SDSS  & 20.89 (1.73), 19.56 (1.02), 19.01 (0.91) & - & -,-, 19.21 (1.98) & 19.35 (1.33), -, 19.35 (4.00) & UKIDSS & 19.08 (1.03), 17.53 (0.94), 16.59 (0.89) & 22.35 (45.41), 19.72 (15.52), 17.98 (7.00) \\
2018hjw  & 22.92 (0.95), 21.25 (0.46), 20.28 (0.30) & 21.88 (0.19), 20.39 (0.17), 19.57 (0.12) & 20.63 (0.98), 19.30 (0.40), 18.59 (0.27) & - & 19.57 (0.34), 18.22 (0.18), 17.53 (0.13) & 19.14 (0.27), 17.77 (0.14), 17.11 (0.10) & 18.86 (0.25), 17.49 (0.12), 16.83 (0.09) & 18.77 (0.25), 17.36 (0.12), 16.69 (0.09) & - & SDSS & - & 18.27 (0.73), 16.72 (0.35), 15.89 (0.24) & 17.28 (0.74), 15.84 (0.39), 15.12 (0.0.30) & 16.85 (0.73), 15.45 (0.40), 14.71 (0.31) & 2MASS  & 17.72 (0.28), 16.18 (0.25), 15.34 (0.25) & 17.83 (0.48), 16.20 (0.40), 15.31 (0.40) \\
2018ilu  & 21.80 (0.54), 20.61 (0.33), 20.12 (0.28) & 22.09 (0.32), 20.81 (0.18), 20.11 (0.13) & 22.60 (5.31), 20.57 (1.18), 19.87 (0.83) & - & 21.71 (1.24), 20.09 (0.51), 19.39 (0.37) & 21.67 (1.25), 19.83 (0.45), 19.18 (0.34) & 21.22 (1.16), 19.53 (0.41), 18.75 (0.28) & 21.25 (1.59), 19.93 (0.95), 19.35 (0.81) & - & SDSS & 20.50 (2.09), 18.91 (0.97), 18.04 (0.65) & 20.53 (3.67), 18.91 (1.65), 17.96 (1.03) & 20.78 (4.25), 18.59 (1.13, 17.60 (0.69)) & 20.23 (4.64), 17.76 (0.96), 16.97 (0.70) & UKIDSS & 18.97 (1.67), 18.12 (2.90), 18.26 (7.16) & - \\
2018jag  & - & 25.27 (1.12), 23.65 (0.56), 22.94 (0.41) & 23.41 (0.93), -, - & - & 23.54 (4.01), 22.08 (1.45), 21.35 (0.97) & 23.57 (3.90), 21.63 (1.08), 20.79 (0.69) & 23.38 (7.61), 21.80 (2.07), 20.48 (0.77) & 22.63 (3.11), 20.62 (0.86), 20.29 (0.88) & - & SDSS & -, -, 21.62 (8.18)  & - & - & 21.12 (5.11), 18.62 (1.02), 18.16 (1.00) & UKIDSS & 20.34 (1.47), 18.72 (1.23), 17.77 (1.14) & 20.44 (3.66), 18.82 (3.08), 17.85 (2.80) \\
2018jky  & 23.94 (0.35), 22.74 (0.21), 21.73 (0.14) & 23.17 (0.17), 21.65 (0.09), 20.78 (0.06) & - & - & - & -,25.03 (4.37), 22.86 (0.88) & - & 23.45 (0.90), 23.36 (1.66), 21.23 (0.35) & - & DES & - & - & -, 16.27 (1.92), 15.51 (1.45) & - & 2MASS & 15.10 (3.88), 18.20 (4.48), 17.19 (3.97) & - \\
2019gf  & - & - & - & - & 24.69 (1.00), 23.63 (0.75), 22.56 (0.42) & -, -, 23.58 (0.64) & 26.05 (1.69), 23.73 (0.40), 23.12 (0.34) & 23.73 (0.52), 23.15 (0.61), - & -, 22.21 (0.60), 22.71 (1.43) & SDSS,Pan-STARRS & -  & 20.81 (4.71), 19.31 (2.36), 18.42 (1.56) & -, 21.75 (46.95), 19.22 (6.49) & 20.34 (13.97), 18.89 (7.36), 18.34 (6.62) & 2MASS & - & - \\
2019jf  & - & - & 22.01 (0.86), 20.78 (0.56), 21.09 (1.11) & -, 22.39 (2.07), 21.41 (1.26) & 16.21 (0.80), 20.36 (0.03), 19.38 (0.02) & 21.42 (0.04), 19.78 (0.02), 18.84 (0.02) & 21.05 (0.04), 19.47 (0.02), 18.54 (0.02) & 21.06 (0.08), 19.36 (0.04), 18.40 (0.03) & 20.97 (0.14), 19.17 (0.06), 18.17 (0.04) & SkyMapper, Pan-STARRS & - & 19.60 (0.53), 18.16 (0.28), 17.28 (0.19) & - & 19.07 (0.65), 17.66 (0.35), 16.66 (0.21) & VISTA & 18.20 (0.36), 16.68 (0.32), 15.80 (0.31) & 18.18 (0.55), 16.62 (0.49), 15.68 (0.46) \\
2019rm  & 19.75 (0.33), 18.04 (0.16), 17.33 (0.12) & 18.78 (0.05), 17.45 (0.04), 16.81 (0.03) & - & - & - & - & - & 16.58 (0.11), 15.48 (0.07), 15.08 (0.07) & - & LegacySurvey & - & 16.17 (0.24), 14.80 (0.14), 14.18 (0.12) & 15.59 (0.35), 14.15 (0.19), 13.49 (0.15) & 15.84 (0.77), 14.08 (0.31), 13.46 (0.26) & 2MASS & 16.14 (0.20), 14.57 (0.19), 13.66 (0.19) & 15.85 (0.20), 14.31 (0.19), 13.42 (0.19) \\
2019so  & 21.16 (0.63), 19.76 (0.34), 18.89 (0.24)  & 20.78 (0.33), 19.58 (0.20), - & - & 18.58 (0.73), 18.07 (0.91), 16.95 (0.49) & 18.28 (0.30), 16.73 (0.15), 15.67 (0.08) & 17.78 (0.19), 16.24 (0.09), 15.12 (0.05) & 16.98 (0.08), 15.54 (0.05), 14.53 (0.03) & 16.66 (0.03), 15.11 (0.02), 14.08 (0.02) & -  & SkyMapper & - & 15.58 (0.02), 14.03 (0.01), 12.96 (0.01) & - & 14.68 (0.04), 13.10 (0.02), 12.04 (0.01) & VISTA & 14.15 (0.17), 12.60 (0.16), 11.63 (0.15) & 14.16 (0.20), 12.60 (0.18), 11.60 (0.17) \\
2019ahi  & 21.31 (0.57), 19.77 (0.29), 19.02 (0.21) & 20.84 (0.25), 19.41 (0.13), 18.76 (0.10) & - & - & 19.00 (0.02), 17.80 (0.02), 17.11 (0.02) & 18.78 (0.02), 17.56 (0.01), 16.86 (0.01) & 18.72 (0.02), 17.47 (0.02), 16.76 (0.02) & 18.65 (0.02), 17.42 (0.02), 16.70 (0.02) & 18.59 (0.03), 17.33 (0.02), 16.58 (0.02) & Pan-STARRS & - & 18.19 (0.22), 16.79 (0.16), 16.05 (0.15) & 17.27 (0.18), 15.95 (0.14), 15.33 (0.13) & 17.20 (0.28), 15.89 (0.17), 15.22 (0.15) & VISTA & 16.62 (0.19), 15.14 (0.18), 14.30 (0.19) & 16.45 (0.25), 14.97 (0.24), 14.13 (0.24) \\
2019akg  &  23.41 (0.28), 21.78 (0.14), 20.85 (0.10) & 22.67 (0.10), 21.22 (0.06), 20.37 (0.04) & 21.56 (2.19), 20.04 (0.70),19.26 (0.43) & - & 20.41 (0.56), 18.89 (0.25), 18.03 (0.17) & 19.97 (0.41), 18.42 (0.19), 17.54 (0.13) & 19.68 (0.41), 18.15 (0.18), 17.26 (0.11) & 19.45 (0.40), 17.89 (0.16), 17.02 (0.11) & - &  SDSS & 19.15 (0.29), 17.62 (0.14), 16.65 (0.09) & 19.39 (0.60), 17.79 (0.28), 16.72 (0.16) & 18.53 (0.28), 16.72 (0.10), 15.73 (0.06) & 18.00 (0.28), 16.40 (0.13), 15.31 (0.07) & UKIDSS & 17.88 (0.30), 16.25 (0.25), 15.36 (0.24) & 17.77 (0.48), 16.10 (0.39), 15.19 (0.37) \\
2019awq  & 25.23 (4.22), 22.97 (1.30), 21.55 (0.64) & 22.84 (0.55), 21.72 (0.34), 21.06 (0.26) & - & - & 20.51 (0.02), 19.06 (0.02), 18.21 (0.02) & 19.98 (0.03), 18.54 (0.02), 17.67 (0.02) & 19.67 (0.02), 18.19 (0.02), 17.32 (0.02) & 19.52 (0.02), 18.03 (0.02), 17.15 (0.02) & 19.28 (0.04), 17.84 (0.03), 16.96 (0.02) & Pan-STARRS & - & 18.49 (0.12), 17.06 (0.07), 16.15 (0.04) & - & 17.43 (0.18), 16.03 (0.11), 15.15 (0.07) & VISTA & 17.94 (0.32), 16.44 (0.29), 15.27 (0.24) & 17.88 (0.49), 16.36 (0.45), 15.19 (0.35) \\
2019bdz  & -, 24.29 (2.22), 22.83 (1.00) & 24.79 (1.07), 23.06 (0.50), 22.13 (0.33) & 22.39 (3.81), 21.36 (1.70), 20.53 (0.96) & - & 22.28 (4.05), 20.43 (0.87), 19.51 (0.45) & 21.48 (0.99), 19.81 (0.39), 18.91 (0.25) & 21.03 (1.04), 19.49 (0.37), 18.59 (0.22) & 21.15 (0.91), 19.50 (0.39) & - & SDSS & 20..59 (1.31), 18.88 (0.54), 18.17 (0.42) & 23.40 (11.91), 20.53 (1.70), 19.70 (1.18) & 22.91 (5.20), 21.36 (2.50), 19.39 (0.61) & 20.65 (1.25), 20.06 (1.46), 18.82 (0.70) & UKIDSS & 19.27 (0.70), 17.75 (0.64), 16.71 (0.57) & 19.03 (1.08), 17.53 (1.02), 16.56 (0.94) \\
2019bus  & -, -, 15.11 (0.01) & 25.61 (1.36), 24.11 (0.72), 23.35 (0.52) & 23.90 (13.29), 22.81 (5.16), 21.89 (2.43) & - & 23.71 (4.66), 22.32 (1.65), 21.11 (0.78) & 22.91 (2.43), 21.47 (0.92), 20.54 (0.55) & 23.16 (4.78), 21.52 (1.27), 20.37 (0.58) & 21.90 (1.17), 20.59 (0.60), 19.74 (0.40) & - &  SDSS & - & - & 21.65 (24.42), 19.88 (1.57), 18.87 (5.64) & - & 2MASS & 21.79 (1.57), 20.08 (1.06), 19.15 (0.97) & 23.72 (19.92), 21.19 (6.49), 20.16 (5.37) \\
2019cvl  & -, 23.08 (1.95), 21.53 (0.81) & 22.34 (0.43), 20.81 (0.22), 19.89 (0.14) & - & - & 22.09 (0.16), 20.08 (0.05), 18.90 (0.03) & 21.22 (0.09), 19.33 (0.03), 18.23 (0.02) & 21.22 (0.11), 19.38 (0.04), 18.23 (0.03) & 21.07 (0.18), 19.19 (0.07), 18.10 (0.04) & 23.92 (5.08), 20.35 (0.46), 18.72 (0.15) & Pan-STARRS & - & -, -, 20.62 (4.61) & - & - & VISTA & 17.41 (0.79), 15.76 (0.68), 14.68 (0.57) & 17.34 (1.15), 15.61 (0.92), 14.54 (0.78) \\
2019cxu  & 24.79 (3.31), 22.91 (1.32), 21.96 (0.85) & 23.50 (0.44), 22.42 (0.28), 21.67 (0.20) & - & - & 22.32 (0.09), 20.75 (0.04), 20.01 (0.03) & 22.03 (0.08), 20.49 (0.04), 19.91 (0.04) & 21.88 (0.08), 17.65 (1.30), 19.71 (0.04) & 21.86 (0.14), 20.25 (0.06), 19.53 (0.05) & 22.08 (0.35), 20.43 (0.15), 19.74 (0.12) & Pan-STARRS & - & 16.17 (10.30), 19.60 (0.51), 19.15 (0.50) & - & 19.57 (0.96), 17.72 (0.35), 17.36 (0.38) & VISTA & -, 20.79 (4.52), 19.62 (3.56) & -, -, 20.00 (13.77)  \\
2019cxx  & - & - & 20.94 (1.23), 19.39 (0.42), 18.56 (0.27) & - & 20.04 (0.43), 18.46 (0.20), 17.55 (0.13) & 19.63 (0.34), 18.06 (0.16), 17.12 (0.10) & 19.43 (0.33), 17.87 (0.15), 16.90 (0.09) &19.40 (0.32), 17-87 (0.16),16.85 (0.10) & -  & SDSS & 18.71 (0.28), 17.18 (0.14), 16.15 (0.08) & 18.56 (0.41), 17.20 (0.23), 16.06 (0.12) & 17.83 (0.20), 16.35 (0.10), 15.33 (0.06) & 17.71 (0.32), 16.06 (0.14), 15.05 (0.09) & UKIDSS & 18.39 (0.67), 16.77 (0.60), 15.83 (0.56) & 18.26 (1.38), 16.65 (1.23), 15.68 (1.13)\\
2019dks  & 23.52 (0.01), 22.90 (1.06),22.60 (1.00) & 25.18 (1.16), 24.34 (0.84), 23.82 (0.69) & - & - & 23.04 (0.12), 21.61 (0.06), 20.87 (0.05) & 22.92 (0.22), 21.44 (0.11), 20.69 (0.09) & 22.70 (8.30), 21.29 (0.06), 20.52 (0.05) & 22.41 (0.17), 21.32 0.12), 20.47 (0.09) & 22.59 (0.48), 21.24 (0.28), 20.56 (0.22) & Pan-STARRS  & - & -, 21.19 (1.66), 20.09 (0.90) & 23.68 (14.62), 20.08 (1.06), 19.21 (0.71) & 20.67 (1.97), 21.05 (5.57), 19.34 (1.73) & VISTA & 22.10 (3.27), 20.60 (2.85), 19.71 (2.76) & 22.00 (8.08), 20.48 (7.05), 19.57 (6.66)\\
2019eim  & -, 24.10 (1.21), 23.08 (0.79) & 24.96 (1.11), 23.30 (0.54), 22.30 (0.34) & - & - & 20.30 (1.43), 18.59 (0.60), 17.93 (0.48) & 19.64 (0.79), 17.85 (0.31), 17.19 (0.25) & - & 19.15 (0.51), 17.34 (0.20), 16.69 (0.16) & - & LegacySurvey  & 17.81 (0.08), 16.38 (0.04), 15.89 (0.04) & 17.28 (0.04), 15.89 (0.02), 15.45 (0.02) & - & 16.52 (0.09), 14.90 (0.06), 14.45 (0.06) & VISTA & 18.29 (0.34), 16.84 (0.32), 15.92 (0.31) &18.14 (0.44), 16.69 (0.42), 15.84 (0.42) \\
2019fcf  & -,-, 23.97 (3.24) & 24.02 (1.06), 22.91 (0.67), 22.36 (0.54) & - & - & 21.19 (0.04), 19.84 (0.03), 19.14 (0.02) & 20.65 (0.02), 19.30 (0.02), 18.57 (0.02) & 20.32 (0.02), 18.95 (0.02), 18.21 (0.02) & 20.17 (0.03), 18.83 (0.02), 18.07 (0.02) & 20.04 (0.06), 18.66 (0.04), 17.89 (0.03) & Pan-STARRS & - & 18.77 (0.83), 17.36 (0.45), 16.57 (0.33) & 18.43 (1.49), 16.91 (0.74), 16.08 (0.52) & 17.82 (1.46), 16.26 (0.70), 15.35 (0.45) & 2MASS & 18.08 (0.23), 16.62 (0.19), 15.76 (0.19) & 17.70 (0.22), 16.23 (0.19), 15.36 (0.183) \\
2019fmr  & 23.45 (0.25), 21.91 (0.13), 20.94 (0.09) & 22.99 (0.09), 21.40 (0.30), 20.43 (0.04) & - & - & 21.03 (0.03), 19.56 (0.02), 18.71 (0.01) & 20.70 (0.02), 19.21 (0.01), 18.32 (0.01) & 20.59 (0.02), 19.10 (0.01), 18.18 (0.01) & 20.51 (0.03), 18.98 (0.02), 18.05 (0.01) & 20.37 (0.08), 18.87 (0.04), 17.94 (0.03) & DES & - & 19.79 (0.45), 18.31 (0.23), 17.29 (0.13) & 18.96 (0.40), 17.63 (0.23), 16.78 (0.16) & 18.96 (0.80), 17.40 (0.38), 16.54 (0.26) & VISTA & 19.07 (1.02), 17.56 (0.97), 16.73 (1.00) & 21.22 (16.49), 17.78 (2.75), 16.57 (2.04)\\
2019gbx  & -, -, 22.06 (2.96) & 22.84 (0.41), 21.32 (0.21), 20.35 (0.13) & - & 19.14 (0.71), 18.64 (0.89), 18.32 (0.99) & 20.31 (1.02), -, 19.75 (1.82) & 21.47 (3.66), -, - & - & -, -, 1934 (1.46) & - & SkyMapper & - & - & 20.77 (4.37), 19.79 (3.53), 19.43 (3.82) & - & VISTA & 18.59 (4.63), 16.92 (3.93), 16.06 (4.01) & 18.93 (11.65), 18.83 (42.25), 16.69 (13.25) \\
2019gwa & - & 22.72 (0.30), 21.29 (0.16), 20.51 (0.12) & 21.14 (1.59), 19.83 (0.61), 19.21 (0.41) & - & 20.21 (0.50), 18.81 (0.25), 18.13 (0.18) & 19.76 (0.37), 18.35 (0.19), 17.65 (0.14) & 19.43 (0.34), 18.02 (0.16), 17.32 (0.12) & 19.27 (0.28), 17.84 (0.14), 17.13 (0.11) & - & SDSS & - & 17.90 (0.51), 16.54 (0.29), 15.86 (0.24) & 17.04 (0.48), 15.78 (0.30), 15.21 (0.27) & 16.90 (0.79), 15.60 (0.46), 14.89 (0.38) & 2MASS & 18.00 (0.23), 16.44 (0.19), 15.62 (0.18) & 17.89 (0.27), 16.33 (0.23), 15.51 (0.23) \\
\hline
\end{tabular}
}
\label{tab:host_table_local}
\end{sidewaystable}

\begin{sidewaystable}

\section{Light curve parameters}
\caption{Light curve parameters from SALT3 ($t_0,x_0,x_1,c$), SNooPy ($t_{max}$,$s_{BV}$,$E(B-V)_{host}$) and BayeSN (Peak$_{MJD}$,$\mu$) along with peak apparent magnitudes in $BJH$. The errors are in the parentheses.}
\resizebox{\textwidth}{!}{ 
\begin{tabular}{c|ccccccc|cccccc|ccccc}
\hline \hline
 & \multicolumn{7}{c|}{SNCosmo} & \multicolumn{6}{c|}{SNooPy} & \multicolumn{5}{c|}{BayeSN} \\
SN&$t_0$&$x_0\cdot 1000$\tablefootmark{a}&$x_1$&$c$&$B_{max}$&$J_{max}$&$H_{max}$&$tmax$& $sBV$&$E(B-V)_{host}$ &$B_{max}$&$J_{max}$&$H_{max}$ &Peak\_MJD& $\mu$  &$B_{max}$&$J_{max}$&$H_{max}$\\
\hline
2018rw   & 58163.51 (0.25) &  0.54 (0.02) & 0.47 (0.17) &  0.50 (0.02)  & 18.99 (0.04) & - & -  & 58161.22 (0.78) & 0.52 (0.06) & $-$0.02 (0.11) & 18.98 (0.13) & - &  - & 58158.59 (0.91) & 36.41 (0.95) & 18.26 (0.10) & - & - \\ 

2018yu   & 58194.86 (0.04) & 75.72 (1.23) &  $-$0.16 (0.06) & $-$0.11 (0.01) & 13.98 (0.02)  & 14.53 (0.02) & 14.68 (0.02) & 58194.58 (0.37) &  1.01 (0.04) & $-$0.04 (0.07) & 14.02 (0.08) & 14.36 (0.15) & 14.79 (0.22) & 58196.17 (0.26) & 32.83 (0.26) & 14.07 (0.04) & 14.29 (0.07) & 14.59 (0.06) \\ 

2018agk  & 58203.96 (0.11) &  3.33 (0.10) & $-$1.27 (0.21) &  $-$0.06 (0.02)  & 16.93 (0.03)  & 17.64 (0.03) & 17.85 (0.03) & 58204.30 (0.60) & 0.90 (0.06) & $-$0.49 (0.11) & 18.25 (0.07) & 17.08 (0.15) & 17.07 (0.11) & 58203.89 (0.51) & 35.51 (0.53) & 17.33 (0.12) & 17.00 (0.07) & 17.04 (0.05)  \\ 

2018aoz  & 58221.93 (0.10) & 99.82 (0.83)& $-$1.49 (0.06) & $-$0.13 (0.01)  & 12.82 (0.02) & 13.64 (0.02) & 13.81 (0.02) & 58222.20 (0.36) & 0.85 (0.03)  & $-$0.05 (0.06) & 12.85 (0.04) & 13.20 (0.13) & 13.65 (0.09)  & 58223.18 (0.15) & 31.71 (0.15) & 12.92 (0.02) & 13.24 (0.06) & 13.54 (0.05) \\ 

2018ayg  & 58238.85 (0.12) & 0.65 (0.02) &  $-$0.58 (0.15) &  0.67 (0.01) & 18.78 (0.10) & - & - & 58237.82 (0.41) & 0.38 (0.03)  & $-$0.38 (0.07)  & 17.21 (0.14) &  - & - & 58237.74 (0.34) & 35.99 (0.35) & 17.54 (0.05) & - & - \\ 

2018bie  & 58265.66	(0.02) &  7.38 (0.05) &  1.06 (0.03) &  0.02 (0.01) & 16.05 (0.01) & 16.51 (0.01) & 16.73 (0.01) & 58265.20 (0.40) & 1.12 (0.04) & 0.13 (0.06) & 16..03 (0.02) & 16.97 (0.32) & 16.62 (0.49) & 58265.32 (0.35) & 34.95 (0.36) & 15.94 (0.03) & 16.40 (0.08) & 16.66 (0.09) \\

2018bta  & 58263.35 (0.22) & 16.34 (0.02) & 0.39 (0.08) & 0.12 (0.01) & 15.58 (0.01) & 15.47 (0.01) & 15.63 (0.01)  & 58267.21 (0.52) & 0.61 (0.05) & $-$0.04 (0.09) & 15.36 (0.06) & 14.98 (0.13) & 15.24 (0.14)  & 58267.00 (0.31) & 33.51 (0.31) & 15.55 (0.03) & 15.20 (0.03) & 15.46 (0.03) \\ 

2018cow  & -               & -            &  -           & -            & - & - & - & -               & -           &  -            & - & - & - &-               & -            & - & - & - \\ 

2018dda  & 58314.28 (0.18) & 9.41 (0.14) &  $-$0.19 (0.10) &  0.17 (0.01) & 15.85 (0.02) & 15.83 (0.02) & 16.03 (0.02) &58312.43 (0.47) & 1.19 (0.04) & 0.17 (0.07) & 15.79 (0.07) & 15.49 (0.26) & 16.49 (0.43)  & 58312.52 (0.23) & 34.21 (0.23) & 15.61 (0.02) & 15.67 (0.06) & 15.94 (0.05)\\ 

2018ebu  & - & - & - &  - & -  & -  & - & 58328.70 (1.10) & 0.95 (0.07) & $-$0.15 (0.08) & 17.98 (0.12) & - & - & 58318.21 (8.60) & 38.03 (9.13) & 18.11 (0.05) & - & -\\ 

2018enc  & 58345.60	(0.05) &  10.00 (0.07) & 1.98 (0.07) & $-$0.09 (0.01) & 16.04 (0.01) & 16.55 (0.01) & 16.76 (0.01) &58344.55 (0.64) & 1.32 (0.05)  & $-$0.15 (0.07) & 16.12 (0.03) & 16.30 (0.07) & 16.78 (0.07) & 58346.18 (0.29) & 35.01 (0.30) & 16.04 (0.02) & 16.35 (0.06) & 16.80 (0.05) \\ 
 
2018eov  & 58342.43 (0.09) & 23.16 (0.30) & $-$1.93 (0.19) & $-$0.16 (0.01) & 15.47 (0.01) & 15.99 (0.01)  & 16.14 (0.01) & 58342.55 (0.61) & 0.68 (0.06)  & $-$0.26 (0.11) & 15.67 (0.08) & 15.71 (0.08) & 15.72 (0.12) & 58341.45 (0.18) & 34.46 (0.19) & 15.59 (0.03) & 15.72 (0.03) & 16.35 (0.05) \\ 

2018evt  & - &  - & - & - & - & - & - &  - & - & - &  - & - &  - & - & - & - & - & - \\ 

2018exb  & 58354.18 (0.04) &  6.45 (0.08) &  0.18 (0.08) & $-$0.29 (0.01) & 16.17 (0.01)  & 17.44 (0.01) & 17.74 (0.01) & 58353.11 (0.51) & 1.12 (0.08) & $-$0.20 (0.08) & 16.39 (0.05) & 17.35 (0.22) & 17.79 (0.20) & 58353.16 (0.20) & 36.32 (0.21) & 16.36 (0.04) & 17.44 (0.06) & 17.71 (0.05)  \\ 

2018exc  & 58355.66 (0.07) &  6.32 (0.01) &  1.34 (0.10) & $-$0.13 (0.01) & 16.22 (0.02) & 16.98 (0.02) & - &   58354.67 (0.41) & 1.09 (0.05) & $-$0.03 (0.07) & 16.27 (0.07) & 17.30 (0.17) & - & 58354.37 (0.34) & 35.61 (0.36) & 16.16 (0.04) & 16.86 (0.06) & 17.21 (0.08)  \\ 

2018feq  & -               & -            &  -           & -            & - & - & - &-               & -           &  -            & - & - &  - &-               & -             & - & - & - \\ 
 
2018hfp  & 58406.97 (0.07) &  5.48 (0.37) & 0.173 (0.44) &  0.04 (0.01) & 16.55 (0.01) & 16.82 (0.01) & 17.02 (0.01) & 58407.59 (0.61) & 0.88 (0.04) & 0.06 (0.07) & 16.63 (0.09) & 16.67 (0.21) & 18.77 (0.13) & 58405.37 (0.21) & 35.62 (0.21) & 16.54 (0.02) & 17.16 (0.07) & 17.75 (0.08) \\ 

2018hgc  & 58412.96 (0.07) &  2.61 (0.02) &  $-$0.01 (0.05) & 0.05 (0.01) & 17.21 (0.01) & 17.46 (0.01) & 17.70 (0.01) & 58412.87 (0.43) & 1.19 (0.04) & 0.17 (0.07) & 17.39 (0.04) & 17.32 (0.20) & 17.61 (0.13) & 58412.50 (0.29) & 36.05 (0.31) & 17.15 (0.03) & 17.51 (0.07) & 17.68 (0.07)  \\ 

2018hhn  & 58417.52 (0.04) &  5.54 (0.06) &  0.56 (0.05) &  $-$0.07 (0.01) & 17.09 (0.01) & 17.31 (0.01) & - & 58417.15 (0.53)) & 1.12 (0.05) & 0.02 (0.07) & 16.59 (0.04) & 16.77 (0.09) & - & 58416.75 (0.28) & 35.49 (0.29) & 16.50 (0.04) & 16.77 (0.05) & 17.04 (0.08) \\ 

2018hjw  & 58412.55 (0.01) &  3.48 (0.08) &  $-$0.40 (0.11) &  0.08 (0.01) & 16.87 (0.02)  & 17.09 (0.02) & 17.31 (0.02) & 58410.30 (0.97) & 1.19 (0.04) & $-$0.04 (0.08) & 16.19 (0.08) & 18.07 (0.46) & 17.49 (0.22) & 58414.87 (0.07) & 35.64 (0.08) & 15.97 (0.05) & 16.47 (0.08) & 17.05 (0.09)  \\ 

2018ilu  & 58450.58 (0.02) & 15.07 (0.10) &  0.73 (0.03) & $-$0.06 (0.01) & 15.42 (0.01) & 16.05 (0.01) & 16.24 (0.01) & 58449.89 (0.35) & 1.08 (0.03) & $-$0.01 (0.06) & 15.45 (0.01) & 15.75 (0.03) & 16.20 (0.04) & 58499.87 (0.20) & 34.43 (0.20) & 15.41 (0.02) & 15.80 (0.03) & 16.19 (0.03) \\ 

2018jag  & 58458.56 (0.72) &  0.40 (0.72) & 1.20 (1.90) & 0.63 (0.08) & 16.55 (0.20) & 16.82 (0.20) & 17.02 (0.20) &  58453.41 (0.87) & 0.47 (0.04) & 0.09 (0.11) & 18.25 (0.17) & 16.95 (0.24) & 17.90 (0.18) & 58453.41 (0.35) & 35.99 (0.37) & 17.90 (0.09) & 18.30 (0.13) & 18.27 (0.13) \\ 

2018jky  & 58468.71 (0.02) & 14.55 (0.09) & $-$2.45 (0.02) & 0.01 (0.01)  & 15.39 (0.01)  & 15.88 (0.01) & 16.05 (0.01) &  58468.67 (0.35) & 0.73 (0.03) & 0.05 (0.06) &  15.46 (0.01) & 15.50 (0.27) & 15.50 (0.21) & 58467.36 (0.14) & 34.08 (0.14) & 15.45 (0.02) & 15.45 (0.09) & 15.74 (0.08)  \\ 

2019gf  & 58499.11 (0.12) &  1.19 (0.02) &  1.06 (0.07) & $-$0.07 (0.01) & 16.55 (0.02) & 16.82 (0.02) & 17.02 (0.02) &  58498.87 (0.38) & 1.03 (0.04) & 0.01 (0.06) & 18.09 (0.02) & 18.79 (0.17) & 18.85 (0.12)  & 58498.60 (0.29) & 37.26 (0.29) & 17.90 (0.02) & 18.56 (0.08) & 18.75 (0.10) \\ 

2019jf   & 58501.48 (0.03) &  1.77 (0.01) & $-$2.00 (0.03) &  0.02 (0.01) & 17.66 (0.01) & 18.04 (0.01)  & - & 58501.79 (0.35) & 0.78 (0.03) & 0.11 (0.06) & 17.69 (0.02) & 18.99 (0.19) & -  & 58501.45 (0.18) & 36.28 (0.19) & 17.62 (0.02) & 17.71 (0.09) & - \\ 

2019rm  & 58503.14	(0.16) & 9.17 (0.11) &  0.59 (0.07) & 0.01 (0.02) & 16.15 (0.01) & 16.38 (0.01) & 16.56 (0.01) &58500.95 (1.03) & 0.45 (0.04) & 0.01 (0.07) & 15.94 (0.05) & 16.90 (0.24) & 16.19 (0.14) & 58503.92 (0.37) & 34.78 (0.37) & 15.99 (0.04) & 16.10 (0.09) & 16.35 (0.09) \\ 

2019so   & 58507.09 (0.02)  & 2.91 (0.16) & $-$1.66 (0.23) & 0.18 (0.01) & 17.51 (0.01) & 17.21 (0.01) & 17.34 (0.01) & 58506.94 (0.38) & 0.45 (0.04) & 0.01 (0.07) & 17.19 (0.03) & 16.10 (0.11) & 16.18 (0.08) & 58507.40 (0.12) & 34.15 (0.13) & 17.18 (0.02) & 16.10 (0.09) & 16.35 (0.09)  \\ 

2019ahi  & 58526.43 (0.08) &  3.67 (0.07) &  1.14 (0.09) &  0.13 (0.01) & 16.88 (0.02) & 16.96 (0.02) & 17.16 (0.02) & 58526.50 (0.47) & 1.11 (0.05) & 0.19 (0.07) &  16.72 (0.09) & 16.69 (0.09) & 16.84 (0.12) & 58525.45 (0.31) & 35.35 (0.32) & 16.69 (0.04) & 16.66 (0.04) & 16.84 (0.04)
\\ 

2019akg  & 58528.61 (0.07) &  2.09 (0.04) &  1.43 (0.12) & $-$0.26 (0.01) & 17.43 (0.02) & 18.62 (0.02) & 18.91 (0.02) & 58529.47 (0.79) & 1.11 (0.06) & 0.06 (0.10) & 17.60 (0.11) & 17.74 (0.75) & 18.28 (0.68) & 58528.89 (0.22) & 36.87 (0.23) & 17.47 (0.04) & 18.05 (0.08) & 18.61 (0.09) \\

2019awq  & 58539.96 (0.06) &  2.61 (0.03) &  $-$0.07 (0.07) &  0.13 (0.01) & 17.27 (0.01) & 17.28 (0.01) & 17.49 (0.01) &  58539.53 (0.42) & 1.19 (0.05) & 0.23 (0.07) & 17.23 (0.03) & 16.71 (0.17) & 17.25 (0.10)  & 58540.03 (0.33) & 35.58 (0.34) & 17.18 (0.02) & 17.07 (0.06) & 17.38 (0.05) \\ 

2019bdz  & 58550.21 (0.05) &  3.77 (0.02) & $-$0.72 (0.03) & $-$0.16 (0.01) & 16.78 (0.01)  & 17.74 (0.01) & 17.99 (0.01) & 58549.80 (0.39) & 0.92 (0.03) & $-$0.04 (0.06) & 16.79 (0.02) & 17.95 (0.28) & 17.93 (0.15) & 58552.13 (0.23) & 36.08 (0.24) & 16.78 (0.02) & 17.63 (0.06) & 17.79 (0.05) \\ 

2019bus  & 58569.40 (0.17) &  0.65 (0.01) &  1.97 (0.19) & $-$0.06 (0.01) & 18.66 (0.03) & - & - & 58568.76 (0.49) & 1.22 (0.04) & $-$0.05 (0.07) & 18.64 (0.07) & - & - & 58568.61 (0.33) & 37.91 (0.36) & 18.31 (0.04) & - & -  \\ 

2019cvl  & 58586.03 (0.07) &  8.27 (0.06) & 0.27 (0.07) & $-$0.06 (0.01) & 16.30 (0.01) & 16.70 (0.01) & 16.89 (0.01) & 58585.61 (0.39) & 0.97 (0.04) &  0.04 (0.06) & 16.25 (0.01) & 16.39 (0.07) & 16.86 (0.14) & 58586.24 (0.35) & 35.05 (0.36) & 16.24 (0.02) & 16.41 (0.04) & 16.73 (0.06)
\\ 

2019cxu  & 58586.35 (0.28)  & 0.55 (0.02) & $-$1.73 (0.40) & 0.37 (0.03) & 18.80 (0.05) & - & - & - & - & - & - & - & - & - & - & - & - & - \\ 

2019cxx  & 58593.66 (0.03) &  6.76 (0.06) & $-$0.14 (0.04) & $-$0.08 (0.01) & 16.12 (0.01) & 16.90 (0.01) & 17.13 (0.01) & 58593.22 (0.37) & 0.98 (0.03) & 0.05 (0.06) & 16.11 (0.04) & 16.85 (0.23) & 17.15 (0.10) & 58593.77 (0.26) & 35.25 (0.26) & 16.11 (0.04) & 16.61 (0.08) & 16.79 (0.08) \\ 

2019dks  & 58599.25 (0.08) &  1.79 (0.01) &  0.99 (0.10) & $-$0.10 (0.01) & 17.53 (0.02) & - & - &58599.03 (0.45) & 1.00 (0.04) & $-$0.01 (0.06) & 17.50 (0.02) & - & - & 58599.07 (0.39) & 36.86 (0.41) & 17.36 (0.02) & - & - \\

2019eim  & - & - &  - &  - & - & - & - & - & - & -  & - & - & - & - & - & - & - & - \\ 

2019fcf  & 58627.88 (0.05) &  1.03 (0.01) & $-$0.18 (0.06) & 0.12 (0.01) & 18.38 (0.01) &  - & - & 58627.55 (0.36) & 0.96 (0.03) & 0.29 (0.06) & 18.33 (0.01) & - & - & 58628.00 (0.27) & 36.69 (0.28) & 18.26 (0.04) &  - & -  \\

2019fmr  & 58632.82 (0.10)  & 19.94 (3.21) & 1.13 (0.08) & $-$3.45 (0.08) & 14.86 (0.37) & 16.75 (0.37) & 17.04 (0.37) & 58631.44 (1.65) & 1.28 (0.14) &  $-$0.88 (0.46) & - & 16.67 (0.30) & 16.81 (0.29) & 58632.25 (0.70) & 35.16 (0.72) & 15.78 (0.08) & 16.54 (0.05) & 16.81 (0.08) \\ 

2019gbx  & 58647.34 (0.01) & 27.57 (0.01) & $-$2.03 (0.01) & $-$0.14 (0.01) & 14.72 (0.01) & 15.65 (0.01) & 15.83 (0.01) & 58646.92 (0.36) & 0.85 (0.03) & $-$0.01 (0.06) & 14.68 (0.03) & 15.22 (0.15) & 15.69 (0.17) & 58647.78 (0.13) & 33.78 (0.13) & 14.84 (0.03) &  15.14 (0.03) & 15.39 (0.05)
 \\ 
 
2019gwa  & 58652.14 (0.03) & 1.69 (0.02) &  $-$0.09 (0.03) &  0.00 (0.01) & 17.76 (0.01) & - & - &  58651.51 (0.34) & 1.11 (0.04) & 0.20 (0.06) & 17.94 (0.03) & - &  - & 58651.78 (0.16) & 36.73 (0.14) & 17.87 (0.06) & - & - \\ 
\hline
\end{tabular}
}
\label{tab:lc_table}
\tablefoottext{a}{Notice that $x_0$ and its error have been multiplied by $10^3$.}
\end{sidewaystable}

\FloatBarrier
\section{Light curves}\label{sec:appendix_light_curves}
Presented here are the light curves for all SNe~Ia in our sample. 
Filters with the subscript $``\_ntt''$ refer to SOFI.

\begin{figure*}[!h]
    \centering
    \includegraphics[trim=3.2cm 4.9cm 3.6cm 6.1cm, clip=True,width=0.92\textwidth]{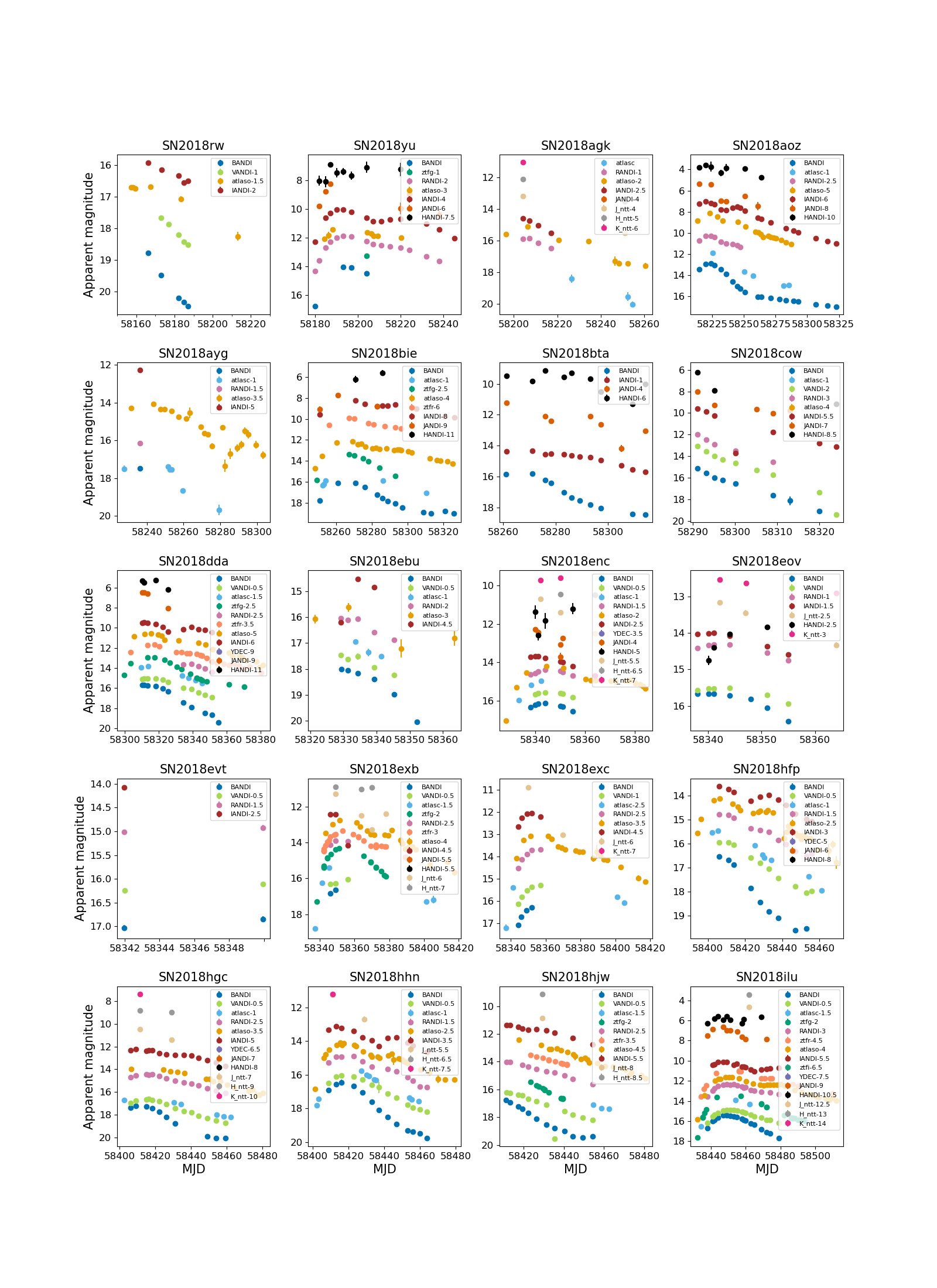}
    \caption{Final light curves.} 
\end{figure*}
\begin{figure*}\ContinuedFloat
    \centering
    \includegraphics[trim=3.2cm 4.9cm 3.6cm 6.1cm, clip=True,width=\textwidth]{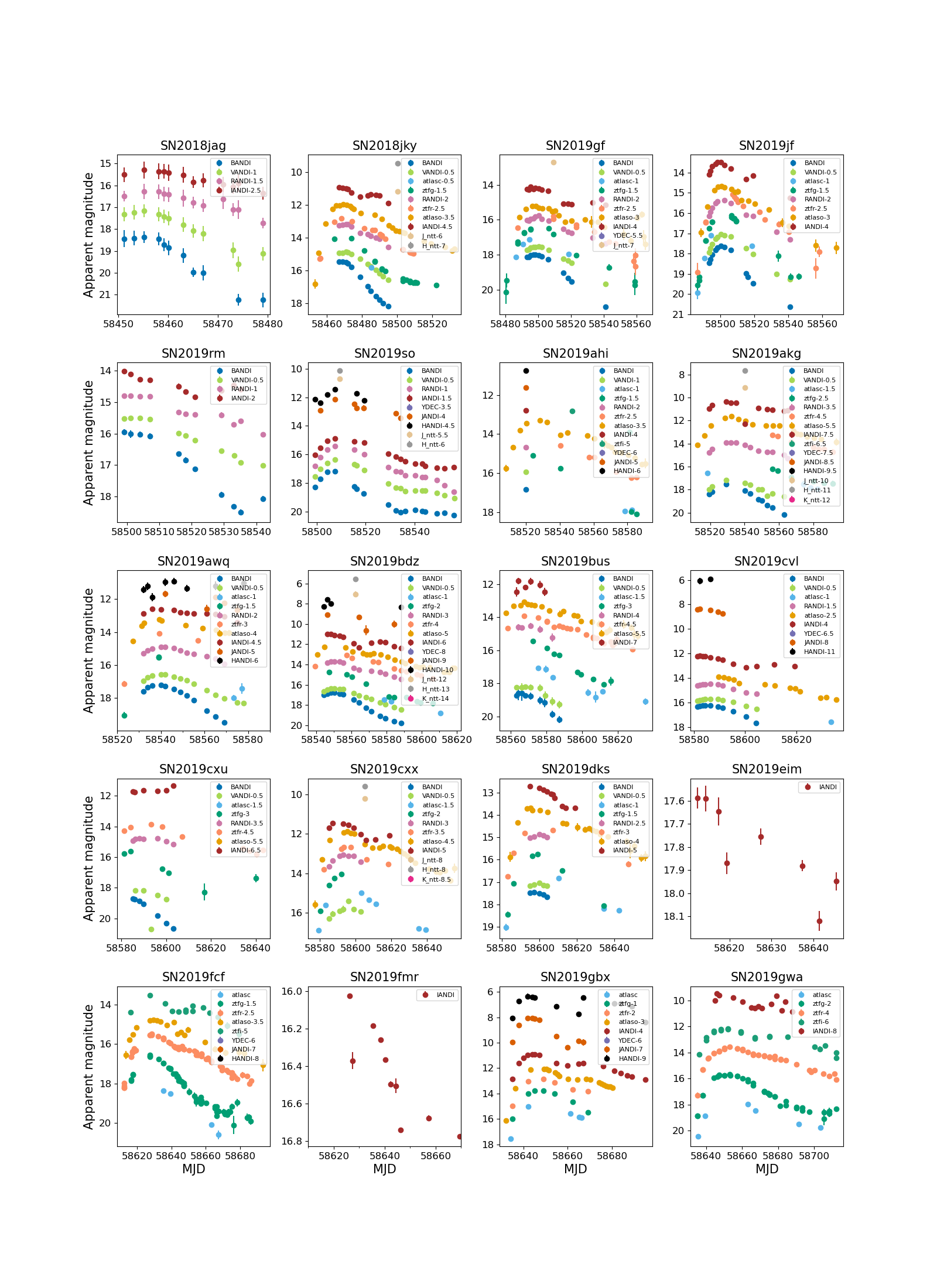}
    \caption{\it (cont.)}
\end{figure*}

\FloatBarrier
\section{Light-curve fitting examples}

\begin{figure*}[!h]
    \centering
    \includegraphics[width=\columnwidth]{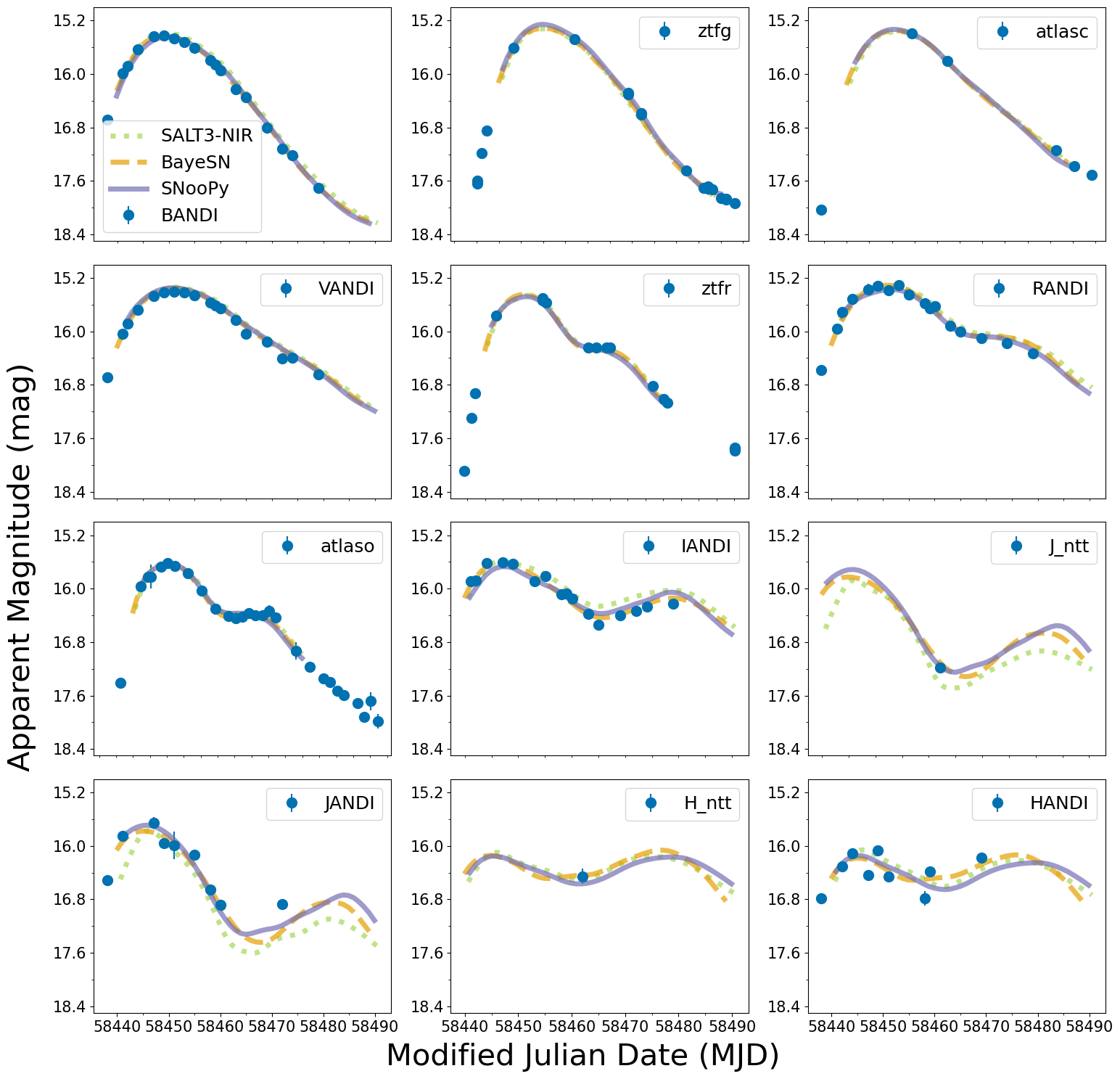}
    \caption{An example of light-curve fitting for SN~2018ilu in multiple filters using SALT3-NIR (green), BayeSN (orange), and SNooPy (purple).}
    \label{fig:lc_fits}
\end{figure*}

\FloatBarrier
\newpage
\section{Color terms}\label{sec:appendix_color_terms}

\begin{figure*}[!h]
    \centering
    \includegraphics[width=0.67\columnwidth]{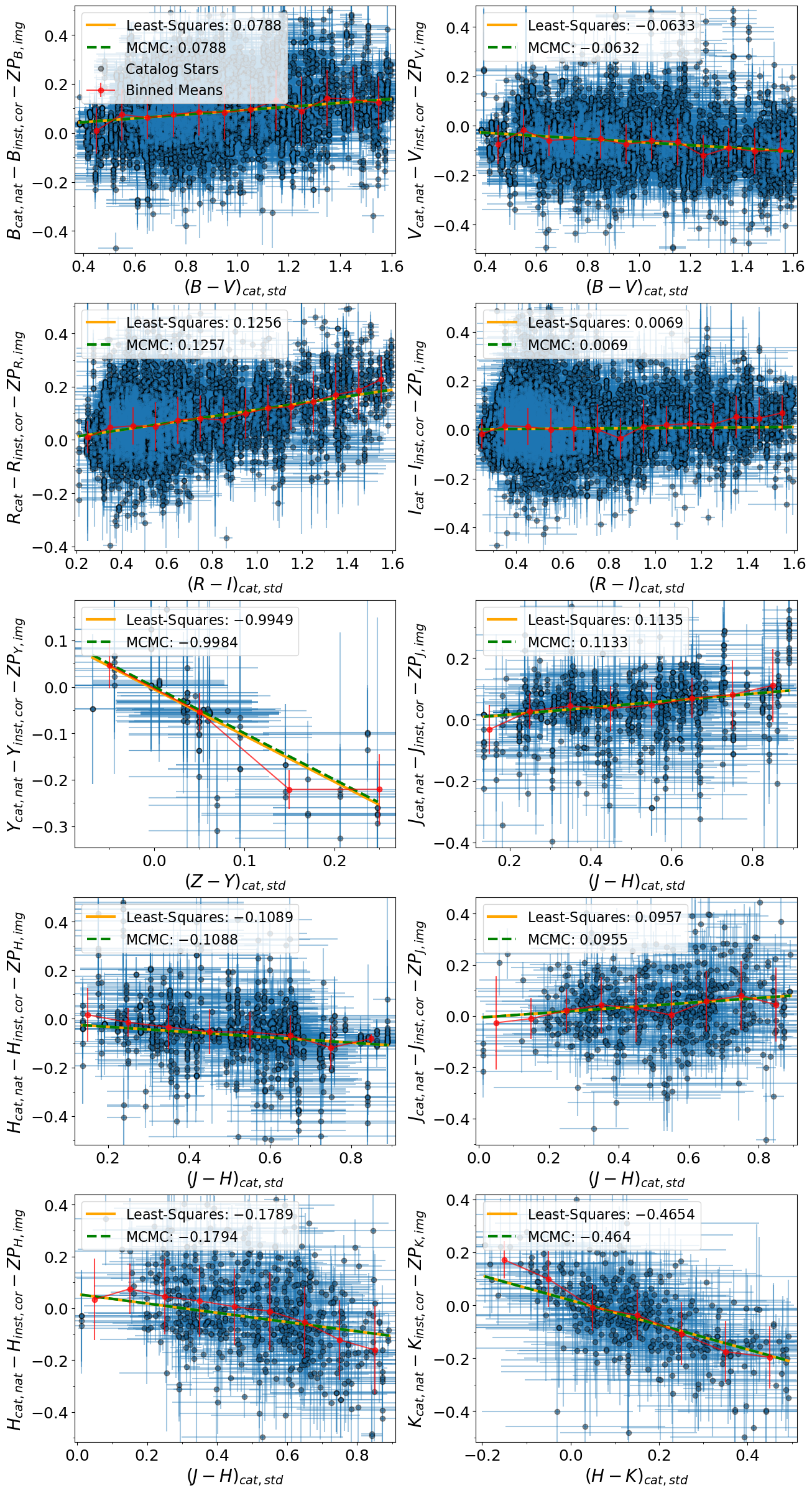}
    \caption{Same as Fig.~\ref{fig:emcee}, but for all ANDICAM and SOFI filters.} 
    \label{fig:color_terms}
\end{figure*}

\end{document}